\documentclass[12pt,a4paper]{article}

\usepackage{mathptmx,times}
\usepackage{amsfonts,amsthm,amsmath,url,cite}

\renewcommand{\frak}{\mathfrak}

\renewcommand{\Im}{\mathop{\rm Im}}
\renewcommand{\Re}{\mathop{\rm Re}}
\renewcommand{\tilde}{\widetilde}
\renewcommand{\det}{{\rm det}}

\newcommand{\Ker}{{\rm ker}}

\newcommand{\ran}{\mathop{\rm ran}}
\newcommand{\spec}{\mathop{\rm spec}}
\newcommand{\res}{\mathop{\rm res}}
\newcommand{\dist}{\mathop{\rm dist}}
\newcommand{\Span}{\mathop{\rm span}}
\newcommand{\supp}{\mathop{\rm supp}}

\newcommand{\rd}{\bot}

\newcommand{\indeg}{\mathop{\mathrm{indeg}}}
\newcommand{\outdeg}{\mathop{\mathrm{outdeg}}}

\newcommand{\ba}{{\mathbf a}}
\newcommand{\cA}{{\mathcal A}}

\newcommand{\ra}{{\mathrm a}}

\newcommand{\cB}{{\mathcal B}}

\newcommand{\rb}{{\mathrm b}}

\newcommand{\mC}{{\mathbb C}}

\newcommand{\rc}{{\mathrm c}}

\newcommand{\bD}{{\mathbf D}}

\newcommand{\cD}{{\mathcal D}}

\newcommand{\bG}{{\mathbf G}}

\newcommand{\cG}{{\mathcal G}}

\newcommand{\cH}{{\mathcal H}}

\newcommand{\cJ}{{\mathcal J}}

\newcommand{\rj}{{\mathrm j}}

\newcommand{\bK}{{\mathbf K}}

\newcommand{\bL}{{\mathbf L}}

\newcommand{\cL}{{\mathcal L}}

\newcommand{\cM}{{\mathcal M}}

\newcommand{\cN}{{\mathcal N}}
\newcommand{\mN}{{\mathbb N}}

\newcommand{\fn}{{\frak n}}

\newcommand{\cO}{{\mathcal O}}

\newcommand{\rp}{{\mathrm p}}

\newcommand{\cQ}{{\mathcal Q}}

\newcommand{\mR}{{\mathbb R}}

\newcommand{\rs}{{\mathrm s}}

\newcommand{\cV}{{\mathcal V}}

\newcommand{\mZ}{{\mathbb Z}}

\newcommand{\dom}{\mathop{\mathrm{dom}}}

\newcommand{\gr}{\mathop{\mathrm{gr\,}}}

\newtheorem{prop}{Proposition}
\newtheorem{lem}[prop]{Lemma}
\newtheorem{thm}[prop]{Theorem}
\newtheorem{corol}[prop]{Corollary}

\theoremstyle{remark}

\newtheorem{rem}[prop]{\bf Remark}

\theoremstyle{definition}

\newtheorem{defin}[prop]{Definition}

\newcommand{\id}{\mathop{\mathrm{id}}}

\numberwithin{prop}{section}
\numberwithin{equation}{section}

\begin{document}


\title{\bf Spectra of self-adjoint extensions and~applications to solvable
Schr\"odinger operators}

\author{\sc Jochen Br\"uning\dag \and \sc Vladimir Geyler\dag\ddag \and \sc Konstantin Pankrashkin\dag\S\\[\bigskipamount]
\dag Institut f\"ur Mathematik, Humboldt-Universit\"at zu Berlin\\
Rudower Chaussee 25, 12489 Berlin, Germany\\
\ddag Mathematical Faculty, Mordovian State University\\
430000 Saransk, Russia\\
\S D\'epartement de Math\'ematiques, Universit\'e Paris Nord\\
99 av. J.-B. Cl\'ement, 93430 Villetaneuse, France\\
Corresponding author, e-mail: const@mathematik.hu-berlin.de}

\date{}

\maketitle

\begin{abstract}
We give a self-contained presentation of the theory of self-adjoint extensions
using the technique of boundary triples. A description of the spectra of self-adjoint
extensions in terms of the corresponding Krein maps (Weyl functions) is given.
Applications include quantum graphs, point interactions, hybrid spaces,
singular perturbations.
\end{abstract}

\tableofcontents

\setcounter{section}{-1}

\section{Introduction}\label{sec1}

In recent two decades, the field of applications of explicitly
solvable models of quantum mechanics based on the operator
extension technique has been expanded considerably. New scopes are
presented e.g. in the Appendix by P.~Exner \cite{Exn1} to the
second edition of the monograph \cite{AGHH}, in the monograph by
S.~Albeverio and P.~Kurasov \cite{AK}, and in the topical issue of
the Journal of Physics~A \cite{DEG}. A review of papers dealing
with the theory of Aharonov--Bohm effects from the point of view
of the operator methods is contained in \cite{GS,MN}; new methods of analyzing singular
perturbations supported by sets with non-trivial geometry are
reviewed in \cite{EK2}. In addition, one should mention the use
of such models in the quantum field theory \cite{GTV, Hua},
including string theory \cite{Kaf}, quantum gravity \cite{Sol},
and quantum cosmology (see S.~P.~Novikov's comment in \cite{Gri2}
to results from \cite{Gri1}). Here the two-dimensional
$\delta$-like potential, which is a point supported perturbation,
is of considerable interest because in this case the Dirac
$\delta$-function has the same dimension as the Laplacian, and
this property leads to an effective non-perturbative
renormalization procedure removing the ultraviolet divergence
\cite{CWWH, Jac1, Jac2}. Another peculiarity of the
two-dimensional case -- so-called dimensional transmutation -- was
observed in \cite{CW,CEFG}. The operator
extension technique allows to build ``toy models'' which help
better understanding some phenomena in various fields of
mathematics and theoretical physics; as typical examples we
mention here the spectral theory of automorphic functions
\cite{BG1} or renormalization group theory \cite{AFG}. This
technique is applicable not only to self-adjoint operators, it can
be used, e.g. in investigating dissipative and accumulative
operators as well \cite{KNW}.

Very important applications of the operator extension theory have
been found recently in the physics of mesoscopic systems like
heterostructures \cite{GZAA}, quantum graphs \cite{KS1, KS2, Kuc,
Nov} and circuits \cite{Ada}, quantum wells, dots, and wires
\cite{Hur}. It should be stressed that in this case, the
corresponding results are not only of qualitative character, but
allow to give a good quantitative explanation of experimental data
(see e.g. \cite{BDGK,HR}) or explain some discrepancy
between experimental data and standard theories \cite{BDGT}.

Among the most popular ways of using singular perturbations in
the physics literature one should mention first of all various renormalization procedures
including the Green function renormalization and cut-off
potentials in the position or momentum representations (see
\cite{AGHH} and an informative citation list in \cite{PY}).
Berezin and Faddeev \cite{BF} were first who showed that the renormalization
approach to singular perturbations is equivalent to searching
for self-adjoint extensions of a symmetric operator related to the
unperturbed operator in question. At the same time, the
mathematical theory of self-adjoint extensions is
reduced as a rule to the classical von Neumann description through unitary operators
in deficiency spaces, which makes its practical use rather
difficult. In many cases self-adjoint operators arise when one
introduces some boundary conditions for a differential expression
(like boundary conditions for the Laplacian in a domain), and it
would be useful to analyze the operators in terms of boundary
conditions directly. Such an approach is common in the physics
literature \cite{BZP, DO}. In the framework of the abstract
mathematical theory of self-adjoint (or, more generally,
dissipative) extensions this approach is widely used in the
differential operator theory (see e.g., \cite{DS,GG,Gru} and the
historical as well as the bibliographical comments therein). Moreover,
there is a series of quantum mechanics problems related to the influence
of topological boundaries, and in this case the above approach is the
most adequate \cite{AIM}.

On the other hand, Berezin and Faddeev pointed out that the
standard expressions for the Green functions of singularly
perturbed Hamiltonians obtained by the renormalization procedure
can be easy derived from the so-called Krein resolvent formula
\cite{BF}. In the framework of the theory of explicitly solvable
models with an internal structure, an elegant way to get the Krein
resolvent formula with the help of abstract boundary conditions
has been proposed by Pavlov \cite{Pav} (see also \cite{APav}),
which was applied to the study of numerous applications, see e.~g. \cite{Pav1,Pav2,Pav3,Pav4,Pav5}.
A machinery of self-adjoint extensions using abstract boundary condisitons
is presented in a rather detailed form
in the monograph \cite{GG}, but only very particular questions
of the spectral theory are adressed.
A systematic theory of self-adjoint extensions in terms of boundary
conditions, including the spectral analysis, was developed by
Derkach and Malamud, who found, in particular, a nice relationship
between the parameters of self-adjoint extensions and the Krein
resolvent formula, and performed the spectral analysis in terms of
the Weyl functions; we refer to the paper \cite{DM} summarizing
this machinery and containing an extensive bibliography.

Nevertheless, one has to admit that the spectral analysis of
self-adjoint extensions in such terms is a rarely used tool in the
analysis of quantum-mechanical Hamiltonians, especially for
operators with infinite deficiency indices. On the other hand,
the authors' experience show that the application of the Krein
resolvent formula in combination with the boundary values for
self-adjoint extensions can advance in solutions of some problems
related to the applications of singular perturbations \cite{BEG,
BGL, BGP}. Therefore, it is useful to give a self-contained
exposition of the abstract technique of boundary value problems
and to analyze some models of mathematical physics using this
machinery. This is the first aim of the present paper.

Using the Krein resolvent formula, it is possible often to reduce
the spectral problem for the considered perturbed operator to a
problem of finding the kernel of an analytic family of operators
-- so-called Krein $\cQ$-function -- with more simple structure in
comparison with the operator in question. Therefore, it would be
useful to find relations between various parts of the spectrum of
the considered operators and the corresponding parts of the
spectrum of $\cQ$-functions. The second aim of the paper is to
describe these relations in a form suitable for applications.
Using the corresponding results, we obtain, in particular, new
properties of the spectra of equilateral quantum graphs and arrays
of quantum dots. Of course, we believe that the technique
presented here can be used to analyze much more general systems.
It is worth noting that this problem was considered in \cite{BrMN},
but the main results were obtained in a form which is difficult to use
for our applications.

In Section~\ref{sec2} we describe the machinery of boundary triples and their applications
to self-adjoint extensions. The most results in this section are not new
(we give the corresponding references in the text), but we do not know any work
where this theory was presented with complete proofs, hence we decided to do it here.
We also relate the technique of boundary triples
with the so-called Krein $\cQ$-functions and $\Gamma$-fields. Some of our definitions
are slightly different from the commonly used ones (although we show later
that the both are equivalent); this is motivated by applied needs.
We conclude the section by several examples showing that the machinery
of boundary triples include the well known situations like singular
perturbations, point perturbation, hybrid spaces.
Section~\ref{sec3} is a summary of a necessary information about the spectra
and spectral measures of self-adjoint operators. In Section~\ref{sec4}
we provide the spectral analysis of self-adjoint extensions with the help
of the Krein $\cQ$-functions. In particular, we analyze the discrete and essential spectra,
and carry out a complete spectral analysis
for a special class of $\cQ$-functions, which includes the recently introduced scalar-type
functions~\cite{ABMN}; these results are new.
Using these results we analyze two classes of quantum-mechanical models:
equilateral quantum graphs and arrays of quantum dots, where
we perform the complete dimension reduction and describe the spectra
of continuous models completely in terms of the associated tight-binding Hamiltonians.
Section~\ref{sec5} is devoted to the study of isolated eigenvalues
of self-adjoint extensions and generalizes previously known results
to the case of operators with infinite deficiency indices.

\bigskip

The second named author, Vladimir Geyler, passed away on April 2, 2007,
several days after the completion and the submission of the manuscript.
His untimely death has become a great loss for us.

\section{Abstract self-adjoint boundary value problems}\label{sec2}

In this section we describe the theory of self-adjoint extensions
using abstract boundary conditions. Some theorems here are not new,
but the existing presentations are spread through the literature,
so we decided to provide here the key ideas with complete proofs.

\subsection{Linear relations}
Here we recall some basic facts on linear relations. For a more
detailed discussion we refer to~\cite{Are}. Let $\cG$ be a Hilbert
space. Any linear subspace of $\cG\oplus\cG$ will be called a
\emph{linear relation} in $\cG$. For a linear relation $\Lambda$
in $\cG$ the sets
\begin{align*}
\dom\Lambda:&=\{x\in \cG:\,\exists y\in \cG
\text{ with } (x,y)\in\Lambda)\},\\
\ran \Lambda:&=\{
x\in \cG:\,\exists y\in \cG \text{ with } (y,x)\in\Lambda) \},\\
\ker\Lambda:&=\{x\in \cG: (x,0)\in\Lambda\}
\end{align*}
will be called the \emph{domain}, the \emph{range}, and the \emph{kernel} of~$\Lambda$,
respectively.  The linear relations
\begin{align*}
\Lambda^{-1}&=\{ (x,y)\in\cG\oplus\cG:\, (y,x)\in\Lambda\},\\
\Lambda^*&=\{(x_1,x_2)\in\cG\oplus\cG:\, \langle x_1|\,y_2\rangle=\langle
x_2|y_1\rangle \quad\forall (y_1,y_2)\in\Lambda\}
\end{align*}
are called \emph{inverse} and \emph{adjoint} to $\Lambda$, respectively.
For $\alpha\in\mC$ we put
\[
\alpha \Lambda=\{(x,\alpha
y):\,(x,y)\in\Lambda\}.
\]
For two linear relations
$\Lambda',\Lambda''\subset \cG\oplus \cG$ one can define their \emph{sum}
\[
\Lambda'+\Lambda''=\{(x,y'+y''):\,(x,y')\in\Lambda',\,
(x,y'')\in\Lambda''\};
\]
clearly, one has $\dom
(\Lambda'+\Lambda'')=\dom\Lambda'\cap\dom\Lambda''$. The graph of
any linear operator $L$ with domain in $\cG$ is a linear relation,
which we denote by $\gr L$. Clearly, if $L$ is invertible, then
$\gr L^{-1}=(\gr L)^{-1}$. For arbitrary linear operators $L',L''$
one has $\gr(\alpha L)=\alpha\gr L$ and $\gr L'+\gr L''=\gr
(L'+L'')$. Therefore, the set of linear operators has a natural
``linear'' imbedding into the set of linear relations. Moreover,
if $L$ is a densely defined closable operator in $\cG$, then $\gr
L^*=(\gr L)^*$, hence, this imbedding commutes with the
star-operation.

In what follows we consider mostly only closed linear relations,
i.e. which are closed linear subspaces in $\cG\oplus\cG$. Clearly,
this generalizes the notion of a closed operator. Similarly
to operators, one introduces the notion of the \emph{resolvent
set} $\res\Lambda$ of a linear relation $\Lambda$. By definition,
$\lambda\in\res\Lambda$ if $(\Lambda-\lambda I)^{-1}$ is the graph
of a certain everywhere defined bounded linear operator (here
$I\equiv \gr \mathop{\mathrm{id}_\cG}=\big\{(x,x):\,
x\in\cG\big\}$); this operator will be also denoted
as $(\Lambda-\lambda I)^{-1}$. Due to the closed graph theorem,
the condition $\lambda\in\res\Lambda$ exactly means that
$\Lambda$ is closed, $\ker(\Lambda-\lambda I)=0$, and $\ran(\Lambda-\lambda I)=\cG$.
The \emph{spectrum} $\spec\Lambda$ of $\Lambda$ is defined as
\[
\spec\Lambda:=\mC\setminus\res\Lambda\,.
\]

A linear relation $\Lambda$ on $\cG$ is called
 \emph{symmetric} if $\Lambda\subset
\Lambda^*$ and is called \emph{self-adjoint} if
$\Lambda=\Lambda^*$. A linear operator $L$ in $\cG$ is symmetric
(respectively, self-adjoint) if and only if its graph is a
symmetric (respectively, self-adjoint) linear relation. A
self-adjoint linear relation
is always maximal symmetric, but the converse in not true;
examples are given by the graphs of maximal symmetric operators
with deficiency indices $(\fn,0)$, $\fn>0$.

To describe all self-adjoint linear relations we need the
following auxiliary result.
\begin{lem}\label{lem-AB}
Let $U$ be a unitary operator in $\cG$. Then the operator
$M:\cG\oplus\cG\to\cG\oplus\cG$,
\begin{equation}
         \label{eq-MU}
M=\frac{1}{2}\begin{pmatrix}
i(1+U) & U-1\\
1-U & i(1+U)
\end{pmatrix}
\end{equation}
is unitary; in particular, $0\in\res M$.
\end{lem}

\begin{proof}
The adjoint operator $M^*$ has the form
\[
M^*=\frac{1}{2}\begin{pmatrix}
-i(1+U^*) & 1-U^*\\
U^*-1 & -i(1+U^*)
\end{pmatrix}\,,
\]
and it is easy to show by direct calculation that $M^*=M^{-1}$.
\end{proof}

\begin{thm}
           \label{prop-iu}
A linear relation $\Lambda$ in $\cG$ is self-adjoint
iff there is a unitary operator $U$ in $\cG$ \textup{(}called the \emph{Cayley transform}
of $\Lambda$\textup{)} such that
\begin{equation}
    \label{eq-LU}
\Lambda=\big\{(x_1,x_2)\in\cG\oplus\cG:\,i(1+U)x_1=(1-U)x_2\big\}.
\end{equation}
\end{thm}
Writing $U$ in the form $U=\exp(-2iA)$ with a self-adjoint operator $A$
one can reformulate theorem~\ref{prop-iu} as follows:
\begin{corol}
A linear relation $\Lambda$ in $\cG$ is self-adjoint
iff there is a self-adjoint operator $A$ acting in $\cG$ such that
$\Lambda=\big\{(x_1,x_2)\in\cG\oplus\cG:\,\cos A\,x_1=\sin A\,x_2\big\}$.
\end{corol}

To prove theorem~\ref{prop-iu} we need the following lemma.
\begin{lem}\label{lem-LL}
Let $U$ be a unitary operator in $\cG$
and $\Lambda$ be defined by \eqref{eq-LU}, then
\begin{equation}
    \label{eq-LU2}
\Lambda=\Big\{\big((1-U)x,i(1+U)x\big):\,x\in\cG\Big\}.
\end{equation}
\end{lem}

\begin{proof}[\bf Proof of lemma~\ref{lem-LL}]
The linear relation $\Lambda$ given by \eqref{eq-LU} is closed
as it is the null space of the bounded operator
\[
\cG\oplus\cG\ni(x_1,x_2)\mapsto i(1+U)x_1-(1-U)x_2\in\cG.
\]
Denote the set on the right-hand side of \eqref{eq-LU2} by $\Pi$.
Clearly, $\Pi\subset\Lambda$. By~lemma~\ref{lem-AB}, the operator
$M^*$ adjoint to $M$ from \eqref{eq-MU} maps closed sets to closed sets. In
particular, the subspace $\Pi\equiv M^*(0\oplus\cG)$ is closed.
Assume that there exists
$(y_1,y_2)\in\Lambda$ such that $(y_1,y_2)\perp \Pi$.
The condition $(y_1,y_2)\in\Lambda$ reads as
$i(1+U)y_1-(1-U)y_2=0$, and $(y_1,y_2)\perp \Pi$ means that
 $\langle y_1|\,(1-U)x\rangle+\langle
y_2|\,i(1+U)x\rangle=0$ for all $x\in\cG$, i.e.
that $(U-1)y_1-i(1+U)y_2=0$. This implies $M(y_1,y_2)=0$. By
lemma~\ref{lem-AB}, $y_1=y_2=0$. The requested equality
$\Lambda=\Pi$ is proved.
\end{proof}

\begin{proof}[\bf Proof of theorem~\ref{prop-iu}]
(1) Let $U$ be a unitary operator in $\cG$
and $\Lambda$ be defined by \eqref{eq-LU}.
By lemma~\ref{lem-LL} one can represent $\Lambda$
in the form \eqref{eq-LU2}. Using this representation
one easily concludes that $\Lambda\subset\Lambda^*$, i.e. that $\Lambda$ is symmetric.

Let $(y_1,y_2)\in\Lambda^*$. The equality $\langle
x_1|\,y_2\rangle=\langle x_1|\,y_2\rangle$ for all
$(x_1,x_2)\in\Lambda$ is equivalent to $\langle
(1-U)x|\,y_2\rangle=\langle i(1+U)x|\,y_1\rangle$ for all
$x\in\cG$, from which $-i(1+U^{-1})y_1=(1-U^{-1})y_2$ and
$i(1+U)y_1=(1-U)y_2$, i.e. $(y_1,y_2)\in\Lambda$. Therefore,
$\Lambda^*\subset\Lambda$, which finally results in
$\Lambda=\Lambda^*$.

(2) Let $\Lambda$ be a self-adjoint linear relation in $\cG$. Set
$L_\pm:=\{x_1\pm ix_2:\,(x_1,x_2)\in\Lambda\}$. Assume that for
some $(x_1,x_2)$ and $(y_1,y_2)$ from $\Lambda$ one has $x_1+i
x_2=y_1+iy_2$, then $(x_1-y_1,x_2-y_2)\in\Lambda$ and
$x_1-y_1=-i(x_2-y_2)$. At the same time, $0=\Im\langle
x_1-y_1|\,x_2-y_2\rangle=\Im \langle
-i(x_2-y_2)|\,(x_2-y_2)\rangle= \Im i\|x_2-y_2\|^2$, therefore,
$x_2=y_2$ and $x_1=y_1$. In the same way one can show that from
$x_1-ix_2=y_1-iy_2$, $(x_1,x_2), (y_1,y_2)\in\Lambda$, it follows
that $x_1=y_1$ and $x_2=y_2$. For $x_1+ix_2$ with
$(x_1,x_2)\in\Lambda$ set $U(x_1+ix_2)=x_1-ix_2$. Clearly,
$U:L_+\to L_-$ is well-defined and bijective. Moreover,
$\|U(x_1+ix_2)\|^2=\|x_1\|^2+\|x_2\|^2=\|x_1+ix_2\|^2$, i.e. $U$
is isometric.

Show that $U$ is actually a unitary operator, i.e. that
$L_\pm=\cG$. We consider only $L_+$; the set $L_-$ can be
considered exactly in the same way. Assume that $y\perp L_+$ for
some $y\in\cG$, then $\langle x_1+ix_2|\,y\rangle=\langle
x_1|\,y\rangle-\langle x_2|\,iy\rangle=0$ for all
$(x_1,x_2)\in\Lambda$. It follows that
$(iy,y)\in\Lambda^*=\Lambda$, which implies $\Im\langle
iy|\,y\rangle=-\Im i\|y\|^2=0$, i.e. $y=0$. Therefore,
$\overline{L_+}=\cG$. To show that $L_+$ is closed we take an
arbitrary sequence $(x_1^n,x_2^n)\in\Lambda$ with $\lim
(x_1^n+ix_2^n)=y$ for some $y\in\cG$, then automatically $\lim
(x_1^n-ix_2^n)=y'$ for some $y'\in\cG$, and
\[
\lim x_1^n=\dfrac{1}{2}\,(y+y') =:y_1\text{ and }
\lim x_2^n=\dfrac{1}{2i}\,(y-y')=:y_2.
\]
As we see, the sequence $(x_1^n,x_2^n)$ converges,
and the limit $(y_1,y_2)$ lies in $\Lambda$ as $\Lambda$
is closed. Therefore, $y=y_1+iy_2$ lies in $L_+$, $L_+$ is closed,
and $U$ is unitary.

Clearly, by construction of $U$,
$\Lambda$ is a subset of the subspace
on the right-hand side of \eqref{eq-LU}.
As shown in item (1), the latter is self-adjoint
as well as $\Lambda$ is, therefore, they
coincide.
\end{proof}

Theorem \ref{prop-iu} gives only one possible way
for parameterizing linear relations with the help
of operators. Let us mention some other ways to to this.

\begin{prop}\label{prop-AB}
Let $A$ and $B$ be bounded linear operators in $\cG$.
Denote $\Lambda:=\big\{(x_1,x_2)\in\cG\oplus\cG:\,Ax_1=Bx_2\big\}$.
$\Lambda$  is self-adjoint iff the following
two conditions are satisfied:
\begin{subequations}
\begin{gather}
   \label{eq-AB1}
AB^*=BA^*,\\
   \label{eq-AB2}
\ker \begin{pmatrix} A & -B\\ B & A \end{pmatrix}=0.
\end{gather}
\end{subequations}
\end{prop}

\begin{proof}
Introduce operators $L:\cG\oplus\cG\ni(x_1,x_2)\mapsto Ax_1-Bx_2\in\cG$
and $J:\cG\oplus\cG\ni(x_1,x_2)\mapsto (-x_2,x_1)\in\cG\oplus \cG$.
There holds $\Lambda^*=J(\Lambda^\perp)$ and $\Lambda=\ker L$.

Let us show first that the condition \eqref{eq-AB1} is equivalent to the inclusion
$\Lambda^*\subset\Lambda$. Note that this inclusion is equivalent to
$J(\Lambda^\perp)\subset \Lambda$ or, due to the bijectivity of $J$, to
\begin{equation}
      \label{eq-JL}
\Lambda^\perp\subset J\Lambda.
\end{equation}
Clearly, $\Lambda\equiv\ker L$ is closed, therefore, by the well known relation,
$\Lambda^\perp=\ker L^\perp=\overline{\ran L^*}$. As $\Lambda$ is closed,
the condition~\eqref{eq-JL} is equivalent to
\begin{equation}
      \label{eq-LsL}
\ran L^*\subset J(\ker L).
\end{equation}
Noting that $L^*$ acts
as $\cG\ni x\mapsto (A^*x,-B^*x)\in\cG\oplus\cG$, we see
that \eqref{eq-LsL} is equivalent to \eqref{eq-AB1}.

Now let $\Lambda$ be self-adjoint, then $J(\Lambda^\perp)=\Lambda$
or, equivalently, $J(\Lambda)=\Lambda^\perp\equiv\ker L^\perp$.
Therefore, the restriction of $L$ to $J(\Lambda)$ is injective.
This means that the systems of equations $Lz=0$, $LJz=0$
has only the trivial solution,
which is exactly the condition \eqref{eq-AB2}.

On the other hand, if \eqref{eq-AB1} and \eqref{eq-AB2} are satisfied, then,
as shown above, $\Lambda^\perp\subset J(\Lambda)$. If
$\Lambda^\perp\ne J(\Lambda)$, then $J(\Lambda)$ contains a non-zero element
of $(\Lambda^\perp)^\perp\equiv \Lambda=\ker L$, i.e.
there exists $z\ne 0$ with $Lz=0$ and $LJz=0$, which contradicts~\eqref{eq-AB2}.
\end{proof}

For a finite-dimensional $\cG$ the condition~\eqref{eq-AB2} simplifies,
and one arrives at
\begin{corol}
Let $\cG$ be finite dimensional, $A,B$ be linear operators in $\cG$.
The linear relation $\Lambda:=\big\{(x_1,x_2)\in\cG\oplus\cG:\,Ax_1=Bx_2\big\}$
is self-adjoint iff the following two conditions are satisfied:
\begin{subequations}
\begin{gather}
   \label{eq-AB1f}
AB^*=BA^*,\\
   \label{eq-AB2f}
\det(AA^*+BB^*)\ne 0 \quad\Leftrightarrow\quad \text{the block matrix }
(A|B) \text{ has maximal rank}.
\end{gather}
\end{subequations}
\end{corol}

The conditions~\eqref{eq-AB1}, \eqref{eq-AB2},    \eqref{eq-AB1f},
\eqref{eq-AB2f} can be rewritten in many equivalent forms, see
e.g.~\cite[Section~125]{AG}, \cite{BG,DM2,RB,Pan1}.

\subsection{Boundary triples for linear operators}

\begin{defin}\label{defin1}
Let $A$ be a closed linear operator in a Hilbert space $\cH$ with the domain $\dom A$.
Assume that there exist
another Hilbert space $\cG$ and two linear maps $\Gamma_1,\Gamma_2:\dom A\to \cG$
such that:
\begin{subequations}
\begin{gather}
        \label{G1}
\langle f|\,Ag\rangle-\langle Af|\,g\rangle=\langle\Gamma_1 f|\,\Gamma_2 g\rangle-
\langle\Gamma_2 f|\,\Gamma_1 g\rangle \text{ for all }f,g\in\dom A,\\
       \label{G2}
\text{the map }(\Gamma_1,\Gamma_2):\dom A\to\cG\oplus\cG \text{ is surjective,}\\
       \label{G3}
\text{the set }\ker\,(\Gamma_1,\Gamma_2) \text{ is dense in }\cH.
\end{gather}
\end{subequations}
A triple $(\cG,\Gamma_1, \Gamma_2)$ with the above properties
is called a \emph{boundary triple} for $A$.
\end{defin}

\begin{rem}
This definition differs slightly from the commonly used one.
In~\cite{Koc,DM,GG} one defines boundary triple only for the case
when $A^*$ is a closed densely defined symmetric operator; the
property~\eqref{G3} holds then automatically. In our opinion, in
some cases it is more convenient to find a boundary triple than to
check whether the adjoint operator is symmetric. Below we will see
(theorem~\ref{corol1}) that these definitions are actually
equivalent if one deals with self-adjoint extensions.
\end{rem}

In Definition~\ref{defin1}, we do not assume any continuity
properties of the maps $\Gamma_1$ and $\Gamma_2$, but they appear
automatically.

\begin{prop}\label{prop-btc}
Let $A$ be a closed linear operator in a Hilbert space
$\cH$ and $(\cG,\Gamma_1,\Gamma_2)$ be its boundary triple, then
the mapping $\dom S\ni g\mapsto (\Gamma_1g,\Gamma_2 g)\in\cG\oplus\cG$
is continuous with respect to the graph norm of $S$.
\end{prop}

\begin{proof} Suppose that a sequence $g_n\in\dom A$, $n\in\mN$,
converges in the graph norm. As $A$ is closed, there holds $g:=\lim g_n\in\dom A$
and $Ag=\lim Ag_n$. Assume that $\lim (\Gamma_1 g_n,\Gamma_2 g_n)= (u,v)$,
where the limit is taken in the norm of $\cG\oplus\cG$. Let us show
that $\Gamma_1 g=u$ and $\Gamma_2 g=v$; this will mean
that the mapping $(\Gamma_1,\Gamma_2)$ is closed and, therefore,
continuous by the closed graph theorem.

For an arbitrary $f\in\dom A$ there holds
\begin{multline*}
\langle\Gamma_1 f|\, \Gamma_2g\rangle-\langle\Gamma_2 f|\,\Gamma_1
g\rangle=
\langle f|\,Ag\rangle-\langle Af|\,g\rangle\\
{}=\lim \langle f|\, Ag_n\rangle-\langle Af|\,g_n\rangle=\lim
\langle\Gamma_1 f|\,\Gamma_2 g_n\rangle
-\langle\Gamma_2f|\,\Gamma_1 g_n\rangle\\
{}=\langle\Gamma_1 f|\, v\rangle-\langle \Gamma_2 f|\,u\rangle.
\end{multline*}
Therefore, $\langle\Gamma_1 f|\, \Gamma_2g\rangle-\langle\Gamma_2
f|\,\Gamma_1 g\rangle= \langle\Gamma_1 f|\, v\rangle-\langle
\Gamma_2 f|\,u\rangle$ and
\begin{equation}
       \label{eq-bct1}
\langle\Gamma_1f|\,\Gamma_2g-v\rangle=\langle\Gamma_2f|\,\Gamma_1g-u\rangle
\end{equation}
for any $f\in\dom A$. Using the property \eqref{G2} from
definition \ref{defin1}, one can take $f\in\dom A$ with $\Gamma_1
f=\Gamma_2 g-v$ and $\Gamma_2f=0$, then \eqref{eq-bct1} reads as
$\|\Gamma_2g-v\|^2=0$ and $\Gamma_2g=v$. Analogously, choosing
$f\in\dom A$ with $\Gamma_1f=0$ and $\Gamma_2f=\Gamma_1g-u$ one
arrives at $\Gamma_1g=u$.
\end{proof}

Our next aim is to describe situations in which boundary triples
exist and are useful. For a symmetric operator $A$ in a Hilbert
space $\cH$ and for $z\in\mC$, we denote throughout the paper
$\cN_z(A):=\Ker(A^*-zI)$ and write sometimes
$\cN_z$ instead of $\cN_z(A)$, if it does not lead to confusion.

Is is well known that $A$ has self-adjoint extensions if and only if
$\dim\cN_i=\dim\cN_{-i}$. The von Neumann theory states a bijection
between the self-adjoint extensions and unitary operators from $\cN_i$
to $\cN_{-i}$. More precisely, if $U$ is a unitary operator
from $\cN_i$ to $\cN_{-i}$, then the corresponding self-adjoint
extension $A_U$ has the domain $\{f=f_0+f_i+Uf_i:\, f_0\in\dom A,\,f_i\in\cN_i\}$
and acts as $f_0+f_i+Uf_i\mapsto Af_0+if_i-iUf_i$. This construction is difficult
to use in practical applications, and our aim is to show that the boundary triples
provide a useful machinery for working with self-adjoint extensions.

The following proposition is borrowed from~\cite{Koc}.
\begin{prop}\label{prop-btp}
Let $A$ be a densely defined closed symmetric operator in a
Hilbert space $\cH$ with equal deficiency indices $(\fn,\fn)$,
then there is a boundary triple $(\cG,\Gamma_1,\Gamma_2)$ for the
adjoint $A^*$ with $\dim\cG=\fn$.
\end{prop}

\begin{proof}
It is well known that $\dom A^*=\dom A+\cN_i+\cN_{-i}$, and this
sum is direct. Let $P_{\pm i}$ be the projector from $\dom A^*$ to
$\cN_{\pm i}$ corresponding to this expansion.

Let $f,g\in\dom A^*$, then $f=f_0+P_i f+P_{-i} f$,
$g=g_0+P_i g+P_{-i} g$, $f_0,g_0\in\dom A$.
Using the equalities $A^*P_i=iP_i$ and $A^*P_{-i}=-iP_{-i}$ one obtains
\begin{multline}
         \label{eq-bt1}
\langle f|\,A^*g\rangle-\langle A^*f|\,g\rangle=
\langle f_0+P_i f+P_{-i}f|\,A g_0+iP_i g-iP_{-i}g\rangle\\
{}=2i\langle P_if|\,P_ig\rangle-2i\langle
P_{-i}f|\,P_{-i}g\rangle.
\end{multline}

As the deficiency indices of $A$ are equal, there is an
isomorphism $U$ from $\cN_{-i}$ onto $\cN_i$. Denote
$\cG:=\cN_{-i}$ endowed with the induced scalar product in $\cH$,
and set $\Gamma_1=iUP_{-i}-iP_i$, $\Gamma_2=P_i+UP_{-i}$, then
\begin{multline}
      \label{eq-gg12}
\langle\Gamma_1f|\,\Gamma_2g\rangle-\langle \Gamma_2 f|\,\Gamma_1
g\rangle= 2i\langle P_if|\,P_ig\rangle-2i\langle
UP_{-i}f|\,UP_{-i}g\rangle\\= 2i\langle
P_if|\,P_ig\rangle-2i\langle P_{-i}f|\,P_{-i}g\rangle.
\end{multline}
Comparing \eqref{eq-bt1} with \eqref{eq-gg12} one shows that
$(\cG,\Gamma_1,\Gamma_2)$ satisfy the property \eqref{G1} of
definition~\eqref{defin1}. Due to $\dom
A\subset\ker(P_i,P_{-i})\subset\ker(\Gamma_1,\Gamma_2)$ the
property \eqref{G3} is satisfied too. To prove \eqref{G2} take any
$F_1,F_2\in \cN_{-i}\equiv\cG$ and show that the system of
equations
\begin{equation}
              \label{eq-btqu}
iUP_{-i}f-iP_if=F_1,\quad UP_{-i}f+P_i f=F_2,
\end{equation}
has a solution $f\in\dom A^*$. Multiplying the second equation by $i$
and adding it to the first one one arrives at $2i U P_{-i}f=F_1+iF_2$.
In a similar way, $2iP_i f=iF_2-F_1$. Therefore,
the funtcion
\[
f=\frac{1}{2i}(iF_2-F_1)+\dfrac{1}{2i}U^{-1}(F_1+iF_2)\in \cN_{i}(A^*)+\cN_{-i}(A^*)\subset\dom A^*
\]
is a possible solution to \eqref{eq-btqu}, and \eqref{G2} is satisfied.
Therefore, $(\cG,\Gamma_1,\Gamma_2)$ is a boundary triple for $A^*$.
\end{proof}

Let $A$ be a closed densely defined linear operator, $A^*$ have a
boundary triple $(\cG,\Gamma_1,\Gamma_2)$, $\Lambda$ be a closed
linear relation in $\cG$. By $A_\Lambda$ in \emph{this subsection} we mean the restriction of
$A^*$ to the domain $\dom A_\Lambda=\{f\in \dom
A^*:\,(\Gamma_1f,\Gamma_2 f)\in\Lambda\}$.

The usefulness of boundary triples is described in the following proposition.
\begin{prop}\label{prop-cladj}
For any closed linear relation $\Lambda$ in
$\cG$ one has $A_\Lambda^*=A_{\Lambda^*}$. In particular, $A_\Lambda$ is symmetric/self-adjoint
if and only if $\Lambda$ is symmetric/self-adjoint, respectively.
\end{prop}

\begin{proof}
Clearly, one has $A\subset A_\Lambda\subset A^*$. Therefore,
$A\subset A_\Lambda^*\subset A^*$. Moreover, one has
\begin{multline*}
\gr A_\Lambda^*=\{
(f,A^*f): \langle f|\,A^*g\rangle=\langle A^*f|\,g\rangle
\quad\forall g\in\dom A_\Lambda \}\\
=\{(f,A^*f): \langle \Gamma_1f|\,\Gamma_2g\rangle-
\langle \Gamma_2 f|\,\Gamma_1g\rangle\quad \forall g\in\dom A_\Lambda\}\\
=\{(f,A^*f): \langle \Gamma_1f|\,x_2\rangle-
\langle \Gamma_2 f|\,x_1\rangle\quad \forall (x_1,x_2)\in\Lambda\}\\
=\{(f,A^*f): (\Gamma_1f,\Gamma_2f)\in \Lambda^*\}=\gr A_{\Lambda^*}.
\end{multline*}
This proves the first part of proposition. The part concerning
the self-adjointness of $A_\Lambda$ is now
obvious, as $A_\Lambda\subset A_{\Lambda'}$ if and only if
$\Lambda\subset\Lambda'$.
\end{proof}

\begin{thm}\label{corol1}
Let $A$ be a closed densely defined symmetric operator.
\begin{enumerate}
\item[(1)] The operator $A^*$ has a boundary triple if and only if
$A$ admits self-adjoint extensions.

\item[(2)] If
$(\cG,\Gamma_1,\Gamma_2)$ is a boundary triple for $A^*$, then
there is a one-to-one correspondence between all self-adjoint
linear relations $\Lambda$ in $\cG$ and all self-adjoint extensions
of $A$ given by $\Lambda\leftrightarrow A_\Lambda$, where $A_\Lambda$
is the restriction of $A^*$ to the vectors $f\in\dom A^*$
satisfying $(\Gamma_1f,\Gamma_2f)\in\Lambda$.
\end{enumerate}
\end{thm}

\begin{proof}
(1) Let $A^*$ have a boundary triple and $\Lambda$ be any
self-adjoint linear relation in $\cG$, then according to
proposition~\ref{prop-cladj} the operator $A_\Lambda$ is
self-adjoint, and $A_\Lambda\supset A$. The converse is exactly
proposition~\ref{prop-btp}.

(2) If $\Lambda$ is a self-adjoint linear relation in $\cG$, then
due to proposition~\ref{prop-cladj} the corresponding operator
$A_\Lambda$ is self-adjoint.

Now let $B$ be a self-adjoint extension of $A$, then $A\subset
B\subset A^*$. Denote $\Lambda=\{(\Gamma_1f,\Gamma_2f),\,f\in\dom
B^*\}$, then $B=A_\Lambda$, and $\Lambda$ is self-adjoint due to proposition~\ref{prop-cladj}.
\end{proof}

\begin{thm}\label{thm-bteq}
Let a closed linear operator $B$  have a boundary triple $(\cG,\Gamma_1,\Gamma_2)$,
and $A:=B|_{\ker(\Gamma_1,\Gamma_2)}$, then
$A\subset B^*$. Moreover, the following three
conditions are equivalent:
\begin{itemize}
\item[(1)] $B$ has at least one restriction which is self-adjoint,
\item[(2)] $B^*$ is symmetric;
\item[(3)] $B^*=A$,
\item[(4)] $A^*=B$.
\end{itemize}
\end{thm}

\begin{proof}
By construction $A$ is densely defined. By definition~\ref{defin1}
for any $f\in \dom A$ one has $\langle f|\,Bg\rangle-\langle
Af|\,g\rangle=0$, which means $A\subset B^*$. In particular, $B^*$
is densely defined. By proposition~\ref{prop-btc}, $A$ is closed,
therefore, (3) and (4) are equivalent.

(1)$\Rightarrow$(2). Let $C$ be a self-adjoint restriction of $B$. From
$C\subset B$ it follows $B^*\subset C^*\equiv C\subset B\equiv (B^*)^*$,
i.e. $B^*$ is symmetric.

(2)$\Rightarrow$(3). Let $D=B^*$ be symmetric,
then $D\subset B$ is closed and $B=D^*$.

Let $f\in\dom D$. According to the definition~\ref{defin1}
there exists $g\in \dom D^*=\dom B$ with $\Gamma_1g=-\Gamma_2 f$
and $\Gamma_2g=\Gamma_1 f$. One has
\begin{multline*}
0=\langle Df|\,g\rangle-\langle D f|\,g\rangle
=\langle f|\,D^*g\rangle-\langle D^*f|\,g\rangle\\
{}\equiv\langle f|\,Bg\rangle-\langle Bf|\,g\rangle =\|\Gamma_1
f\|^2+\|\Gamma_2 f\|^2,
\end{multline*}
from which $\Gamma_1f=\Gamma_2f=0$.
Therefore, $\dom D\subset \ker (\Gamma_1,\Gamma_2)\equiv \dom A$.
At the same time, as shown above, $A\subset B^*$, which means $A=D=B^*$.

(4)$\Rightarrow$(1). Let $B=A^*$. By theorem \ref{corol1}(1)
the operator $A$ has self-adjoint extensions, which are at the same time
self-adjoint restrictions of $A^*=B$.
\end{proof}

The proof of proposition~\ref{prop-btp} gives a possible
construction of a boundary triple. Clearly, boundary triple is not
fixed uniquely by definition~\ref{defin1}. For a description of
all possible boundary triple we refer to \cite{MS,Mik}. We
restrict ourselves by the following observations.

\begin{prop}\label{prop-gam1}
Let $A$ be a closed densely defined symmetric operator
with equal deficiency indices.
For any self-adjoint extension $H$ of $A$ there exists
a boundary triple $(\cG,\Gamma_1,\Gamma_2)$ for $A^*$
such that $H$ is the restriction of $A^*$ to $\ker\Gamma_1$.
\end{prop}

\begin{proof}
Let $(\cG,\Gamma'_1,\Gamma'_2)$ be an arbitrary boundary triple
for $A^*$. According to theorem~\ref{corol1}(2), there exists a
self-adjoint linear relation $\Lambda$ in $\cG$ such that $H$ is
the restriction of $A^*$ to the vectors $f\in\dom A^*$ satisfying
$(\Gamma'_1f,\Gamma'_2f)\in\Lambda$. Let $U$ be the Cayley
transform of $\Lambda$ (see theorem~\ref{prop-iu}). Set
\[
\Gamma_1:=\dfrac{1}{2}\Big(\,i(1+U)\Gamma'_1+(U-1)\Gamma'_2\Big),
\quad
\Gamma_2:=\dfrac{1}{2}\Big((1-U)\Gamma'_1+i(1+U)\Gamma'_2\Big).
\]
By lemma~\ref{lem-AB} the map $(\Gamma_1,\Gamma_2):\dom
A^*\to\cG\oplus\cG$ is surjective and
$\ker(\Gamma_1,\Gamma_2)=\ker(\Gamma'_1,\Gamma'_2)$. At the same
time one has
$\langle\Gamma_1f|\,\Gamma_2g\rangle-\langle\Gamma_2f|\,\Gamma_1g\rangle
\equiv\langle\Gamma'_1f|\,\Gamma'_2g\rangle-\langle\Gamma'_2f|\,\Gamma'_1g\rangle$,
which means that $(\cG,\Gamma_1,\Gamma_2)$ is a boundary triple
for $A^*$. It remains to note that the conditions
$(\Gamma'_1f,\Gamma'_2f)\in\Lambda$ and $\Gamma_1f=0$ are
equivalent by the choice of $U$.
\end{proof}

\begin{prop}\label{prop-btb}
Let $(\cG,\Gamma_1,\Gamma_2)$ be an arbitrary boundary triple for $A^*$,
and $L$ be a bounded linear self-adjoint operator in $\cG$, then
$(\cG,\widetilde\Gamma_1,\widetilde\Gamma_2)$ with
$\widetilde\Gamma_1=\Gamma_1$ and $\widetilde\Gamma_2=\Gamma_2+L\Gamma_1$
is also a boundary triple for $S^*$.
\end{prop}

\begin{proof} The conditions of~definition~\ref{defin1} are verified directly.
\end{proof}

An explicit construction of boundary triples is a rather difficult
problem, see e.g.~\cite{Vis} for the discussion of elliptic
boundary conditions. In some cases there are natural boundary
triples reflecting some specific properties of the problem, like in the
theory of singular perturbations, see~\cite{Pos2} and
subsection~\ref{ss-sing} below.

\subsection{Krein's resolvent formula}

In this subsection, if not specified explicitly,
\begin{itemize}
\item $S$ is a densely defined symmetric operator with equal deficiency indices  $(\fn,\fn)$,
$0<\fn\le\infty$, in a Hilbert space $\cH$,
\item $\cN_z:=\ker(S^*-z)$,
\item $\cG$ is a Hilbert
space of dimension $\fn$,
\item $H^0$ is a certain self-adjoint
extension of $S$,
\item for $z\in\res H^0$ denote $R^0(z):=(H^0-z)^{-1}$, the resolvent of $H^0$.
\end{itemize}
For $z_1,z_2\in\res H^0$ put
\[
U(z_1,z_2)=(H^0-z_2)(H^0-z_1)^{-1}\equiv 1+(z_1-z_2)R^0(z_1).
\]
It
is easy to show that $U(z_1,z_2)$ is a linear topological
isomorphism of $\mathcal{H}$ obeying the following properties:
\begin{subequations}
         \label{eq-UUU}
\begin{align}
U(z,z)&=I,\label{eq-U1}\\
U(z_1,z_2)U(z_2,z_3)&=U(z_1,z_3),\label{eq-U2}\\
U^{-1}(z_1,z_2)&=U(z_2,z_1),\label{eq-U3}\\
U^*(z_1,z_2)&=U(\bar z_1,\bar z_2),\label{eq-U4}\\
U(z_1, z_2){\mathcal{N}}_{z_2}(S)&=
\mathcal{N}_{z_1}(S)\label{eq-U5}.
\end{align}
\end{subequations}

\begin{subequations}

\begin{defin}\label{defin2}
A map $\gamma:\res H^0\to \bL(\cG,\cH)$ is called a \emph{Krein
$\Gamma$-field} for $(S,H^0,\cG)$ if the following two
conditions are satisfied:
\begin{gather}
      \label{Gam1}
\begin{minipage}{120mm}
$\gamma(z)$ is a linear topological
isomorphism of $\cG$ and $\cN_z$ for all $z\in\res H^0$,
\end{minipage}\\[\smallskipamount]
      \label{Gam2}
\begin{minipage}{120mm}
for any $z_1,z_2\in\res H^0$ there holds
$\gamma(z_1)=U(z_1,z_2)\gamma(z_2)$ or, equivalently,
$\gamma(z_1)-\gamma(z_2)=(z_1-z_2)R^0(z_1)\gamma(z_2)=
(z_1-z_2)R^0(z_2)\gamma(z_1)$.
\end{minipage}
\end{gather}
\end{defin}
\end{subequations}
Let us discuss questions concerning the existence
and uniqueness of $\Gamma$-fields.

\begin{prop}\label{prop-gamma}
For any triple $(S,H^0,\cG)$ there
exists a Krein $\Gamma$-field $\gamma$.
If $\tilde\gamma(z)$ is another Krein $\Gamma$-field
for $(S,H^0,\widetilde\cG)$ with a certain Hilbert space
$\widetilde \cG$, then there exists a linear topological
isomorphism $N$ from $\widetilde\cG$ to $\cG$ such that
$\tilde\gamma(z)=\gamma(z)N$.
\end{prop}

\begin{proof}
Fix any $z_0\in\res H^0$, choose any linear topological isomorphism
$L:\cG\rightarrow\cN_{z_0}$, and set $\gamma(z_0):=L$. Then property
\eqref{Gam2} forces to set
\begin{equation}
        \label{pr-gamma}
\gamma(z)=U(z,z_0)L\equiv L+(z-z_0)R^0(z)L.
\end {equation}
On the other hand, the properties \eqref{eq-UUU} of
$U(z_1,z_2)$ show that $\gamma(z)$ defined by \eqref{pr-gamma} is
a $\Gamma$-field for $(S,H^0,\cG)$.

If $\tilde\gamma(z):\widetilde\cG\rightarrow \cH$, $z\in\res H^0$,
is another $\Gamma$-field for $(S,H^0,\cG)$, then setting
$N=\tilde\gamma(z_0)\gamma^{(-1)}(z_0)$ where $\gamma^{(-1)}(z_0)$
is the inverse to $\gamma(z_0):\,\cG\rightarrow\cN_{z_0}$, and
using \eqref{Gam2} again, we see that $\tilde\gamma(z)=\gamma(z)N$
for all $z\in\res H^0$.
\end{proof}

The following propositions gives a characterization of all
Krein $\Gamma$-fields.

\begin{prop}\label{prop-gamma1}
Let $H^0$ be a self-adjoint operator in a Hilbert space $\cH$,
$\cG$ be another Hilbert space, and $\gamma$
be a map from $\res H^0$ to $\bL(\cG,\cH)$,
then the following assertions are equivalent:
\begin{itemize}
\item[(1)]\it there is a closed densely defined symmetric
restriction $S$ of $H^0$ such that $\gamma$ is the $\Gamma$-field
for $(S,H^0,\cG)$.
\item[(2)] $\gamma$ satisfies the condition
\eqref{Gam2} above and the following additional condition:
\begin{equation}
\label{Gamp}
\parbox{100mm}{for some $\zeta\in\res H^0$
the map $\gamma(\zeta)$ is a linear topological
isomorphism of $\cG$ on a subspace $\cN\subset\cH$ such that
${\cN}\cap\dom H^0=\{0\}$.}
\end{equation}
\end{itemize}
\end{prop}

\begin{proof}
Clearly, any $\Gamma$-field satisfies~\eqref{Gamp}.

Conversely, let the conditions \eqref{Gamp} and \eqref{Gam2} be
fulfilled for a map $\gamma:\,\res H^0 \rightarrow \bL(\cG,\cH)$.
Then, in particular, $\gamma(z)$ is a linear topological
isomorphism on a subspace of $\cH$ for any $z\in\res H^0$. Denote
$\cD_z=\Ker\, \gamma^*(z)(H^0-\bar z)$. According to \eqref{Gam2}
we have for any $z_1,z_2\in\res H^0$
\[
\gamma^*(z_2)=\gamma^*(z_1)U^*(z_2,z_1)= \gamma^*(z_1)(H^0-\bar
z_1)(H^0-\bar z_2)^{-1}\,.
\]
Hence $\gamma^*(z_2)(H^0-\bar z_2)=\gamma^*(z_1)(H^0-\bar z_1)$,
therefore $\cD_z$ is independent of $z$. Denote $\cD:=\cD_z$ and
define $S$ as the restriction of $H^0$ to $\cD$. Show that $\cD$
is dense in $\cH$. Let $\varphi\perp\cD$. Since
$\cD=\cD_\zeta=R^0(\bar\zeta)(\cN^\bot)$, this means that $\langle
R^0(\zeta)\varphi|\,\psi\rangle=0$ for each $\psi\in\cN^\bot$,
i.e. we have $R^0(\zeta)\varphi\in\cN$. Hence,
$R^0(\zeta)\varphi=0$, therefore $\varphi=0$. Thus, $S$ is densely
defined. Let us show that
\begin{equation}
                      \label{aa1}
\ran (S-\overline z)=\ker \gamma^*(z)
\end{equation}
for any $z\in\res H^0$. Let $\gamma^*(z)\varphi=0$; set
$\psi:=(H^0-\overline z)^{-1}\varphi$, then $\psi\in \dom S\equiv\cD$,
therefore, $\varphi\in \ran(S-\overline z)$. Conversely, if
$\varphi\in\ran(S-\overline z)$, then $\varphi=(S-\overline z)\psi$ where
$\gamma^*(z)(H^0-\bar z)\psi=0$, and \eqref{aa1} is proven. In
particular, \eqref{aa1} implies that $S$ is closed. Moreover, we
have from \eqref{aa1}
\[
\cN_z=\ran(S-\bar z)^\bot =\ker \gamma^*(z)^\bot=\overline{\ran\gamma(z)}=\ran\gamma(z).
\]
Thus, $\gamma$ is a $\Gamma$-field for $(S,H^0,\cG)$.
\end{proof}

Let now the triple $(S,H^0,\cG)$ be endowed with a $\Gamma$-field
$\gamma$, $\gamma:\,\res H^0\rightarrow\bL(\cG,\cH)$.

\begin{defin}\label{defin2q}
A map $Q:\res H^0\to \bL(\cG,\cG)$ is called a \emph{Krein
$\cQ$-function} for $(S,H^0,\cG,\gamma)$, if
\begin{equation}
        \label{Q1}
Q(z_1)-Q^*(\Bar z_2)=(z_1-z_2)\gamma^*(\Bar z_2)\gamma(z_1) \text{
for any } z_1,z_2\in\res H^0.
\end{equation}
\end{defin}

\begin{prop}\label{prop-qu}
For any $(S,H^0,\cG)$ endowed with a Krein $\Gamma$-field
$\gamma$ there exists a Krein $\cQ$-function $Q:\res H^0\to
\bL(\cG,\cG)$. If $\widetilde Q(z):\cG\rightarrow \cG$, $z\in\res
H^0$, is another $\cQ$-function for $(S,H^0,\cG,\gamma)$, then
$\widetilde Q(z)=Q(z)+M$, where $M$ is a bounded self-adjoint
operator in $\cG$.
\end{prop}

\begin{proof}
Fix as any $z_0\in\res H^0$ and denote $x_0:=\Re\,z_0$,
$y_0:=\Im\,z_0$, $L:=\gamma(z_0)$.
If a $\cQ$-function exists, then by \eqref{Q1} one has
$Q(z)=Q^*(z_0)+(z-\Bar z_0)L^*\gamma(z)$. On the other hand
\[
Q^*(z_0)=\frac{Q(z_0)+Q^*(z_0)}{2}-\frac{Q(z_0)-Q^*(z_0)}{2}\,.
\]
Clearly, $Q(z_0)+Q^*(z_0)$ is a bounded self-adjoint operator in
$\cG$, denote it by $2C$. According to \eqref{Q1},
$Q(z_0)-Q^*(z_0)=2iy_0L^*L$, and therefore
\begin{equation}
            \label{qqq}
Q(z)=C-iy_0L^*L+(z-\Bar z_0)L^*\gamma(z)\,.
\end{equation}
We have from \eqref{qqq} that if $\widetilde Q(z)$ is another
$\cQ$-function for $(S,H^0,\cG,\gamma)$, then $\widetilde
Q(z)-Q(z)=M$ where $M$ is a bounded self-adjoint operator which is
independent of $z$.

It remains to show that a function of the form \eqref{qqq} obeys \eqref{Q1}.
Take arbitrary $z_1,z_2\in\res H^0$.
We have $Q^*(z_2)=C+iy_0L^*L+(\Bar z_2-z_0)\gamma^*(z_2)L$.
Therefore,
\begin{equation}
     \label{qqqq}
Q(z_1)-Q^*(z_2)=
(\Bar z_0-z_0)L^*L+(z-\Bar z_0)L^* \gamma(z_1)+(z_0-\Bar
z_2)\gamma^*(z_2)L.
\end{equation}
By \eqref{Gam2}, $L=\gamma(z_0)=\gamma(z_1)+(z_0-z_1)R^0(z_0)\gamma(z_1)$
and $L^*=\gamma^*(z_0)=\gamma^*(z_2)+ (\Bar z_0-\Bar
z_2)\gamma^*(z_2)R^0(\Bar z_0)$.
Substituting these expressions in (\ref{qqqq}) we obtain
\begin{multline*}
Q(z_1)-Q^*(z_2)=(z_1-\Bar z_2)\gamma^*(z_2)\gamma(z_1)\\
+\gamma^*(z_2)\,\Big\{ (\Bar z_0-z_0)\big[(\Bar z_0-\Bar z_2)R^0(\Bar
z_0)+ (z_0-z_1)R^0(z_0)\\
+(\Bar z_0-\Bar z_2)(z_0-z_1)R^0(\Bar z_0)R^0(z_0)\big]+ (z_1-\Bar
z_0)(\Bar z_0-\Bar z_2)R^0(\Bar z_0)\\
+(z_0-\Bar z_2)(z_0-z_1)R^0(z_0)\Big\} \, \gamma(z_1)\,.
\end{multline*}
The expression in the curly brackets is equal to
\begin{multline*}
(\Bar z_0-z_0)(\Bar z_0-\bar z_2)R^0(\Bar z_0)+(z_1-\Bar z_0)(\Bar
z_0-\Bar z_2) R^0(\Bar z_0)\\
+(z_0-z_1)(\Bar z_0-\Bar z_2)R^0(\Bar z_0)+(\Bar
z_0-z_0)(z_0-z_1)R^0(z_0)\\
+(z_0-\Bar z_2)(z_0-z_1)R^0(z_0)-(\Bar z_0-\Bar
z_2)(z_0-z_1)R^0(z_0)\,.
\end{multline*}
It is easy to see that the latter expression is equal to zero, and
we get the result.
\end{proof}

Below we list some properties of $\Gamma$-fields and $\mathcal Q$-functions
which follow easily from the definitions.
\begin{prop}
Let $\gamma$ be a Krein $\Gamma$-field for $(S,H^0,\cH)$, then
$\gamma$ is holomorphic in $\res H^0$ and satisfies
\begin{subequations}
\begin{gather}
         \label{Gam3}
\dfrac{d}{dz} \gamma(z)=R^0(z)\gamma(z),\\
         \label{Gam4}
S^*\gamma(z)=z\gamma(z),\\
         \label{Gam5a}
\text{$\gamma^*(z)$ is a bijection from $\cN_z$ onto $\cG$,}\\
         \label{Gam5}
\text{$\gamma^*(z)f=0$ iff $f\perp\cN_z$,}\\
         \label{Gam6}
\gamma^*(\overline z_1)\gamma(z_2)=\gamma^*(\overline z_2)\gamma(z_1),\\
         \label{Gam7}
\ran\big[\gamma(z_1)-\gamma(z_2)\big]\subset\dom H^0 \text{ for any } z_1,z_2\in\res H^0.
\end{gather}
\end{subequations}
Let in addition $Q$ be a Krein $\cQ$-function for $(S,H^0,\cG)$
and $\gamma$, then $Q$ is holomorphic in $\res H^0$, and the
following holds:
\begin{subequations}
\begin{gather}
\label{Q2}
\dfrac{d}{d z}Q(z)=\gamma^*(\bar z)\gamma(z),\\
\label{Q3}
Q^*(\bar z)=Q(z),\\
\label{Q4} \text{for any $z\in\mC\setminus\mR$ there is $c_z>0$
with } \dfrac{\Im Q(z)}{\Im z}\ge c_z.
\end{gather}
\end{subequations}
\end{prop}

\begin{rem}
The property~\eqref{Q4} means that $\cQ$--function is an
operator-valued Nevanlinna function (or Herglotz function). This
implies a number of possible relations to the measure theory,
spectral theory etc., and such functions appear in many areas
outside the extension theory, see e.g. \cite{DM,DM2,GT,GKMT,Na1,Na2} and
references therein.
\end{rem}

Our next aim is to relate boundary triples in definition
\ref{defin1} to Krein's maps from definition \ref{defin2}.
\begin{thm}\label{thm-btgf}
Let $S$ be a closed densely defined symmetric operator in a Hilbert space $\cH$
with equal deficiency indices.
\begin{enumerate}
\item[(1)] For any self-adjoint extension $H$ of $S$ and any $z\in
\res H$ there holds $\dom S^*=\dom H+\cN_z$, and this sum is
direct. \item[(2)] Let $(\cG,\Gamma_1,\Gamma_2)$ be a boundary
triple for $S^*$ and $H^0$ be the restriction of $S^*$ to
$\ker\Gamma_1$ which is self-adjoint due to theorem~\ref{corol1}.
Then:
\begin{enumerate}
\item[(2a)] for any $z\in\res H^0$ the restriction of $\Gamma_1$
to $\cN_z$ has a bounded inverse
$\gamma(z):\cG\to\cN_z\subset\cH$ defined everywhere,
\item[(2b)] this map $z\mapsto\gamma(z)$ is a Krein $\Gamma$-field for
$(S,H^0,\cG)$,
\item[(2c)] the map $\res H^0\ni z\mapsto
Q(z)=\Gamma_2\gamma(z)\in\bL(\cG,\cG)$ is a Krein $\cQ$-function
for $(S,H^0,\cG)$ and $\gamma$.
\item[(2d)] for any $f\in\dom H^0$
and $z\in\res H^0$ there holds $\gamma^*(\Bar z)(H^0-z)f=\Gamma_2
f$.
\end{enumerate}
\end{enumerate}
\end{thm}

\begin{proof}
(1) Let $f\in\dom S^*$, Denote $f_0:=(H-z)^{-1}(S^*-z)f$.
Clearly, $f_0\in\dom H$. For $g:=f-f_0$ one has
$(S^*-z)g=(S^*-z)f-(S^*-z)(H-z)^{-1}(S^*-z)f \equiv
(S^*-z)f-(H-z)(H-z)^{-1}(S^*-z)f=0$, therefore,
$g\in\ker(S^*-z)\equiv \cN_z$.

Now assume that for some $z\in\res H$ one has $f_0+g_0=f_1+g_1$
for some $f_0,f_1\in\dom H$ and $g_0,g_1\in\cN_z$, then
$f_0-f_1=g_1-g_0\in\cN_z$ and
$(H-z)(f_0-f_1)=(S^*-z)(f_0-f_1)=0$. As $H-z$ is invertible,
one has $f_0=f_1$ and $g_0=g_1$.

(2a) Due to condition \eqref{G2},
$\Gamma_1(\dom S^*)=\cG$. Due to $\Gamma_1(\dom H^0)=0$ and item (1)
one has $\Gamma_1(\cN_z)=\cG$. Assume that $\Gamma_1 f=0$ for some
$f\in\cN_z$, then $f\in\dom H^0\cap\cN_z$ and $f=0$ by item (1).
Therefore, $\Gamma_1:\cN_z\to\cG$ is a bijection and,
moreover, $\Gamma_1$ is continuous in the graph norm of $S^* $ by
proposition~\ref{prop-btc}. At the same time, the graph
norm of $S^*$ on $\cN_z$ is equivalent to the usual norm in $\cH$,
which means that the restriction of $\Gamma_1$ to $\cN_z$
is a bounded operator. The graph of this map is
closed, and the inverse map is bounded by the closed graph
theorem.

(2b) The property \eqref{Gam1} is already proved in item (2a). Take
arbitrary $z_1,z_2\in\res H^0$ and $\xi\in\cG$. Denote
$f=\gamma(z_1)\xi$ and $g=U(z_2,z_1)f\equiv f+(z_2-z_1)R^0(z_2)f$.
As $R^0(z_2)f\in\dom H^0$, there holds $\Gamma_1R^0(z_2)f=0$ and
$\Gamma_1g=\Gamma_1f$. Clearly, $f\in\cN_{z_1}$, and to prove property~\eqref{Gam2}
it is sufficient to show that $(S^*-z_2)g=0$. But this follows from the
chain $(S^*-z_2)g=(S^*-z_2)f+(z_2-z_1)(S^*-z_2)(H^0-z_2)^{-1}f=
(S^*-z_2)f+(z_2-z_1)(H^0-z_2)(H^0-z_2)^{-1}f=(S^*-z_1)f=0$.

Therefore, $\gamma$ satisfies both properties \eqref{Gam1} and \eqref{Gam2}
in definition~\ref{defin2}.

(2c) As $\gamma(z)$ is bounded by item (2a) and $\Gamma_2$
is bounded by proposition~\ref{prop-btc}, the map $Q(z)$
is a bounded linear operator on $\bL(\cG,\cG)$. To prove
property \eqref{Q1} take arbitrary $z_1,z_2\in\res H$, $\phi,\psi\in\cG$, and set
$f:=\gamma(\Bar z_2)\phi$, $g:=\gamma(z_1)\psi$.
Clearly,
\begin{multline}
           \label{eq-Sfg}
\langle f|\,S^*g\rangle-\langle f|\,S^*g\rangle-(z_1-z_2)\langle f|\,g\rangle\\
{}= \langle f|\,(S^*-z_1)g\rangle-\langle (S^*-\Bar
z_2)f|\,g\rangle=0.
\end{multline}
At the same time one has
\begin{equation}
     \label{eq-ggg}
\langle f|\,g\rangle=\langle \gamma(\Bar
z_2)\phi|\,\gamma(z_1)\psi\rangle=\langle \phi|\,\gamma^*(\Bar
z_2)\gamma(z_1)\psi\rangle.
\end{equation}
Moreover, using the equality
$\Gamma_1 \gamma(z)\xi=\xi$, which holds for all $\xi\in\cG$ and
$z\in\res H^0$, one obtains
\begin{multline*}
\langle f|\,S^*g\rangle-\langle
f|\,S^*g\rangle=\langle\Gamma_1f|\,\Gamma_2g\rangle-
\langle\Gamma_2f|\,\Gamma_1g\rangle\\
{}=\langle\Gamma_1\gamma(\Bar
z_2)\phi|\,\Gamma_2\gamma(z_1)\psi\rangle
-\langle \Gamma_2\gamma(\Bar z_2)\phi|\,\Gamma_1\gamma(z_1)\psi\rangle\\
{}=\langle \phi|\,Q(z_1)\psi\rangle-\langle Q(\Bar
z_2)\phi|\,\psi\rangle =\langle \phi\,|\,\big[Q(z_1)-Q^*(\Bar
z_2)\big]\psi\rangle.
\end{multline*}
Therefore, Eqs. \eqref{eq-Sfg} and~\eqref{eq-ggg} read as
\[
\langle \phi|\,\big[Q(z_1)-Q^*(\Bar z_2)\big]\psi\rangle=\langle
\phi|\, (z_1-z_2)\gamma^*(\Bar z_2)\gamma(z_1)\psi\rangle,
\]
which holds for any $\phi,\psi\in\cG$. This implies \eqref{Q1}.

(2d) For any $\phi\in\cG$ one has
\begin{multline*}
\langle\phi|\,\gamma^*(\Bar z)(H^0-z) f\rangle=\langle \gamma(\Bar
z)\phi|\,(H^0-z)f\rangle =\langle \gamma(\Bar z)\phi|\,S^*f\rangle
-z\langle \gamma(\Bar z)\phi|\,f\rangle\\
{}=\langle S^*\gamma(\Bar z)\phi|\,f\rangle -z\langle \gamma(\Bar
z)\phi|\,f\rangle+ \langle\Gamma_1\gamma(\Bar z)\phi|\,\Gamma_2
f\rangle
-\langle\Gamma_2\gamma(\Bar z)\phi|\,\Gamma_1 f\rangle\\
=\langle (S^*-\Bar z)\gamma(\Bar z)\phi|\,f\rangle
+\langle\phi|\,\Gamma_2 f\rangle=\langle \phi|\,\Gamma_2f\rangle,
\end{multline*}
i.e. $\Gamma_2f=\gamma^*(\Bar z)(H^0-z)f$.
\end{proof}

\begin{defin}
The Krein $\Gamma$-field and $\cQ$-function
defined in theorem~\ref{thm-btgf} will be called
\emph{induced} by the boundary triple $(\cG,\Gamma_1,\Gamma_2)$.
\end{defin}

\begin{rem}
The $\cQ$-function induced by a boundary triple is often called
the \emph{Weyl function}~\cite{DM,ABMN}.
\end{rem}

Conversely, starting with given Krein maps one can construct
a boundary triple.

\begin{prop}\label{prop-gamma5}
Let $\gamma$ be a Krein $\Gamma$-field for $(S,H^0,\cG)$. For any
$z\in\res H^0$, represent $f\in\dom S^*$ as
\begin{equation}
      \label{eq-fzF}
f=f_z+\gamma(z)F,
\end{equation}
where $f_z\in\dom H^0$, $F\in\cG$. For a \emph{fixed} $z\in\res
H^0$ define
\[
\Gamma_1 f:=F,\quad \Gamma_2 f:=\dfrac{1}{2}\,\Big( \gamma^*(\Bar
z)(H^0-z)f_z+\gamma^*(z)(H^0-\Bar z)f_{\Bar z}\Big),
\]
then $(\cG,\Gamma_1,\Gamma_2)$ is a boundary triple for $S^*$,
and $\gamma(z)$ is the induced $\Gamma$-field.
\end{prop}

For further references we formulate a simplified version of proposition~\ref{prop-gamma5}
for the case when $H^0$ has gaps.
\begin{corol} \label{corol-gamma5}
Let $\gamma$ be a Krein $\Gamma$-field for $(S,H^0,\cG)$. Assume
that $H^0$ has a gap, and $\lambda\in\res H^0\cap\mR$.
Represent $f\in\dom S^*$ as
$f=f_\lambda+\gamma(\lambda)F$, where $f_\lambda\in\dom H^0$, $F\in\cG$.
Define
\[
\Gamma_1 f:=F,\quad \Gamma_2 f:=\gamma^*(\lambda)(H^0-\lambda)f_\lambda,
\]
then $(\cG,\Gamma_1,\Gamma_2)$ is a boundary triple for $S^*$.
\end{corol}

\begin{proof}[\bf Proof of proposition~\ref{prop-gamma5}]
First of all note that the component $F$ in \eqref{eq-fzF} is
independent of $z$. To see that it is sufficient to write $f$ as
$f_z+\big(\gamma(z)-\gamma(\lambda)\big)F+\gamma(\lambda)F$ and to
use the uniqueness of this expansion and the inclusion
$\big(\gamma(z)-\gamma(\lambda)\big)F\in\dom H^0$ following
from~\eqref{Gam7}.

The property \eqref{G2} of boundary triples is obvious. From the
equality $(H^0-z)\dom S^\bot=\ker(S^*-\Bar z)$ and \eqref{Gam5} it
follows that $\dom S\subset\ker(\Gamma_1,\Gamma_2)$, which proves
\eqref{G3}. To show \eqref{G1} we write
\begin{multline*}
2\langle f|\,S^*g\rangle-2\langle S^*f|\,g\rangle\\
= \langle f|\,(S^*-z)g\rangle+\langle f|\,(S^*-\Bar z)g\rangle-
\langle (S^*-z)f|\,g\rangle-\langle (S^*-\Bar z)f|\,g\rangle\\
=\langle f_{\Bar z}+\gamma(\Bar z)\Gamma_1f|\,(H^0- z)g_z\rangle+
\langle f_z+\gamma(z)\Gamma_1f|\,(H^0-\Bar z)g_{\Bar z}\rangle\\
-\langle (H^0-z)f_z|\,g_{\Bar z}+\gamma(\Bar z)\Gamma_1g\rangle
-\langle (H^0-\Bar z)f_z|\,g_z+\gamma(z)\Gamma_1 g\rangle\\
=\langle f_{\Bar z}|\,(H^0-z)g_z\rangle+\langle f_z|\,(H^0-\Bar
z)g_{\Bar z}\rangle
-\langle (H^0-\Bar z)f_{\Bar z}|\,g_z\rangle-\langle (H^0-z)f_z|\,g_{\Bar z}\rangle\\
+\langle \Gamma_1f|\,\gamma^*(\Bar z)(H^0-z)g_z\rangle
+\langle \Gamma_1f|\,\gamma^*(z)(H^0-\Bar z)g_{\Bar z}\rangle\\
-\langle \gamma^*(\Bar z)(H^0-z)f_z|\,\Gamma_1 g\rangle
-\langle \gamma^*(z)(H^0-\Bar z)f_z,\Gamma_1 g\rangle\\
=2\langle\Gamma_1f|\,\Gamma_2g\rangle-2\langle
\Gamma_2f|\,\Gamma_1g\rangle.
\end{multline*}
To show that this boundary triple induces $\gamma$
it is sufficient to note that $\Gamma_1 \gamma(z)=\id_\cG$
and $\gamma(z)\Gamma_1 =\id_{\dom S^*}$.
\end{proof}

Proposition~\ref{prop-gamma5} does not use any information on $\cQ$-functions,
and $\cQ$-functions can be taken into account as follows.

\begin{prop}
Let $\gamma$ be a $\Gamma$-field for $(S,H^0,\cG)$
and $Q$ be an associated $\cQ$-function, then
there exists a boundary triple $(\cG,\Gamma_1,\Gamma_2)$
for $S^*$ which induces $\gamma$ and $Q$.
\end{prop}

\begin{proof}
Let $(\cG,\Gamma'_1,\Gamma'_2)$ be the boundary triple for $S^*$
defined in proposition~\ref{prop-gamma5} and  $\widetilde Q$ be the induced
$\cQ$-function. By proposition~\ref{prop-qu}, there exists
a bounded self-adjoint operator $M$ on $\cG$
with $Q(z)=\widetilde Q(z)+M$. Clearly,
$(\cG,\Gamma_1,\Gamma_2)$ with $\Gamma_1=\Gamma'_1$
and $\Gamma_2=\Gamma'_2+M\Gamma'_1$ is another boundary triple
for $S^*$ by proposition~\ref{prop-btb}. On the other hand, $\gamma$ is still
the $\Gamma$-field induced by this new boundary triple,
and the induced $\cQ$-function, which is $\Gamma_2\gamma(z)\equiv
\Gamma'_2\gamma(z)+M\Gamma'_1\gamma(z)\equiv \widetilde Q(z)+M$,
coincides with $Q(z)$.
\end{proof}

One of the most useful tools for the spectral analysis
of self-adjoint extensions is the Krein resolvent formula
described in the following theorem.

\begin{thm}\label{krein}
Let $S$ be a closed densely defined symmetric operator with equal
deficiency indices in a Hilbert space $\cH$,
$(\cG,\Gamma_1,\Gamma_2)$ be a boundary triple for $S^*$, $H^0$ be
the self-adjoint restriction of $S^*$ to $\ker \Gamma_1$, $\gamma$
and $Q$ be the Krein $\Gamma$-field and $\cQ$-function induced by
the boundary triple. Let $\Lambda$ be a self-adjoint linear
relation in $\cG$ and $H_\Lambda$ be the restriction of $S^*$ to
the functions $f\in\dom S^*$ satisfying
$(\Gamma_1f,\Gamma_2f)\in\Lambda$.
\begin{itemize}
\item[(1)] For any $z\in\res H^0$ there holds
$\ker(H_\Lambda-z)=\gamma(z)\ker \big(Q(z)-\Lambda\big)$.
\item[(2)] For any $z\in\res H^0\cap\res H_\Lambda$ there holds
$0\in\res\big(Q(z)-\Lambda\big)$ and
\[
(H^0-z)^{-1}-(H_\Lambda-z)^{-1}=\gamma(z)\big(Q(z)-\Lambda\big)^{-1}\gamma^*(\Bar
z).
\]
\item[(3)] There holds
$\spec H_\Lambda\setminus\spec H^0=\big\{
z\in\res H^0:\,0\in\spec \big(Q(z)-\Lambda\big)\big\}$.
\end{itemize}
\end{thm}

\begin{proof}
(1) Assume that $\phi \in\ker\big(\Lambda-Q(z)\big)$ then
there exists $\psi\in\cG$ such that $(\phi,\psi)\in\Lambda$
and $\psi-Q(z)\phi=0$. This means the inclusion
$(\phi,Q(z)\phi)\in\Lambda$. Consider the vector
$F=\gamma(z)\phi$. Clearly, $(S^*-z)F=0$. The condition $(\Gamma_1
F,\Gamma_2 F)\equiv(\phi,Q(z)\phi)\in\Lambda$ means that $F\in\dom
H_\Lambda$ and $(H_\Lambda-z)F=0$. Therefore,
$\gamma(z)\ker\big(Q(z)-\Lambda\big)\subset\ker(H_\Lambda-z)$.

Conversely, let $F\in\ker(H_\Lambda-z)$, $z\in\res H^0$. Then also
$(S^*-z)F=0$ and by theorem~\ref{thm-btgf}(1) there exists
$\phi\in\cG$ with $F=\gamma(z)\phi$. Clearly,
$\big(\phi,Q(z)\phi\big)\equiv(\Gamma_1F,\Gamma_2 F)\in\Lambda$, i.e. there
exist $\psi\in\cG$ with $(\phi,\psi)\in\Lambda$ and
$Q(z)\phi=\psi$. But this means
$\phi\in\ker\big(Q(z)-\Lambda\big)$.

(2) Let $z\in\res H^0\cap\res H_\Lambda$. Take any $h\in\cH$ and
denote $f:=(H_\Lambda-z)^{-1}h$; clearly, $f\in\dom H_\Lambda$,
and by theorem~\ref{thm-btgf}(1) there exist uniquely determined
functions $f_z\in\dom H^0$ and $g_z\in\cN_z$ with $f=f_z+g_z$.
There holds
$h=(H_\Lambda-z)f=(S^*-z)f=(S^*-z)f_z+(S^*-z)g_z=(S^*-z)f_z=(H^0-z)f_z$
and $f_z=(H^0-z)^{-1}h$. Moreover, from $\Gamma_1 f_z=0$ one has
$\Gamma_1 f=\Gamma_1 g_z$, $g_z=\gamma(z)\Gamma_1f$, and, therefore,
\begin{equation}
         \label{eq-gfz}
(H_\Lambda-z)^{-1}h=(H^0-z)^{-1}h+\gamma(z)\Gamma_1 f.
\end{equation}

Applying to the both sides of the equality
$f=f_z+\gamma(z)\Gamma_1f$ the operator $\Gamma_2$
one arrives at $\Gamma_2 f=\Gamma_2 f_z+Q(z)\Gamma_1 f$
and
\begin{equation}
        \label{eq-ggfz}
\Gamma_2 f-Q(z)\Gamma_1 f=\Gamma_2 f_z.
\end{equation}
When $h$ runs through the whole space $\cH$, then $f_z$ runs
through $\dom H^0$ and the values $\Gamma_2 f_z$ cover the whole
space $\cG$. At the same time, if $f$ runs through $\dom H_\Lambda$, then
the values $(\Gamma_1f,\Gamma_2f)$ cover the whole $\Lambda$.
It follows then from \eqref{eq-ggfz} that
$\ran\big(\Lambda-Q(z)\big)=\cG$. On the other hand, by (1) one
has $\ker\big(\Lambda-Q(z)\big)=0$ and
$0\in\res\big(\Lambda-Q(z)\big)$. From \eqref{eq-ggfz} one obtains
\begin{equation}
        \label{eq-gammas}
\Gamma_1 f=\big(\Lambda-Q(z)\big)^{-1}\Gamma_2 f_z.
\end{equation}
By theorem~\ref{thm-btgf}(2d) there holds $\Gamma_2f_z=\gamma^*(\Bar z)h$.
Substituting this equality
into \eqref{eq-gammas} and then into \eqref{eq-gfz}
one arrives at the conclusion.

The item (3) follows trivially from the item (2).
\end{proof}

\begin{rem}\label{disj}
Note that the operators $H_\Lambda$ and $H^0$ satisfy
$\dom H_\Lambda\cap\dom H^0=\dom S$ iff $\Lambda$ is a self-adjoint
operator (i.e. is a single-valued); one says that $H_\Lambda$
and $H^0$ are \emph{disjoint} extensions of $S$.
In this case the resolvent formula
conains only operators and has the direct meaning.
As we will see below, in this case one can obtain slightly more
spectral information in comparison with the case when
$\Lambda$ is a linear relation, so it is useful to understand
how to reduce the general case to the disjoint one.
\end{rem}

Let $T$ be the maximal common part of $H^0$ and $H_\Lambda$,
i.e. the restriction of $S^*$ to $\dom H^0\cap \dom H_\Lambda$.
Clearly, $T$ is a closed symmetric operator,
\begin{equation}
             \label{eq-domt}
\dom T=\{f\in\dom S^*:\, \Gamma_1 f=0,\quad \Gamma_2f\in \cL\}
\end{equation}
where $\cL=\ker (\Lambda^{-1})$ is a closed linear subspace of $\cG$.

\begin{lem} Let $\cL$ be a closed linear subspace of $\cG$
and $T$ be defined by \eqref{eq-domt}, then $\dom T^*=\{f\in\dom S^*:\, \Gamma_1 f \in \cL^\perp\}$.
\end{lem}

\begin{proof}
It is clear that both $T$ and $T^*$ are restrictions of $S^*$.
Hence, for any $f\in\dom T$ and $g\in\dom S^*$ one has
\[
W(f,g):=
\langle f| S^*g\rangle-\langle T f| g\rangle=\langle\Gamma_1 f|\Gamma_2 g\rangle
-\langle\Gamma_2 f|\Gamma_1 g\rangle=\langle\Gamma_2 f|\Gamma_1 g\rangle.
\]
As $\Gamma_2(\dom T)=\cL$, one has $W(f,g)=0$ for all $f$ iff $\Gamma_1g\perp \cL$.
\end{proof}

Now one can construct a boundary triple for $T^*$ starting from the boundary triple for $S^*$.

\begin{thm} \label{krein2q}
Let the assumptions of theorem \ref{krein} be satisfied.
Let $\cL$ be a closed subset of $\cG$ and an operator $T$
be defined by \eqref{eq-domt}.
Then $(\Tilde\cG,\Tilde\Gamma_1,\Tilde\Gamma_2)$ is a boundary triple for $T^*$,
where $\Tilde\cG:=\cL^\perp$ with the induced scalar product,
$\Tilde\Gamma_j:= P \Gamma_j$, $j=1,2$, and $P$ is the orthogonal projection
onto $\Tilde \cG$ in $\cG$.
The induced $\Gamma$-field $\Tilde \gamma$
and $\cQ$-function $\Tilde Q$ are  $\Tilde \gamma(z):=\gamma(z)P$,
$\Tilde Q(z):=P Q(z)P$ considered as maps from $\Tilde \cG$ to $\cN_z$
and in $\Tilde \cG$, respectively.
\end{thm}

\begin{proof} Direct verification.
\end{proof}

Returning to the operators $H^0$ and $H_\Lambda$ one sees that, by construction,
they are disjoint extensions of $T$, and in the notation of theorem \ref{krein2q}
they are given by the boundary conditions $\Tilde\Gamma_1 f =0$ and
$\Tilde\Gamma_2 f=L\Tilde\Gamma_1 f$, respectively, where $L$ is a certain self-adjoint operator
in $\Tilde \cG$. Using theorem \ref{krein} one can relate the resolvents of
$H^0$ and $H_\Lambda$ by
\begin{multline}
           \label{eq-ndis}
(H^0-z)^{-1}-(H_\Lambda-z)^{-1}=\Tilde\gamma(z)\big(\Tilde Q(z)-L\big)^{-1}\Tilde\gamma^*(\Bar
z)\\
=\gamma(z)P \big(P Q(z)P-L\big)^{-1}P\gamma^*(\Bar z),
\end{multline}
and $\spec H_\Lambda\setminus\spec H^0=\big\{ z\in\res H^0:\,0\in\spec \big(PQ(z)P-L\big)\big\}$.

The operator $L$ can be calculated, for example, starting from the Cayley
tranform of $\Lambda$ (see proposition \ref{prop-iu}).
Namely, let $U_\Lambda$ be the Cayley transform of $\Lambda$,
then, obviously, $\Tilde\cG=\ker (1-U_\Lambda)^\perp$. 
The Cayley transform of $L$ is then of the form $U_L:=PU_\Lambda P$
considered as a unitary operator in $\Tilde\cG$, and
$L=i(1-U_L)^{-1}(1+U_L)$.

\begin{rem}\label{simp}
For the case of a \emph{simple} symmetric operator
(that is, having no nontrivial invariant subspaces) one can describe
the \emph{whole} spectrum in terms of the limit values of the Weyl function,
and not only the spectrum lying in gaps of a fixed self-adjoint extensions,
see \cite{BrMN,BLu} for discussion. We note that, neverthless, the simplicity
of an operator is a quite rare property in multidimensional problems
which is quite difficult to check.
\end{rem}

\begin{rem} It seems that the notion of boundary value triple
appered first in the papers by Bruk~\cite{bruk} and Kochubei~\cite{Koc},
although the idea goes back to the paper by Calkin~\cite{cal}.
The notion of a $\Gamma$-field and a $\cQ$-function appeared first
in~\cite{KL,LT}, where they were used to describe the generalized
resolvents of self-adjoint extensions. The relationship between
the boundary triples and the resolvent formula in the form
presented in theorems~\ref{thm-btgf} and~\ref{krein} was found by
Derkach and Malamud, but it seems that the only existing
discussion was in~\cite{DM1}, which is hardly available, so we
preferred to provide a complete proof here. The same scheme of the
proof works in more abstract situations, see e.g.~\cite{Der}.
The forumula \eqref{eq-ndis} is borrowed from \cite{Pos3}, but
we give a different proof.
\end{rem}

\begin{rem}
Theorem~\ref{krein} shows that one can express the resolvents of
\emph{all} self-adjoint extensions of a certain symmetric operator
through the resolvent of a \emph{fixed} extension, more precisely,
of the one corresponding to the boundary condition $\Gamma_1 f=0$.
On the other hand, proposition~\ref{prop-gam1} shows that by a
suitable choice of boundary triple one start with any extension.
Formulas expressing $\cQ$-functions associated with different extensions of
the same operator can be found e.g. in \cite{DM,GMT,KK}.
\end{rem}

In view of proposition~\ref{prop-AB} on the parameterization of linear relations
it would be natural to ask whether one can rewrite the Krein resolvent formula
completely in terms of operators without using linear relations. Namely,
if a self-adjoint linear relation $\Lambda$ is given in the form
$\Lambda=\{(x_1,x_2)\in\cG\oplus\cG;\,Ax_1=Bx_2\}$, where $A$ and $B$ are bounded
linear operators satisfying \eqref{eq-AB1} and \eqref{eq-AB2}, can one write
an analogue of the Krein resolvent formula for $H_\Lambda$ in terms of $A$
and $B$? We formulate only here the main result referring to
the recent work~\cite{Pan1} for the proof.

\begin{thm}
Let the assumptions of theorem~\ref{krein} be satisfied
and $A$, $B$ be bounded linear operators in $\cG$
satisfying \eqref{eq-AB1} and \eqref{eq-AB2}. Denote
by $H^{A,B}$ the self-adjoint extension of $S$
corresponding to the boundary conditions $A\Gamma_1 f=B\Gamma_2 f$, then
\begin{itemize}
\item[(1)] For any $z\in\res H^0$ there holds
$\ker(H^{A,B}-z)=\gamma(z)\ker \big(BQ(z)-A\big)$.
\item[(2)] For any $z\in\res H^0\cap\res H^{A,B}$ the operator $BQ(z)-A$
is injective and
\begin{equation}
        \label{eq-kab}
(H^0-z)^{-1}-(H^{A,B}-z)^{-1}=\gamma(z)\big(BQ(z)-A\big)^{-1}B\gamma^*(\Bar
z).
\end{equation}
\item[(3)] If $A$ and $B$ satisfy additionally the stronger condition
\begin{equation}
        \label{eq-AB3}
        0\in\res\begin{pmatrix}
        A & -B\\
        B &A
        \end{pmatrix},
\end{equation}
then $0\in\res \big(BQ(z)-A\big)$ for all $z\in\res H^0\cap\res H^{A,B}$, and, respectively,
$\spec H^{A,B}\setminus\spec H^0=\big\{
z\in\res H^0:\,0\in\spec \big(BQ(z)-A\big)\big\}$.
\end{itemize}
\end{thm}
Note that the condition \eqref{eq-AB3} is satisfied if one uses
the parameterization by the Cayley transform (theorem
\ref{prop-iu}), i.e. $A=i(1+U)$, $B=1-U$ with a unitary $U$, see
proposition~\ref{lem-AB}. Therefore, one can perform a ``uniform''
analysis of all self-adjoint extensions using the single unitary
parameter $U$. Note that the above normalization condition is
trivial for finite deficiency indices, hence the Krein formula has
a particularly transparent form~\cite{AP}.

We note in conclusion that the resolvent formulas \eqref{eq-ndis}
and \eqref{eq-kab} provide two different ways of working
with non-disjoint extensions, and thay can be obtained one from another \cite{Pos3}.

\subsection{Examples}

Here we consider some situations in which boundary triples arise.

\subsubsection{Sturm-Liouville problems}\label{ss-sl}

A classical example comes from the theory of ordinary differential operators.
Let $V\in L^2_\text{loc}(0,\infty)$ be real valued and, for simplicity,
semibounded below. Denote by $S_0$ the closure of the operator
$-\dfrac{d^2}{dx^2}+V$ with the domain $C_0^\infty(0,\infty)$
in the space $\cH:=L^2(0,\infty)$. It is well-known that the deficiency
indices of $S_0$ are $(1,1)$. Using the integration by parts one can easily show
that for the adjoint $S:=S_0^*$ as a boundary triple one can take
$(\mC,\Gamma_1,\Gamma_2)$, $\Gamma_1 f= f(0)$, $\Gamma_2 f=f'(0)$.
Denoting for $z\ne\mC$ by $\psi_z$ the unique $L^2$-solution to
$-\psi''_z+V\psi_z=z\psi_z$ with $\psi_z(0)=1$ we arrive to the induced
Krein $\Gamma$-field, $\gamma(z)\xi=\xi \psi_z$, and the induced
$\cQ$-function $Q(z)=\psi'_z(0)$, which is nothing but
the Weyl-Titchmarsh function. Determining the spectral properties
of the self-adjoint extensions of $S_0$ with the help
of this function is a classical problem of the spectral analysis.

An analogous procedure can be done for Sturm-Louville operators
on a segment. In $\cH:=L^2[a,b]$, $-\infty<a<b<\infty$ consider an operator
$S$ acting by the rule $f\mapsto -f''+Vf$ with the domain
$\dom S=H^2[a,b]$; here we assume that $V\in L^2[a,b]$
is real-valued. It is well-known that $S$ is closed.
By partial integration one easily sees that $(\cG,\Gamma_1,\Gamma_2)$,
\[
\cG=\mC^2,\quad \Gamma_1 f=\begin{pmatrix} f(a)\\f(b)
\end{pmatrix},
\quad
\Gamma_2 f=\begin{pmatrix} f'(a)\\ -f'(b)
\end{pmatrix},
\]
is a boundary triple for $S$. The distinguisged extension $H^0$
corresponding to the boundary condition $\Gamma_1 f=0$ is nothing
but the operator $-d^2/dx^2+V$ with the Dirichlet boundary conditions.

Let two functions $s(\cdot;z),c(\cdot;z)\in\ker(S-z)$ solve
the equation
\begin{equation}
         \label{eq-SL}
-f''+Vf=zf,\quad z\in\mC,
\end{equation}
and satisfy $s(a;z)=c'(a;z)=0$ and $s'(a;z)=c(a;z)=1$.
Clearly, $s$, $c$ as well as their derivatives are entire functions of $z$;
these solutions are linearly independent, and their Wronksian
$w(z)\equiv s'(x;z)c(x;z)-s(x;z)c'(x;z)$ is equal to $1$.
For $z\notin\spec H^0$ one has $s(b;z)\ne0$, and any solution $f$
to~\eqref{eq-SL} can be written as
\begin{equation}
            \label{eq-fxz}
f(x;z)=\frac{f(b)-f(a) c(b;z)}{s(b;z)}\,s(x;z)+f(a)c(x;z),
\end{equation}
which means that the $\Gamma$-field induced by the above boundary triple is
\[
\gamma(z)\begin{pmatrix} \xi_1\\
\xi_2 \end{pmatrix}=
\frac{\xi_2-\xi_1 c(b;z)}{s(b;z)}\,s(x;z)+\xi_1c(x;z).
\]
The calculation of $f'(a)$ and $-f'(b)$ gives
\begin{equation}
               \label{eq-rsz}
\begin{pmatrix}
f'(a;z)\\-f'(b;z)
\end{pmatrix}
=Q(z)
\begin{pmatrix}
f(a;z)\\f(b;z)
\end{pmatrix},
\quad
Q(z)=
\frac{1}{s(b;z)}\,\begin{pmatrix}
-c(b;z) & 1\\
1 & -s'(b;z)
\end{pmatrix},
\end{equation}
and $Q(z)$ is the induced $\cQ$-function.

A number of examples of boundary triples in problems concerning
ordinary differential equations as well as their applications to
scattering problems can be found e.g. in~\cite{BMN,DM}.

The situation becomes much more complicated when dealing with
elliptic differential equations on domains (or manifolds) with
boundary. In this case the construction of a boundary triple
involves certain information about the geometry of the domain,
namely, the Dirichlet-to-Neumann map, see e.g. the recent works~\cite{BL,Pos3}
and the classical paper by Vishik~\cite{Vis}, and the
question on effective description of all self-adjoint boundary
value problems for partial differential equations is still open,
see the discussion in~\cite{EM,EM2} and historical comments
in~\cite{Gru}; an explicit construction of boundary triples
for the Laplacian in a bounded domain is presented in Example 5.5 in~\cite{Pos3}.
We remark that boundary triples provide only one
possible choice of coordinates in the defect subspaces. Another
possibility would be to use some generalization of boundary
triples, for example, the so-called boundary relations resulting
in unbounded Weyl functions~\cite{DHMS,wood}, but it seems that this technique is rather
new and not developed enough for applications.

\subsubsection{Singular perurbations}\label{ss-sing}

Here we discuss the construction of self-adjoint extensions in the
context of the so-called singular perturbations; we follow in part
the constriction of~\cite{Pos2}.

Let $H^0$ be a certain self-adjoint operator in a separable Hilbert space
$\cH$; its resolvent will be denoted by $R^0(z)$, $z\in\res H^0$.
Denote by $\cH_1$ the domain $\dom H_0$ equiped
with the graph norm, $\|f\|_1^2=\|H^0 f\|^2+\|f\|^2$; clearly,
$\cH_1$ is a Hilbert space. Let $\cG$ be another Hilbert space.
Consider a bounded linear map $\tau:\cH_1\to \cG$. We assume that
$\tau$ is surjective and that $\ker\tau$ is dense in $\cH$.

By definition, by a singular perturbation of $H^0$
supported by $\tau$ we mean any self-adjoint extension
of the operator $S$ which is the restriction of $H^0$ to
$\dom S:=\ker\tau$. Due to the above restrictions, $S$
is a closed densely defined symmetric operator.

It is wortwhile to note that singular perturbations just provide
another language for the general theory of self-adjoint extensions.
Namely, let $S$ by any closed densely defined symmetric
operator with equal deficiency indices and $H^0$ be some
its self-adjoint extension. Construct the space $\cH_1$ as above.
Clearly, $\cL:=\dom S$ is a closed subspace of $\cH_1$,
therefore, $\cH_1=\cL\oplus \cL^\perp$. Denoting $\cL^\perp$ by $\cG$
and the orthogonal projection from $\cH_1$ to $\cL^\perp$ by $\tau$,
we see the self-adoint extensions of $S$ are exactly
the singular perturbations of $H^0$ supported by $\tau$.
At the same time, knowing explicitly the map $\tau$ gives
a possibility to construct a boundary triple for $S$.

\begin{prop}\label{prop-gsing}
The maps $\gamma(z)$, $\gamma(z)=\big(\tau R^0(\overline z)\big)^*$,
$z\in\res H^0$, form a Krein $\Gamma$-field for $(S,H^0,\cG)$.
\end{prop}

\begin{proof}
Note that the operator $A:=\tau R^0(z):\cH\to\cG$ is surjective, therefore,
$\ran A^*=\ker A^\perp$. In other words,
\begin{multline}
\ran \gamma(z)=\ker \tau R^0(\overline z)^\perp=\big\{
f\in\cH:\, \tau R^0(\overline z) f=0\big\}^\perp\\
=\big\{(H^0-\overline z) g:\, \tau g=0\big\}^\perp
=\big\{(S-\overline z) g:\, g\in\dom S\big\}^\perp\\
=\ran (S-\overline z)^\perp =\ker(S^*-z)=:\cN_z.
\end{multline}
Let us show that $\gamma(z)$ is an isomorphism of $\cG$ and $\cN_z$.
First note that $\gamma(z)$ is bounded and, as we have shown above, surjective.
Moreover, $\ker\gamma(z)=\ran A^\perp=\cG^\perp=\{0\}$.
Therefore, $\gamma(z):\cG\to\cN_z$ has a bounded inverse defined everywhere
by the closed graph theorem, and the condition~\eqref{Gam1}
is satisfied.

The condition~\eqref{Gam2} is a corollary of the Hilbert resolvent identity.
\end{proof}

Now one can construct a boundary triple for the operator $S^*$.
\begin{prop}\label{prop-sbt}
Take any $\zeta\in\res H^0$ and represent any $f\in\dom S^*$ in the form
$f=f_\zeta+\gamma(\zeta)F$, $f_\zeta\in\dom H^0$, $F\in\cG$, where
$\gamma$ is defined in~proposition~\ref{prop-gsing}. Then $(\cG,\Gamma_1,\Gamma_2)$,
$\Gamma_1f=F$,
$\Gamma_2 f=\dfrac12 \tau\big(f_{\zeta}+f_{\overline \zeta}\big)$,
is a boundary triple for $S^*$. The induced $\Gamma$-field
is $\gamma(z)$, and the induced $\cQ$-function $Q(z)$ has the form
\[
Q(z)=\dfrac12 z \tau R^0(z)\big(\gamma(\zeta)+\gamma(\overline\zeta)\big)
-\dfrac12\tau R^0(z)\big(\zeta\gamma(\zeta)+\overline{\zeta}\gamma(\overline\zeta)\big).
\]
\end{prop}
\begin{proof}
The major part follows from proposition~\ref{prop-gamma5}.
To obtain the formula for $Q(z)$ it is sufficient to see that
for the function $f=\gamma(z)\varphi$, $\varphi\in\cG$,
one has $f_\zeta=\big(\gamma(z)-\gamma(\zeta)\big)\varphi$
and to use the property~\eqref{Gam2}.
\end{proof}

Let us consider in greater detail a special type of the above construction, the so-called
finite rank perturbations \cite{akur}.

Let $H^0$ be as above. For $\alpha\ge0$ denote by $\cH_\alpha$ the domain of the operator $((H^0)^2+1)^{\alpha/2}$
equiped with the norm $\|f\|_\alpha=\big\|\big((H^0)^2+1\big)^{\alpha/2} f\big\|$. The space
$\cH_\alpha$ becomes a Hilbert space, and this
notation is compatible with the above definition of $\cH_1$, i.e. $\cH_1$ is the domain of $H^0$
equiped with the graph norm, and $\cH_0=\cH$. Moreover, for $\alpha<0$ we denote
the completion of $\cH$ with respect to the norm $\|f\|_\alpha=\big\|(H^0)^2+1)^{\alpha/2} f\big\|$.
Clearly, $\cH_\alpha\subset\cH_\beta$ if $\alpha>\beta$.

Take $\psi_j\in\cH_{-1}$, $j=1,\dots,n$. In many problems
of mathematical physics one arrives at operators given by formal expressions of the form
\begin{equation}
           \label{eq-htau}
H=H^0+\sum_{j,k=1}^n \alpha_{jk}\langle \psi_j|\cdot\rangle\psi_k,
\end{equation}
where $\alpha_{jk}$ are certain numbers (``coupling constants'').
This sum is not defined directly, as generically $\psi_j\notin\cH$.
At the same time, the operator $H$ given by this expression is usually supposed to be self-adjoint
(and then one has formally $\alpha_{jk}=\overline{\alpha_{kj}}$).
Denote by $S$ the restriction of $H^0$ to the functions $f\in\dom H^0$ with $\langle \psi_j|f\rangle=0$
for all $j$; we additionally assume that $\psi_j$ are linearly indepedent modulo $\cH$ (otherwise
$S$ might become nondensely defined).
Clearly, for any reasonable definition, the operators $H^0$ and $H$ must coincide
on the domain of $S$.
Therefore, by definition, under an operator given by the right-hand side of~\eqref{eq-htau}
we understand the whole family of self-adjoint extensions of $S$. The boundary triple
for $S^*$ can be easily obtained using the above constructions if one set
\[
\tau f:=\begin{pmatrix}
\langle \psi_1|f\rangle\\
\dots\\
\langle \psi_n|f\rangle
\end{pmatrix}\in\mC^n.
\]
The corresponding $\Gamma$-field from proposition~\ref{prop-gsing} takes the form
\[
\gamma(z)\xi=\sum_{j=1}^n \xi_j h_j(z),\quad
h_j(z):=R^0(z)\psi_j\in\cH,\quad \xi=(\xi_1,\dots\xi_n)\in\mC^n,
\]
and the boundary triples and the $\cQ$-function are obtained using the formulas
of proposition~\ref{prop-sbt}.

Unfortunately, the above construction has a severe disadvantage,
namely, the role of the coefficients $\alpha_{jk}$
in~\eqref{eq-htau} remains unclear. The definition of $H$ using
self-adjoint extensions involves self-adjoint linear relations in
$\mC^n$, and it is difficult to say what is the relationship
between these two types of parameters. In some cases, if both $H$
and $H^0$ have certain symmetries, this relationship can be found
using a kind of renormalization technique~\cite{Kur,KP}. The
situation becomes more simple if in the above construction one has
$\psi_j\in\cH_{-1/2}$ and $H^0$ is semibounded. In this case one
can properly define $H$ given by \eqref{eq-htau} using the
corresponding quadratic form,
\[
h(f,g)=h^0(f,g)+\sum_{j,k=1}^n \alpha_{jk}\langle f|\psi_j\rangle\,\langle\psi_k|g\rangle,
\]
where $h^0$ is the quadratic form associated with $H^0$, see
\cite{Kos}. Also in this case one arrives at boundary triples and
resolvent formulas. A very detailed analysis of rank-one
perturbations of this kind with an extensive bibliography list
is given in~\cite{Sim}.

We also remark that one can deal with operator of the
form~\eqref{eq-htau} in the so-called supersingular case
$\psi_j\notin\cH_{-1}$; the corresponding operators $H$ must be
constructed then in an extended Hilbert or Pontryagin space, see
e.g.~\cite{Kur,Sho,DKS} and references therein.

\subsubsection{Point interactions on manifolds}\label{ss-ppm}

Let $X$ be a manifold of bounded geometry of dimension $\nu$,
$\nu\le 3$.  Let $A=A_j\,dx^j$ be a 1-form on $X$, for simplicity we suppose
here $A_j\in {C}^\infty(X)$. The functions $A_j$ can be considered
as the components of the vector potential of a magnetic field on
$X$. On the other hand, $A$ defines a connection $\nabla_A$ in the
trivial line bundle $X\times \mC\rightarrow X$,
$\nabla_Au=du+iuA$; by $-\Delta_A=\nabla_A^*\nabla_A$ we denote
the corresponding Bochner Laplacian. In addition, we consider a
real-valued scalar potential $U$ of an electric field on $X$. This
potential will be assumed to satisfy the following conditions:
\begin{gather*}
U_+:=\max(U,0)\in {L}^{p_0}_\text{loc}(X),
\quad U_-:=\max(-U,0)\in \sum_{i=1}^n {L}^{p\,{}_i}(X),\\
\quad 2\le p_i\le\infty,\quad 0\le i\le n;
\end{gather*}
we stress that $p_i$ as well as $n$ are not fixed and depend on
$U$. The class of such potentials will be denoted by
$\mathcal{P}(X)$.  For the case $X=\mR^n$ one can study Schr\"odinger
operators with more general potentials from the Kato class \cite{BHL,Simbull}.

We denote by $H_{A,U}$ the operator acting on functions $\phi\in
{C}_0^\infty(X)$ by the rule $H_{A,U}\phi=-\Delta_A\phi+U\phi$.
This operator is essentially self-adjoint in ${L}^2(X)$ and
semibounded below~\cite{BGP0}; its closure will be also denoted by
$H_{A,U}$. It is also known~\cite{BGP0} that
\begin{equation}
          \label{eq-dhc}
\dom H_{A,U}\subset C(X).
\end{equation}
In what follows, the Green function $G_{A,U}(x,y;\zeta)$
of $H_{A,U}$, i.e. the integral kernel of the resolvent $R_{A,U}(\zeta):=(H_{A,U}-\zeta)^{-1}$,
$\zeta\in\res H_{A,U}$, will be of importance.

The most important its
properties for us are the following ones:
\begin{subequations}
\begin{gather}
\label{C1}
\parbox{100mm}{for any $\zeta\in\res H_{A,U}$,
$G_{A,U}$ is continuous in $X\times X$ for $\nu=1$
and in $X\times X\setminus \{(x,x),\,x\in X\}$ for $\nu=2,3$;}\\
\label{C2}
\parbox{100mm}{for $\zeta\in\res H^0$ and $y\in X$ one has
$G_{A,U}(\cdot, y;\zeta)\in L^2(X)$;}\\
\label{C3}
 \parbox{100mm}{for any $f\in L^2(X)$ and $\zeta\in\res H_{A,U}$, the function
$x\mapsto \displaystyle\int_X G_{A,U}(x,y;\zeta)\,f(y)\,dy$
is continuous.}
\end{gather}
\end{subequations}
We remark that for any $f\in\dom H_{A,U}$ and $\zeta\in\res H_{A,U}$
one has $f=R_{A,U}(\zeta)(H_{A,U}-\zeta)f$. Using the Green function
we rewrite this as
\[
f(x)=\int_X G_{A,U}(x,y;\zeta)(H_{A,U}-\zeta)f(y)\,dy \quad \text{a.e.;}
\]
by \eqref{C3} and \eqref{eq-dhc} the both sides are continuous
functions of $x$, therefore, they coincide everywhere, i.e.
\begin{equation}
          \label{eq-fxgx}
f(x)=\int_X G_{A,U}(x,y;\zeta)(H_{A,U}-\zeta)f(y)\,dy,\quad f\in\dom H_{A,U},\text{ for all } x\in X.
\end{equation}

Fix points $a_1,\dots, a_n\in X$, $a_j\ne a_k$ if $j\ne k$,
and denote by $S$ the restriction of $H_{A,U}$
on the functions vanishing at all $a_j$, $j=1,\dots,n$. Clearly, due to~\eqref{eq-dhc}
this restriction is well defined, and $S$ is a closed densely defined
symmetric operator. By definition, by a point perturbation of the operator $H_{A,U}$
supported by the points $a_j$, $j=1,\dots,n$, we mean any self-adjoint
extension of $S$. Now we are actually in the situation of subsubsection~\ref{ss-sing}.
To simplify notation, we denote $H^0:=H_{A,U}$ and change respectively the indices
for the resolvent and the Green function.
Denote by $\tau$ the map
\[
\tau:\dom H^0\ni f\mapsto \begin{pmatrix} f(a_1)\\ \dots\\ f(a_n)
\end{pmatrix}\in\mC^n.
\]
By~\eqref{eq-fxgx} and \eqref{C2}, $\tau$ is bounded in the graph norm of $H^0$. Now let
us use proposition~\ref{prop-gsing}. The map $\tau R^0(\overline z)$ is of the form
\[
f\mapsto
\begin{pmatrix}
\displaystyle \int_X G^0(a_1,y;\overline z)f(y)dy\\
\dots\\
\displaystyle \int_X G^0(a_n,y;\overline z)f(y)dy
\end{pmatrix}.
\]
Calculating the adjoint operator and taking into account the identity
$G^0(x,y;z)=\overline{G^0(y,x;\overline z)}$ we arrive at
\begin{lem}\label{prop-gz}
The map
\begin{equation}
   \label{eq-ppgamma}
\gamma(\zeta):\mC^n\ni (\xi_1,\dots,\xi_n)\mapsto
\sum_{j=1}^n \xi_j G^0(\cdot,a_j;\zeta)\subset L^2(X)
\end{equation}
is a Krein $\Gamma$-field for $(S,H^0,\mC^n)$.
\end{lem}

Let us construct a boundary triple corresponding to the problem.
Use first corollary~\ref{corol-gamma5}.
Choose $\zeta\in\res H^0\subset\mR$; this is possible because $H^0$
is semibounded below. For any $f\in\dom S^*$
there are $F_j\in \mC$ such that $f_\zeta:=f-\sum_j F_j G^0(\cdot,a_j;\zeta)\in\dom H^0$.
The numbers $F_j$ are $\zeta$-independent, and by corollary~\ref{corol-gamma5},
the maps
\begin{equation}
         \label{eq-wtg}
\widetilde \Gamma_1 f:= (F_1,\dots,F_n),\quad
\widetilde \Gamma_2 f=\big(
f_\zeta(a_1),\dots,f_\zeta(a_n)\big)
\end{equation}
form a boundary triple for $S^*$. Nevertheless, such a construction is rarely used in practice
due to its dependence on the energy parameter.
We modifiy the above considerations using some information
about the on-diagonal behavior of $G^0$.

Consider the case $\nu=2$ or $3$. As shown in~\cite{BGP1}, there exists
a function $F(x,y)$ defined for $x\ne y$ such that
for any $\zeta\in\res H^0$ there exists another
function $G^0_\text{ren}(x,y;\zeta)$ continuous in $X\times X$
such that
\begin{equation}
         \label{eq-ren}
G^0(x,y;\zeta)=F(x,y)+G^0_\text{ren}(x,y;\zeta),
\end{equation}
and we additionally request $F(x,y)=\overline F(y,x)$.
It is an important point that under some assumptions
the function $F$ can be chosen independent of the magnetic
potential $A_j$ and the scalar potential $U$.
For example, if $\nu=2$ one can always set
$F(x,y)=\log\dfrac{1}{d(x,y)}$.
In the case $\nu=3$ the situation becomes more complicated.
For example, for two scalar potentials $U$ and $V$
satisfying the above conditions one can take the same
function $F$ for the operators $H_{A,U}$ and $H_{A,V}$
provided $U-V\in L^q_\text{loc}(X)$ for some $q>3$.
In paritucular, for any $U$ satisying the above conditions and,
additionally, $U\in L^q_\text{loc}(X)$, for the operator
$H_{0,U}$ one can always put $F(x,y)=\dfrac{1}{4\pi d(x,y)}$.

For the Schr\"odinger operator with a uniform magnetic field in $\mR^3$,
$H^0=(i\nabla+A)^2$, where $\nabla\times A=:B$ is constant,
one can put $F(x,y):=\dfrac{e^{iBxy/2}}{4\pi|x-y|}$.
For a detailed discussion of on-diagonal singularities we refer to our paper~\cite{BGP1}.

Let us combine the representation~\eqref{eq-ren} for the Green function
and the equality $\dom S^*=\dom H^0+\cN_\zeta$.
Near each point $a_j$, any function $f\in\dom S^*$ has the following asymptotics:
\[
f(x)=f_j+ F_j F(x,a_j)+o(1),\quad f_j,F_j\in\mC.
\]
\begin{prop} The triple $(\mC^n,\Gamma_1,\Gamma_2)$ with $\Gamma_1 f=(F_1,\dots,F_n)\in\mC^n$
and $\Gamma_2 f=(f_1,\dots,f_n)\in\mC^n$ is a boundary triple for $S^*$.
\end{prop}

\begin{proof} Let us fix some $\zeta\res H^0\cap\mR$.
Comparing the maps $\Gamma_j$ with the maps $\widetilde\Gamma_j$
from \eqref{eq-wtg} one immediately see $\Gamma_1\equiv \widetilde\Gamma_1$.
Furthermore, $\Gamma_2 f=\widetilde\Gamma_2 f+B\widetilde \Gamma_1$, where
$B$ is a $n\times n$ matrix,
\[
B_{jk}=\begin{cases} G^0(a_j,a_k;\zeta) & \text{if } j\ne k,\\
G^0_\text{ren}(a_j,a_j;\zeta) & \text{otherwise.}
\end{cases}
\]
As $B=B^*$, it remains to use proposition~\ref{prop-btb}.
\end{proof}

Clearly, the map~\eqref{eq-ppgamma} is the Krein $\Gamma$-field
induced by the boundary triple $(\mC^n,\Gamma_1,\Gamma_2)$. The calculation
of the corresponding $\cQ$-function $Q(\zeta)$ gives
\[
Q_{jk}(\zeta)=\begin{cases} G^0(a_j,a_k;\zeta) & \text{if } j\ne k,\\
G^0_\text{ren}(a_j,a_j;\zeta) & \text{otherwise.}
\end{cases}
\]
We note that the calculating of the $\cQ$-function needs \emph{a
priori} the continuity of the Green function (otherwise the values
of the Green function at single points would not be defined). A
bibliography concerning the analysis of operators of the above
type for particular Hamiltonians $H^0$ can be found e.g. in
\cite{AGHH}.

The above construction can generalized to the case of point
perturbations supported by non-finite (but countable) sets
provided some uniform discreteness, we refer to \cite{GMC} for the
general theory, to \cite{AG1,BG2,Gey} for the analysis of
periodic configurations, and to~\cite{BMG,d1,d2,d3} for multidimensional models
with random interactions.

For analysis of interactions supported by submanifolds of higher
dimension we refer to \cite{EK1,EY,Pos1,CDP,CDP2,BDE} and references
therein.

\subsubsection{Direct sums and hybrid spaces}

Assume that we have a countable family of closed linear operators $S_\alpha$
in some Hilbert spaces $\cH_\alpha$, $\alpha\in\cA$, having boundary triples
$(\cG^\alpha,\Gamma^\alpha_1,\Gamma^\alpha_2)$.
Denote by $H^0_\alpha$ the corresponding distinguished extensions,
$H^0_\alpha:=S_\alpha|_{\ker\Gamma^\alpha_1}$.
We impose some additional regularity conditions, namely, that:
\begin{itemize}
\item there exist constants $a$ and $b$ such that for any $\alpha\in\cA$
and $f_\alpha\in\dom S_\alpha$ there holds
$\|\Gamma^\alpha_{1/2} f_\alpha\|\le a \|S_\alpha f_\alpha\|+b\|f_\alpha\|$,
\item for any $(\xi^{1/2}_\alpha)\in\bigoplus_{\alpha\in\cA}\cG^\alpha$ there is
$(f_\alpha)\in\bigoplus_{\alpha\in\cA}\cH_\alpha$, $f_\alpha\in\dom S_\alpha$,
with $\Gamma_{1/2}^\alpha f_\alpha=\xi^{1/2}_\alpha$.
\end{itemize}
The above conditions are obviously satisfied if, for example,
the operators $S_\alpha$ are copies of a finite set of operators,
and the same holds for the boundary triples. Another situation
where the conditions are satisfied, is provided
by the operators $S_\alpha=-\dfrac{d^2}{dx^2}+U_\alpha$ acting in $L^2[a_\alpha,b_\alpha]$
with the domains $H^2[a_\alpha,b_\alpha]$ provided that there are constants
$l_1,l_2,C$ such that $l_1\le|a_\alpha-b_\alpha|<l_2$ and $\|U_\alpha\|_{L^2}<C$
and that the boundary triples are taken as in subsection~\ref{ss-sl}, see~\cite{Pan3}
for details.

Under the above conditions, the operator $S:=\bigoplus_{\alpha\in\cA} S_\alpha$
acting in $\cH:=\bigoplus_{\alpha\in\cA} \cH_\alpha$
is closed and has a boundary triple $(\cG,\Gamma_1,\Gamma_2)$,
\[
\cG:=\bigoplus_{\alpha\in\cA} \cG_\alpha,
\quad \Gamma_j:=\bigoplus_{\alpha\in\cA} \Gamma_j^\alpha,\quad j=1,2.
\]
Moreover, the corresponding distinguished extension $H^0$
and the induced Krein maps $\gamma$ and $Q$ are also direct sums, i.e., at least
\[
H^0:=\bigoplus_{\alpha\in\cA} H^0_\alpha,\quad
\gamma(z)=\bigoplus_{\alpha\in\cA} \gamma^\alpha(z),
\quad
Q(z)=\bigoplus_{\alpha\in\cA} Q^\alpha(z).
\]
Note that $\gamma(z)$ and $Q(z)$ are defined only for $z\notin\spec H^0\equiv
\overline{\bigcup_{\alpha\in\cA} \spec H^0_\alpha}$. Let us show
how this abstract construction can be used to define Schr\"odinger
operators on hybrid spaces, i.e. on configurations consisting
of pieces of different dimensions.

Let $M_\alpha$, $\alpha\in\cA$, be  a countable family of manifolds
as in subsubsection~\ref{ss-ppm}. 
Fix
several points $m_{\alpha j}\in\cM_\alpha$, $j=1,\dots,n_\alpha$.
We interpret these points as points of glueing. More precisely,
we consider a matrix $T$ with the entries $T_{(\alpha j)(\beta k)}$
such that $T_{(\alpha j)(\beta k)}=1$ if the point $m_{\alpha j}$
is identified with $m_{\beta k}$ (i.e. point $m_{\alpha j}$ of $M_\alpha$
is glued to the point $m_{\beta k}$ of $M_\beta$),
and $T_{(\alpha j)(\beta k)}=0$ otherwise. The obtained topological space
is not a manifold as it has singularities at the points of glueing;
we will refer it to as a \emph{hybrid manifold}. Our aim is to show how to define
a Schr\"odinger operator in such a structure.

On of the manifolds $M_\alpha$ consider Schr\"odinger operators
$H_\alpha$ as in subsubsection~\ref{ss-ppm}.
To satisfy the above regularity conditions
we request that these operators are copies of a certain finite family.
For $\alpha\in\cA$ denote by $S_\alpha$ the restriction of $H_\alpha$
to the functions vanishing at all the points $m_{\alpha j}$
and construct a boudary triple $(\mC^{m_\alpha},\Gamma_{\alpha 1},\Gamma_{\alpha 2})$
for $S_\alpha^*$ as in  in subsubsection~\ref{ss-ppm}. Clearly,
as a boundary triple for the operator $S^*$, $S:=\bigoplus_{\alpha\in\cA} S_\alpha$,
in the space $L^2(M):=\bigoplus_{\alpha\in\cA} L^2(M_\alpha)$
one take $\big(\cG,\Gamma_1,\Gamma_2\big)$ with
$\cG:=\bigoplus_{\alpha\in\cA} \mC^{n_\alpha}$, $\Gamma_j (f_\alpha)=
(\Gamma_{\alpha j}f_\alpha)$, $j=1,2$. Under a Schr\"odinger
operator on $L^2(M)$ one can mean any self-adjoint extension of $S$.
To take into account the way how the manifolds are glued with each other,
one should restrict the class of possible boundary conditions.
A reasonable idea would be to consider boundary conditions
of the form $A\Gamma_1=B\Gamma_2$ such that $A_{(\alpha j)(\beta k)}=B_{(\alpha j)(\beta k)}=0$ if $T_{(\alpha j)(\beta k)}=0$, i.e.
assuming that each boundary condition involves only points glued to each other.

The analysis of generic Schr\"odinger operators on hybrid
manifolds is hardly possible, as even Schr\"odinger operators on a
single manifold do not admit the complete analysis. One can say
some more about particular configuration, for example, if one has
only finitely many pieces $M_\alpha$ and they all are
compact~\cite{Exn2}. Some additional information can be obtained
for periodic configurations~\cite{BEG, BG1}.

One can extend the above construction by combining operators
from subsection~\ref{ss-sl} and \ref{ss-ppm}; in this way one arrive
at a space with consists of manifolds connected with each other through
one-dimensional segments.

One can also take a direct sum of operators from subsubsection~\ref{ss-sl}
to define a Schr\"odinger operator on a configuration consisting
of segments and halflines connected with each other; such operators
are usually referred to as \emph{quantum graphs}, and their
analysis becomes very popular in the last decades, see e.g.~\cite{QG3}
for the review and recent developments.

\section{Classification of spectra of self-adjoint operators}\label{sec3}

\subsection{Classification of measures}

Here we recall briefly some concepts of the measure theory.

Let $\cB$ be the set of all the Borel subsets of a locally compact
separable metric space $X$. A mapping $\mu:\cB\rightarrow
[0,+\infty]$ is called a {\it positive Borel measure} on $X$ if
it is $\sigma$-additive (i.e. $\displaystyle
\mu(\bigcup\limits_k\,B_k)=\sum\limits_k\,\mu(B_k)$ for every
countable family $(B_k)$ of  mutually not-intersecting sets from
$\cB$) and has the following regularity properties:
\begin{itemize}
\item $\mu(K)<\infty$ for every compact $K\subset X$;
\item for every $B\in\cB$ there holds
$\mu(B)=\sup\{\mu(K):\,K\subset B\,,\,K\text{ is compact}\}=
\inf\{\mu(G):\,G\supset B\,,\,G\text{ is open}\}$\,.
\end{itemize}

A complex valued Borel measure on $X$ is a
$\sigma$-additive mapping $\mu:\cB\rightarrow\mC$ such that the
variation $|\mu|$ of $\mu$ defined on $\cB$ by
\[
|\mu|(B)=\sup\,\sum|\mu(B_k)|\,,
\]
where the supremum is taken over all finite families $(B_k)$ of
mutually non-intersecting sets $B_k$ from $\cB$ such that $\bigcup
B_k\subset B$, is a Borel measure. For a positive measure $\mu$ one has
$|\mu|=\mu$. If $|\mu|(X)<\infty$, then $\mu$ is called
{\it finite} (or {\it bounded}) and $|\mu|(X)$ is denoted also by
$\|\mu\|$. We will denote by $\cM(X)$ (respectively, by
$\cM^+(X)$) the set of all complex Borel measures (respectively,
the set of all positive Borel measures) on $X$; if $X=\mR$ we
write simply $\cM$ and $\cM^+$. It is clear that $\cM(X)$ is a
complex vector space (even a complex vector lattice) and the
subset $\cM^{\rb}(X)$ of all bounded measures from $\cM(X)$ is a
vector subspace of $\cM(X)$ which is a Banach space with respect
to the norm $\|\mu\|$.

Ona says that a measure $\mu$ is {\it concentrated} on a Borel set $S\in\cB$, if
$\mu(B)=\mu(B\cap S)$ for all $B\in\cB$. Let $\mu_1$ and $\mu_2$
be two measures; they are called {\it disjoint} or {\it mutually
singular}, if there exists two disjoint Borel set $S_1$ and $S_2$
such that $\mu_j$ is concentrated on $S_j$ $(j=1,2)$; we will
write $\mu_1\rd\mu_2$ if $\mu_1$ and $\mu_2$ are disjoint. The
measure $\mu_1$ is said to be {\it subordinated} to $\mu_2$ (or
{\it absolutely continuous with respect to $\mu_2$}) if every
$|\mu_2|$-negligible Borel set is simultaneously
$|\mu_1|$-negligible. According to the Radon--Nikodym theorem, the
following assertions are equivalent: (1) $\mu_1$ is subordinated
to $\mu_2$; (2) there exists a Borel function $f$ such that
$\mu_1=f\mu_2$ (in this case $f\in L^1_{\rm loc}(X,\mu_2)$ and $f$
is called the \emph{Radon--Nikodym derivative} of $\mu_1$ with
respect to $\mu_2$). If $\mu_1$ is subordinated to $\mu_2$ and
simultaneously $\mu_2$ is subordinated to $\mu_1$ (i.e., if
$\mu_1$ and $\mu_2$ have the same negligible Borel sets), then
$\mu_1$ and $\mu_2$ are called {\it equivalent} (in symbols:
$\mu_1\sim \mu_2$). For a subset $M\subset\cM(X)$ we denote
$M^{\rd}=\{\mu\in\cM(X):\,\mu\rd\nu\,\,\forall\,\nu\in M\}$;
$M^{\rd}$ is a vector subspace of $\cM(X)$. A {\it subspace} $M
\subset\cM(X)$ is called a {\it band} (or a {\it component}) in
$\cM(X)$, if $M=M^{\rd\rd}$. For every subset $L\in\cM(X)$ the set
$L^{\rd}$ is a band; the band $L^{\rd\rd}$ is called the {\it band
generated by $L$}. In particular, if $\mu\in\cM(X)$, then the band
$\{\mu\}^{\rd\rd}$ consists of all $\nu$ which are subordinated to
$\mu$. Moreover, $\mu_1$ is subordinated to $\mu_2$ if and only if
$\{\mu_1\}^{\rd\rd}\subset\{\mu_2\}^{\rd\rd}$; in particular,
$\mu_1\sim\mu_2$ if and only if
$\{\mu_1\}^{\rd\rd}=\{\mu_2\}^{\rd\rd}$. The bands $M$ and $N$ are
called {\it disjoint}, if $\mu\rd\nu$ for every pair $\mu\in M$
and $\nu\in N$.

The family $(L_\xi)_{\xi\in\Xi}$ of bands in $\cM(X)$ such that
$\mu\in\displaystyle\bigg(\bigcup\limits_{\xi\in\Xi}\,L_\xi\bigg)^{\rd}$
implies $\mu=0$ is called {\it complete}. Let a complete family of
mutually disjoint bands $L_\xi$, $\xi\in\Xi$, is given. Then for
every $\mu\in\cM(X)$, $\mu\ge0$, there exists a unique family
$(\mu_\xi)_{\xi\in\Xi}$, $\mu_\xi\in L_\xi$, such that
$\displaystyle\mu=\sup\limits_{\xi\in\Xi}\,\mu_\xi$, where the supremum
is taken in the vector lattice $\cM(X)$; $\mu_\xi$ is called the
{\it component} of $\mu$ in $L_\xi$. If, in addition, the family
$(L_\xi)$ is finite, then $\cM(X)$ is the direct sum of $(L_\xi)$ and
$\mu$ is the sum of its components $\mu_\xi$.

In particular, if $L$ is a band, then the pair $(L,L^\rd)$ is a
complete family of mutually disjoint bands; the component of a
measure $\mu$ in $L$ coincides in this case with the projection of
$\mu$ onto $L$ parallel to $L^\rd$ and denoted by $\mu^L$. The
measure $\mu^L$ is completely characterized by the following two properties:
\begin{itemize}
\item $\mu^L\in L$;
\item $(\mu-\mu_L)\rd L$.
\end{itemize}

A Borel measure $\mu$ is called a {\it
point} or {\it atomic} measure, if it is concentrated on a
countable subset $S\subset X$. A point $s\in S$ such that
$\mu(\{s\})\ne0$ is called an atom for $\mu$. For every set
$B\in\cB$ there holds
\[
\mu(B)=\sum_{s\in B\cap S}\,\mu(\{s\})\,.
\]
The set of all complex point Borel measures on $X$ we will denote
by $\cM_{\rp}(X)$, this is a band in $\cM(X)$.

A Borel measure $\mu$ is called a {\it
continuous} measure, if $\mu(\{s\})=0$ for every $s\in X$. The set
of all continuous Borel measures on $X$ we will denote by
$\cM_{\rc}(X)$. It is clear that
$\cM_{\rc}^{\rd}(X)=\cM_{\rp}(X)$,
$\cM_{\rp}^{\rd}(X)=\cM_{\rc}(X)$, and $\cM(X)$ is the direct sum
of the bands $\cM_{\rp}(X)$ and $\cM_{\rc}(X)$.

Let now $X$ be a locally compact separable metric group with the
continuous Haar measure. We fix the left Haar measure
$\lambda$; if $X$ is a compact space, we choose $\lambda$ to be
normalized, in the case $X=\mR$ we choose $\lambda$ to be the
Lebesgue measure. A measure $\mu$ on $X$ is called {\it absolutely
continuous}, if it is subordinated to $\lambda$ and {\it
singular}, if it is disjoint to $\lambda$ (it is clear that these
definitions are independent on the particular choice of
$\lambda$). The set of all absolutely continuous Borel measures on
$X$ (respectively, the set of all singular Borel measures on $X$)
will be denoted by $\cM_{\ra\rc}(X)$ (respectively, by
$\cM_{\rs}(X)$). In particular, $\cM_{\rp}(X)\subset\cM_{\rs}(X)$.
It is clear that $\cM_{\rs}^{\rd}(X)=\cM_{\ra\rc}(X)$,
$\cM_{\ra\rc}^{\rd}(X)=\cM_{\rs}(X)$, and $\cM(X)$ is the direct
sum of the bands $\cM_{\ra\rc}(X)$ and $\cM_{\rs}(X)$.

A Borel measure $\mu$ on $X$ is called a {\it singular continuous}
measure, if it is simultaneously continuous and singular. The set
of all singular continuous Borel measures on $X$ we will denote by
$\cM_{\rs\rc}(X)$; this is a band in $\cM(X)$. By definition
$\mu\in\cM_{\rs\rc}$ if and only if $\mu$ is concentrated on a
Borel set of zero Haar measure and $\mu(S)=0$ for every countable
set $S$.

According to the {\it Lebesgue
decomposition theorem} each Borel measure $\mu$ on the group $X$
is decomposable in a unique way into the sum
\[
\mu=\mu^{\rm p}+\mu^{\rm ac}+\mu^{\rm sc}\,,
\]
where $\mu^{\rp}\in\cM_{\rp}(X)$,
$\mu^{\ra\rc}\in\cM_{\ra\rc}(X)$,
$\mu^{\rs\rc}\in\cM_{\rs\rc}(X)$. We will denote also $\mu^{\rc}=
\mu^{\ra\rc}+\mu^{\rs\rc}$ and $\mu^{\rs}=
\mu^{\rp}+\mu^{\rs\rc}$. It is clear that
$\mu^{\rc}\in\cM_{\rc}(X)$, $\mu^{\rs}\in\cM_{\rs}(X)$.

\subsection{Spectral types and spectral measures}

In this section, $A$ denotes a
self-adjoint operator in a Hilbert space $\cH$, $\res A$ is the resolvent set of
$A$, $\spec A:=\mC\setminus\res A$ is the spectrum of $A$. For
$z\in\res A$ we denote $R(z;A):=(A-z)^{-1}$ (the resolvent of
$A$).

The first classification of spectra is related to the stability
under compact perturbations of $A$. By definition, the {\it
discrete} spectrum of $A$ (it is denoted by $\spec_{\rm dis} A$)
consists of all {\it isolated} eigenvalues of {\it finite}
multiplicity, and the {\it essential} spectrum of $A$, $\spec_{\rm
ess}A$, is the complement of the discrete spectrum in the whole spectrum:
$\spec_{\rm
ess}A=\spec A\setminus\spec_{\rm dis} A$. By the famous Weyl
perturbation theorem, for a point $x_0\in\spec A$ the following
assertions are equivalent
\begin{itemize}
\item $\zeta\in\spec_{\rm ess}A$,
\item for every compact operator $K$ in $\cH$ there holds
$\zeta\in\spec_\text{ess}(A+K)$.
\end{itemize}

The second classification is related to the
transport and scattering properties of a quantum mechanical system
with the Hamiltonian $H=A$.  For
$\Omega\in\cB$ denote $P_\Omega(A)=\chi_\Omega(A)$, where
$\chi_\Omega$ is the indicator function of the subset
$\Omega\subset\mR$; $P_\Omega(A)$ is the spectral projector for
$A$ on the subset $\Omega$. The mapping  $\cB\ni\Omega\mapsto
P_\Omega(A)$ is called the projection valued measure associated
with $A$ (the resolution of identity). For every pair
$\varphi,\psi\in\cH$, the mapping
\[
\cB\ni\Omega\mapsto\big\langle\varphi\big|P_\Omega(A)\psi\big\rangle= \big\langle
P_\Omega(A)\varphi\big|P_\Omega(A)\psi\big\rangle
\]
is a complex Borel measure on the real line $\mR$ which is called
the {\it spectral measure} associated with the triple
$(A,\varphi,\psi)$ and denoted by $\mu_{\varphi,\psi}$ (or more
precisely, by $\mu_{\varphi,\psi}(\cdot\,;A))$. If $\varphi=\psi$,
then $\mu_\varphi\equiv\mu_{\varphi,\varphi}$ is a bounded
positive Borel measure on $\mR$,
\[
\cB\ni\Omega\mapsto\langle\varphi|P_\Omega(A)\varphi\rangle=
\|P_\Omega(A)\varphi\|^2\,,
\]
with the norm $\|\mu_\varphi\|=\|\varphi\|^2$. Therefore,
$\mu_{\varphi,\psi}$ is bounded and
\[
|\mu_{\varphi,\psi}|(\Omega)\le\left[\mu_\varphi(\Omega)\mu_\psi(\Omega)
\right]^{1/2}\,.
\]
Moreover, ${\rm supp}\,\mu_{\varphi,\psi}\subset\spec A$.

According to the Riesz--Markov theorem, for a bounded complex
Borel measure $\mu$ on $\mR$ the following three conditions are
equivalent:
\begin{itemize}
\item $\mu=\mu_{\varphi,\psi}$ for some $\varphi,\psi\in\cH$;
\item for every continuous function $f$ on $\mR$ with compact
support
\[
\langle\varphi|f(A)\psi\rangle=\int_{\mR}\,f(x)\,d\mu(x)\,;
\]
\item for every bounded Borel function $f$ on $\mR$
\[
\langle\varphi|f(A)\psi\rangle=\int_{\mR}\,f(x)\,d\mu(x)\,.
\]
\end{itemize}

The following proposition is obvious.
\begin{prop}\label{prop1133} For a Borel subset $\Omega\subset\mR$ the following
assertions hold:
\begin{enumerate}
\item[(1)] $\mu_\varphi(\Omega)=0$ if and only if
$P_\Omega\varphi=0$.
\item[(2)] $\mu_\varphi$ is concentrated on $\Omega$
if and only if $P_\Omega\varphi=\varphi$.
\end{enumerate}
\end{prop}

\begin{prop}\label{prop1134}
The following assertions take place.
\begin{enumerate}
\item[(1)] $\mu_{\varphi,\psi}$ {\it and $\mu_{\varphi+\psi}$ are
subordinated to $\mu_\varphi+\mu_\psi$};
\item[(2)] $\mu_{a\varphi}=|a|^2\mu_{\varphi}$ {\it for every
$a\in\mC$};
\item[(3)] {\it if $\mu_{\varphi}\rd\mu_{\psi}$, then
$\varphi\perp\psi$};
\item[(4)] {\it if $\mu_{\varphi}\rd\mu_{\psi}$, and $B=f(A)$
where $f$ is a bounded Borel function, then
$\mu_{B\varphi}\rd\mu_{\psi}$};
\item[(5)] {\it if $\mu_{\varphi_n}\rd\mu_{\psi}$ for a sequence
$\varphi_n$ from $\cH$, and $\varphi_n\to\varphi$ in $\cH$, then
$\mu_{\varphi}\rd\mu_{\psi}$}.
\end{enumerate}
\end{prop}

\begin{proof}

(1) For $B\in\cB$ we have:
\begin{align*}
|\mu_{\varphi,\psi}|(B)&\le\|P_B(\varphi)\|\|P_B(\psi)\|,\\
\big[\mu_{\varphi+\psi}(B)\big]^{1/2}=\|P_B(\varphi+\psi)\|&\le
\|P_B(\varphi\|+\|P_B(\psi)\|,
\end{align*}
hence $|\mu_{\varphi,\psi}|(B)=\mu_{\varphi+\psi}(B)=0$, if
$\mu_{\varphi+\psi}(B)=0$.

(2) Trivial.

(3) Let $S,T\in\cB$, $S\cap T=\emptyset$, $\mu_{\varphi}$ be
concentrated on $S$ and $\mu_{\psi}$ be concentrated on $T$. Then,
according to proposition \ref{prop1133},
$\langle\varphi,\psi\rangle=\langle P_S\varphi,P_T\psi\rangle=
\langle \varphi,P_SP_T\psi\rangle=0\,$.

(4) Let $S$ and $T$ be as in item (3). Then
$P_S\varphi=\varphi$, $P_T\psi=\psi$. Hence
$P_Sf(A)\varphi=f(A)P_S\varphi=f(A)\varphi$ and we can refer to
proposition \ref{prop1133}

(5) Let $S_n,T_n\in\cB$, $S_n\cap T_n=\emptyset$,
$\mu_{\varphi_n}$ be concentrated on $S_n$ and $\mu_{\psi}$ be
concentrated on $T_n$. Set $T=\bigcap T_n$, $S=\mR\setminus T$.
Then $\mu_{\varphi_n}$ is concentrated on $S$ for every $n$ and
$\mu_{\psi}$ is concentrated on $T$. By proposition
\ref{prop1133}, $P_S\varphi_n=\varphi_n$, $P_T\psi=\psi$. As a
result, we have $P_S\varphi=\varphi$, hence
$\mu_\varphi\rd\mu_\psi$ by proposition \ref{prop1133}.
\end{proof}

Let $L$ be a band in $\cM$. Denote
$\cH_L\equiv\{\psi\in\cH:\,\mu_\psi\in L\}$. Then by proposition \ref{prop1134}
$\cH_L$ is a closed $A$-invariant subspace of $\cH$. Moreover, let
$(L_\xi)_{\xi\in\Xi}$ be a complete family of bands in $\cM$. Then
$\cH$ is the closure of the linear span of the family of closed
$A$-invariant subspaces $\cH_{L_\xi}$. If, in addition, $L_\xi$
are mutually disjoint then $\cH$ is the orthogonal sum of
$\cH_{L_\xi}$. In particular, $\cH_L^\bot=\cH_{L^\rd}$. Moreover,
the following proposition is true.

\begin{prop}
Let $\varphi\in\cH$ and $\varphi_L$ is the
orthogonal projection of $\varphi$ onto $\cH_L$. Then
\begin{enumerate}
\item[(1)] {\it $\mu_\varphi-\mu_{\varphi_L}\ge0$ and is
subordinated to $\mu_{\varphi-\varphi_L}$};

\item[(2)] $\mu_{\varphi_L}=\mu^L_\varphi$.
\end{enumerate}
\end{prop}

\begin{proof}
(1) First of all we show that
$\mu_\varphi-\mu_{\varphi_L}\ge0$. Let $B\in\cB$, then
$(\mu_\varphi-\mu_{\varphi_L})(B)=\|P_B\varphi\|^2-
\|P_BP_{\cH_L}\varphi\|^2$. Since $\cH_L$ is $A$-invariant,
$P_BP_{\cH_L}=P_{\cH_L}P_B$, therefore
$(\mu_\varphi-\mu_{\varphi_L})(B)=\|P_B\varphi\|^2-
\|P_{\cH_L}P_B\varphi\|^2\ge0$.

Further we have for $B\in\cB$
\begin{multline*}
(\mu_\varphi-\mu_{\varphi_L})(B)\\=\|P_B\varphi\|^2-
\|P_B\varphi_L\|^2=(\|P_B\varphi\|+
\|P_B\varphi_L\|)(\|P_B\varphi\|-\|P_B\varphi_L\|)\\
\le
2\|\varphi\|\|P_B(\varphi-\varphi_L)\|=
2\|\varphi\|\,\left[\mu_{\varphi-\varphi_L}(B)\right]^{1/2}\,,
\end{multline*}
and the proof of the item is complete.

(2) $\mu_{\varphi_L}\in L$, and according to item (1)
$\mu_\varphi-\mu_{\varphi_L}\in L^\rd$.
\end{proof}

Since $\cH_L$ is $A$ invariant, the
restriction of $A$ to $\cH_L$ is a self-adjoint operator in
$\cH_L$. The spectrum of this restriction is denoted by
$\spec_L A$ and is called {\it $L$-part} of the spectrum of $A$.

It is clear that for a point $x_0\in\mR$ the following assertions
are equivalent:
\begin{itemize}
\item $x_0\in\spec A$;
\item for any $\varepsilon>0$ there exists $\varphi\in\cH$ such that
$\mu_{\varphi}(x_0-\varepsilon,x_0-\varepsilon)>0$.
\end{itemize}

Therefore, we have
\begin{prop}
The following assertions are equivalent:
\begin{itemize}
\item $x_0\in\spec_L A$;
\item
for any $\varepsilon>0$ there exists $\varphi\in\cH$ with
$\mu^L_{\varphi}(x_0-\varepsilon,x_0-\varepsilon)>0$;
\item
for any $\varepsilon>0$ there exists $\varphi\in\cH_L$ with
$\mu_{\varphi}(x_0-\varepsilon,x_0-\varepsilon)>0$.
\end{itemize}
\end{prop}

Let $(L_\xi)_{\xi\in\Xi}$ be a complete
family of mutually disjoint bands in $\cM$. Then
\[
\spec A=\overline{\bigcup\limits_{\xi\in\Xi}\,{\spec}_{L_\xi} A}\,.
\]

Let $L$ be a band in $\cM$, $N=L^\rd$ and
$\Omega\in\cB$. If $B\cap\spec_M A=\emptyset$, then we say that
$A$ has only $L$-spectrum on $\Omega$ (or the spectrum of $A$ on
$\Omega$ is purely $L$).

Denote now $\cH_\rj$, where
$\rj\in\{\rp,\,\ra\rc,\,\rs\rc,\,\rs,\,\rc\}$, the subspace
$\cH\equiv\cH_{\cM_\rj}$. Then the spectrum of the restriction of
$A$ to $\cH_\rj$ is denoted $\spec_\rj A$. In particular,
\begin{itemize}
\item $\cH=\cH_\rp\oplus\cH_{\ra\rc}\oplus\cH_{\rs\rc}$,
therefore
$\spec A=\spec_\rp A\cup\spec_{\ra\rc} A\cup\spec_{\rs\rc} A$.
The part $\spec_\rp A$ is called the {\it point} spectrum of $A$,
$\spec_{\ra\rc} A$ is called the {\it absolutely continuous}
spectrum of $A$ and $\spec_{\rs\rc} A$ is called the {\it
singular continuous} spectrum~of~$A$.

\item $\cH=\cH_\rp\oplus\cH_{\rc}$, therefore
$\spec A=\spec_\rp A\cup\spec_{\rc} A$. The part $\spec_\rc A$
is called the {\it continuous} spectrum of $A$.

\item $\cH=\cH_{\ra\rc}\oplus\cH_{\rs}$, therefore
$\spec A=\spec_{\ra\rc} A\cup\spec_{\rs} A$. The part
$\spec_\rs A$ is called the {\it singular} spectrum of $A$.
\end{itemize}

Consider the point part of the spectrum
in detail. The set of all eigenvalues of $A$ is denoted by
$\spec_{\rp\rp} A$ and is called {\it pure point} spectrum of
$A$. In particular, for a point $x_0\in\mR$ the following
assertions are equivalent:
\begin{itemize}
\item $x_0\in\spec_{\rp\rp} A$;
\item $\mu_{\varphi}(\{x_0\})>0$ for some
$\varphi\in\cH$.
\end{itemize}

\begin{prop}
Let $\delta_a$, where $a\in\mR$, be the Dirac measure concentrated
on $a$. Then for $a\in\mR$ and $\varphi\in\cH$ the following
conditions are equivalent
\begin{enumerate}
\item[(1)] $P_{\{a\}}\varphi=\varphi$\,; \item[(2)]
$\mu_\varphi=\|\varphi\|^2\delta_a$\,; \item[(3)]
$A\varphi=a\varphi$\,.
\end{enumerate}
\end{prop}

\begin{proof}

(1)$\Rightarrow$(2). For $\Omega\in\cB$ we have
$\mu_\varphi(\Omega)=\|P_\Omega\varphi\|^2= \|P_\Omega
P_{\{a\}}\varphi\|^2$. Therefore,
$\mu_\varphi(\Omega)=\|\varphi\|^2$, if $a\in\Omega$ and
$\mu_\varphi(\Omega)=0$ otherwise.

(2)$\Rightarrow$(3). We have for a $z\in\res(A)$
\[
\langle\varphi|R(z;A)\varphi\rangle=\int\limits_\mR\,
\frac{d\mu_{\varphi}(x)}{x-z}=\frac{\|\varphi\|^2}{a-z}\,,
\]
hence, by polarization, $R(z;A)\varphi=(a-z)^{-1}\varphi$.

(3)$\Rightarrow$(1).
Indeed, $P_{\{a\}}=\chi_{\{a\}}(A)$ and
$\chi_{\{a\}}(a)=1$.
\end{proof}

As a corollary we have that if $a$ is an atom for a spectral
measure $\mu_\psi$, then $a\in\spec_{\rp\rp} A$. Indeed, if
$\mu_\psi(\{a\})>0$, then $\varphi=P_{\{a\}}\ne0$. On the other
hand, $P_{\{a\}}\varphi=\varphi$.

\begin{prop} $\cH_\rp$  is the
orthogonal direct sum $\cH_{\rp\rp}$ of the eigensubspaces of $A$,
and $\spec_\rp A=\overline{\spec_{\rp\rp} A}$.
\end{prop}

\begin{proof}
 It is clear that
$\cH_{\rp\rp}\subset\cH_{\rp}$. To show that
$\cH_{\rp\rp}\supset\cH_{\rp}$ it is sufficient to prove that if
$\psi\perp\cH_{\rp\rp}$, then $\mu_\psi$ has no atoms. Suppose that
$\psi\perp\cH_{\rp\rp}$ but $\mu_\psi(\{a\})>0$. Then
$\varphi=P_{\{a\}}\psi\ne0$. Further, $\varphi=P_{\{a\}}\varphi$,
therefore $\varphi\in\cH_{\rp\rp}$. On the other hand
$\langle\psi,\varphi\rangle=\langle\psi,P_{\{a\}}\psi\rangle=
\mu_\psi(\{a\})>0$.

It is clear that $\spec_{\rp\rp} A\subset\spec_\rp A$. Suppose
that $a\in\spec_\rp A$. Take $\varepsilon>0$, then $\mu_\psi(a-
\varepsilon,a+\varepsilon)>0$ for some $\psi\in\cH_\rp$. Hence,
there is an atom $s$ for $\mu_\psi$ such that $s\in (a-
\varepsilon,a+\varepsilon)$. It remains to remark that
$s\in\spec_{\rp\rp} A$.
\end{proof}

The considered classifications of spectra are related as follows.

\begin{itemize}
\item $\spec_{\rm dis} A\subset\spec_{\rp\rp}A$;

\item $\spec_{\rm ess} A$ is the union of the following three sets:

\begin{itemize}

\item[(1)] $\spec_{\rc} A$,

\item[(2)] $\{x\in\mR:\,x \text{ is a limiting point of
}\spec_{\rp\rp}A\}$,

\item[(3)] $\{x\in\spec_{\rp\rp}A:\,x \,\,\,\text{is of infinite
multiplicity}\}$.
\end{itemize}
\end{itemize}

\subsection{Spectral projections}

Let $x,y\in\mR$. In what follows we will use often the identities
\begin{equation}
      \label{eq-imR}
\begin{aligned}
\Im\big\langle \varphi\big|R(x+iy;A)\varphi\big\rangle&=\dfrac{1}{2i}
\Big[\langle\varphi|R(x+iy;A)\varphi\rangle-\langle R(x+iy;A)\varphi|\varphi\rangle
\Big]\\
&=\dfrac{1}{2i}\,\big\langle\varphi\big|\big[R(x+iy;A)-R(x-iy;A)\big]\varphi\big\rangle\\
&= y\,\langle\varphi|R(x-iy;A)R(x+iy;A)\big]\varphi\rangle\\
&= y\big\|R(x+iy;A)\varphi\big\|^2.
\end{aligned}
\end{equation}

The following Stone formulas for spectral
projections will be very useful, cf. Theorem 42 in~\cite{Jak}.
Let $-\infty<a<b<+\infty$ and
$\varphi\in\cH$, then
\begin{equation}
                        \label{SP1}
\begin{aligned}
\frac{1}{2}\left[P_{[a,b]}\varphi+P_{(a,b)}\varphi\right]
&=\lim\limits_{y\to+0}\,\frac{1}{2\pi
i}\int_a^b\big[R(x+iy;A)-R(x-iy;A)\big]\varphi\,dx\\
&=\lim_{y\to+0}\,\frac{1}{\pi}
\int\limits_a^b\big[\Im\,R(x+iy;A)\big]\,\varphi\,dx\\
&=\lim\limits_{y\to+0}\,\frac{y}{\pi} \int_a^b
R(x-iy;A)R(x+iy;A)\varphi\,dx\,.
\end{aligned}
\end{equation}

Since
$\mu_\varphi(\Omega)=\langle\varphi|P_\Omega(A)\varphi\rangle=\|
P_\Omega(A)\varphi\|^2$, we have for $a,b\in\mR\setminus{\rm
spec}_{\rp\rp}(A)$
\begin{equation}
                        \label{SP1m}
\begin{aligned}
\mu_\varphi((a,b))= \mu_\varphi([a,b])&=\lim\limits_{y\to+0}\,\frac{1}{\pi}
\int_a^b\Im\,\langle\varphi|\,R(x+iy;A)\varphi\rangle\,dx\\
&=\lim\limits_{y\to+0}\,\frac{1}{\pi}\,
\Im\int_a^b\,\langle\varphi|\,R(x+iy;A)\varphi\rangle\,dx\\
&=\lim\limits_{y\to+0}\,\frac{y}{\pi} \int_a^b
\|R(x+iy;A)\varphi\|^2\,dx\,.
\end{aligned}
\end{equation}

If $a\in\mR$ and $\varphi\in\cH$, then
$P_{\{a\}}(A)\varphi=-i\lim\limits_{y\to+0}yR(a+iy;A)\varphi$,
therefore $\mu_\varphi(\{a\})=\|P_{\{a\}}(A)\varphi\|^2=\lim\limits_{y\to+0}
y^2\|R(a+iy;A)\varphi\|^2$.

The following statement is known \cite{Jak, AM}
\begin{thm}\label{prop-jak} Let
$\varphi\in\cH$. For Lebesgue a.e. $x\in\mR$ there exists the
limit
\[
\langle\varphi|R(x+i0;A)\varphi\rangle:= \lim\limits_{y\to0+}
\langle\varphi|R(x+iy;A)\varphi\rangle;
\]
this limit is is finite and non-zero a.e. and, additionally,
using \eqref{eq-imR},
\begin{enumerate}
\item[(1)] $\mu^{\ra\rc}_\varphi=\pi^{-1}F_\varphi\,dx$\,, where
\[
F_\varphi(x)= \Im\,\langle\varphi|R(x+i0;A)\varphi\rangle=
\lim\limits_{y\to0+}\,y\|R(x+iy)\varphi\|^2\,.
\]
\item[(2)] $\mu^{\rs}_\varphi$ is concentrated on the set
$\{x\in\mR:\,\Im\,\langle\varphi|R(x+i0;A)\varphi\rangle=\infty\}$.
\end{enumerate}
Additionally, for $-\infty<a\le b<+\infty$ one has:
\begin{enumerate}

\item[(3)] $\mu_\varphi^{\ra\rc}([a,b])=0$ if and only if for
some $p$, $0<p<1$,
\[
\lim\limits_{y\to0+}\,\int_a^b\,
\left[\Im\,\langle\varphi|R(x+iy;A)\varphi\rangle\right]^p\,dx=0\,.
\]

\item[(4)] Assume that for some $p$, $1<p\le\infty$ one has
\[
\sup\,\{\|\Im\,\langle\varphi|R(\cdot\,+iy;A)\varphi\rangle\|_p:\,0<y<1\}
<\infty\,,
\]
where $\|\cdot\|_p$ stands for the standard norm in the space
$L^p([a,b])$. Then $\mu_\varphi^{\rs}((a,b))=0$.

\item[(5)] Let $(a,b)\cap\spec_{\rs}A=\emptyset$. Then
there is a dense subset $D\subset\cH$ such that
$\sup\{\|\Im\,\langle\varphi|R(\cdot\,+iy;A)\varphi\rangle\|_p:\,0<y<1\}
<\infty$ for every $p$, $1<p\le+\infty$, and every $\varphi\in D$.

\item[(6)] $\mu_\varphi^{\rp}((a,b))=0$ if and only if
\[
\lim\limits_{y\to0+}\,y\,\int_a^b\,
\left[\Im\,\langle\varphi|R(x+iy;A)\varphi\rangle\right]^2\,dx=0\,.
\]
\end{enumerate}
\end{thm}

\begin{lem}\label{lem30} Let $\theta$ be a smooth strictly
positive function on $[a,b]$ and $a,b\notin\spec_{\rp\rp}A$.
Then
\begin{multline}
        \label{eq-yt}
\lim_{y\to+0}\,\frac{1}{\pi}\,
\Im\int_a^b\,\langle\varphi|\,R(x+iy;A)\varphi\rangle\,dx\\
=\lim_{y\to+0}\,\frac{1}{\pi}\,
\Im\int_a^b\,\langle\varphi|\,R(x+iy\theta(x);A)\varphi\rangle\,dx\\
=\lim_{y\to+0}\,\frac{y}{\pi} \int_a^b
\|R(x+iy\theta(x);A)\varphi\|^2\theta(x)\,dx\,.
\end{multline}
\end{lem}

\begin{proof}
The second equality in \eqref{eq-yt} follows from \eqref{eq-imR},
so its is sufficient to prove the first equality only.

Rewrite the left-hand side of \eqref{eq-yt} as
\begin{equation}
                         \label{path}
\int_a^b\,\langle\varphi|\,R(x+iy;A)\varphi\rangle\,dx=
\int_{\ell(y)}\,\langle\varphi|\,R(\zeta;A)\varphi\rangle\,d\zeta\,,
\end{equation}
where the path  $\ell(y)$ is given in the coordinates
$\zeta=\xi+i\eta$ by the equations: $\xi=t$, $\eta=y$,
$t\in[a,b]$.
Consider another path $\lambda(y)$ given by $\xi=t$,
$\eta=y\theta(t)$, $t\in[a,b]$ and two vertical intervals:
$v_1(y)$: $\xi=a$, $\eta$ between $y$ and $y\theta(a)$ and
$v_2(y)$: $\xi=b$, $\eta$ between $y$ and $y\theta(b)$. Since the
integrand in \eqref{path} is an analytic function, we can choose
the orientation of the intervals $v_1(y)$ and $v_2(y)$  in such a
way that
\begin{multline}
                         \label{path1}
\int_{\ell(y)}\,\langle\varphi|\,R(\zeta;A)\varphi\rangle\,d\zeta=
\int_{\lambda(y)}\,\langle\varphi|\,R(\zeta;A)\varphi\rangle\,d\zeta\\
+\int_{v_1(y)}\,\langle\varphi|\,R(\zeta;A)\varphi\rangle\,d\zeta+
\int_{v_2(y)}\,\langle\varphi|\,R(\zeta;A)\varphi\rangle\,d\zeta\,.
\end{multline}
Suppose $\theta(a)\ge1$ (the opposite case is considered
similarly). Then
\[
\int_{v_1(y)}\,\langle\varphi|\,R(\zeta;A)\varphi\rangle\,d\zeta=
\int_{y}^{y\theta(a)}\,\langle\varphi|\,R(a+i\eta;A)\varphi\rangle\,d\eta\,.
\]
Let $\nu_\varphi$ be the spectral measure associated with $A$ and
$\varphi$, then by Fubini
\begin{multline*}
\Im\,\int_{y}^{y\theta(a)}\,\langle\varphi|\,R(a+i\eta;A)\varphi\rangle\,d\eta=
\Im\,\int_{y}^{y\theta(a)}\,\int_{\mR} \dfrac{d\nu_\varphi(t)}{t-a-i\eta}\,
d\eta\\
{}=\int_{y}^{y\theta(a)}d\eta\,\eta\int_{\mR}\frac{d\nu_\varphi(t)}
{(t-a)^2+\eta^2}= \frac{1}{2}\int_{\mR}\ln\frac{(t-a)^2+y^2\theta(a)^2}
{(t-a)^2+y^2}\,d\nu_\varphi(t)\,.
\end{multline*}
Using the estimate
\[
\ln\frac{(t-a)^2+y^2\theta(a)^2} {(t-a)^2+y^2}=
\ln\left(1+\frac{y^2(\theta(a)^2-1)}{(t-a)^2+y^2}\right)\le
2\ln\theta(a)\,.
\]
and the boundedness of $\nu_\varphi$ we obtain by the Lebesgue
majorization theorem
\begin{subequations}
\begin{equation}
       \label{eq-se1}
\lim_{y\to0+}\,\Im
\int_{v_1(y)}\,\langle\varphi|\,R(\zeta;A)\varphi\rangle\,d\zeta=
0\,.
\end{equation}
Exactly in the same way there holds
\begin{equation}
       \label{eq-se2}
\lim_{y\to0+}\,y
\int_{v_2(y)}\,\langle\varphi|\,R(\zeta;A)\varphi\rangle\,d\zeta=
0\,.
\end{equation}
\end{subequations}
On the other hand,
\begin{multline}
         \label{eq-I1I2}
\Im\int_{\lambda(y)}\,\langle\varphi|\,R(\zeta;A)\varphi\rangle\,d\zeta\\
= \Im\int_{a}^{b} \langle\varphi|\,R(x+iy\theta(x);A)\varphi\rangle \big(1+iy\theta'(x)\big)dx
=I_1(y)+iyI_2(y),
\end{multline}
where, by \eqref{eq-imR},
\begin{align*}
I_1(y)&:=\Im\int_{a}^{b} \langle\varphi|R(x+iy\theta(x);A)\varphi\rangle dx\\
&\equiv y\int_{a}^{b} \big\|R(x+iy\theta(x);A)\varphi\big\|^2dx,\\
I_2(y)&:=\Im\int_{a}^{b} \langle\varphi|R(x+iy\theta(x);A)\varphi\rangle\, \theta'(x)\,dx\\
&\equiv y\int_{a}^{b} \big\|R(x+iy\theta(x);A)\varphi\big\|^2\theta'(x)dx.
\end{align*}
Denoting $c=\max_{x\in[a,b]}\big|\theta'(x)\big|$ one immediately obtains
$|I_2(y)|\le c|I_1(y)|$. Therefore, passing to the limit $y\to0+$ in \eqref{eq-I1I2}
we arrive at
\[
\lim_{y\to0+}I_1(y)=
\lim_{y\to0+}\Im\int_{\lambda(y)}\,\langle\varphi|\,R(\zeta;A)\varphi\rangle\,d\zeta.
\]
Substituting the latter equality, \eqref{eq-se1}, and \eqref{eq-se2}
in \eqref{path1} results in \eqref{eq-yt}.
\end{proof}

\section{Spectra and spectral measures for self-adjoint extensions}\label{sec4}

\subsection{Problem setting and notation}\label{sec-spec}

In this section we return to self-adjoint extensions. Below
\begin{itemize}
\item $S$ is a densely defined symmetric operator in
$\cH$ with equal deficiency indices in a Hilbert space $\cH$,
\item $\cN_z:=\ker(S^*-z)$,
\item $(\cG,\Gamma_1,\Gamma_2)$ is a boundary triple for $S^*$,
\item $\Lambda$ is a self-adjoint \emph{operator} in $\cG$,
\item $H^0$ is the self-adjoint restriction of $S^*$ to $\ker\Gamma_1$,
\item $H_\Lambda$ is the self-adjoint restriction of $S^*$ to $\ker(\Gamma_2-\Lambda\Gamma_1)$; due the the condition on $\Lambda$,
$H_\Lambda$ and $H^0$ are disjoint, see Remark~\ref{disj}.
\item $R^0(z):=(H^0-z)^{-1}$ for $z\in\res H^0$,
\item $R_\Lambda(z):=(H_\Lambda-z)^{-1}$ for $z\in\res H_\Lambda$,
\item $\gamma$ is the Krein $\Gamma$-field induced by the boundary triple,
\item $Q$ is the Krein's $\cQ$-function induced by the boundary triple.
\end{itemize}
Recall that the resolvent are connected by the Krein resolvent formula
(theorem \ref{krein}):
\begin{equation} \label{res-krein}
R_\Lambda(z)=R^0(z)-\gamma(z)\big[Q(z)-\Lambda\big]^{-1}\gamma^*(\Bar z).
\end{equation}
We are interested in the spectrum of $H_\Lambda$ assuming that
the spectrum of $H^0$ is known. Theorem~\ref{thm-btgf}
and Eq.~\eqref{res-krein} above show
the equality
\begin{equation}
            \label{eq-spH}
\spec H_\Lambda\setminus\spec H^0=\Big\{ E\in\res H^0:\,0\in\spec
\big(Q(E)-\Lambda\big) \Big\}.
\end{equation}
We are going to refine this correspondence in order to distinguish between
different spectral types of $H_\Lambda$ in gaps of $H^0$.
Some of our results are close to that obtained in~\cite{BrMN}
for simple operators, but are expressed in different terms.

\subsection{Discrete and essential spectra}

The aim of the present subsection is to relate the discrete
and essential spectra for $H_\Lambda$ with those for $Q(z)-\Lambda$.

\begin{lem}\label{lem22b}
Let $A$ and $B$ be self-adjoint operators in $\mathcal{G}$, and $A$
be bounded and strictly positive, i.e. $\langle\phi,A\phi
\rangle\ge c \langle\phi,\phi\rangle$ for all $\phi\in\dom A$ with
some $c>0$. Then $0$ is an isolated eigenvalue of $B$ if and only
if $0$ is an isolated eigenvalue of $ABA$.
\end{lem}

\begin{proof}
Denote $L:=ABA$. Let $0$ is a non-isolated point of
the spectrum of $B$. Then there is $\phi_n\in \dom B$
such that $B\phi_n\to0$ and $\dist (\ker B,\phi_n)\ge
\epsilon>0$. Set $\psi_n=A^{-1}\phi_n$. Then $L\psi_n\to0$.
Suppose that $\lim\inf \dist (\ker L,\psi_n)=0$.
Then there are $\psi'_n\in \ker L$ such that
$\lim\inf\|\psi_n-\psi'_n\|=0$. It is clear that
$\phi'_n=A\psi'_n\in\ker  B$ and
$\lim\inf\|\phi_n-\phi'_n\|=\lim\inf\|A\psi_n-A\psi'_n\|=0$.
This contradiction shows that $\dist (\ker L,\psi_n)\ge
\epsilon'>0$ and $0$ is a non-isolated point of the spectrum of
$L$.

The converse follows by symmetry, as $A^{-1}$ is also positive definite.
\end{proof}

\begin{thm}\label{th22b}

For $E\in \res H^0$ the following assertions are equivalent:
\begin{enumerate}
\item[(1)] $E$ is an isolated point of the spectrum of
$H_\Lambda$; \item[(2)] $0$ is an isolated point of the spectrum
of $Q(E)-\Lambda$.
\end{enumerate}
Moreover, if one of these conditions is satisfied, then for $z$ in
a punctured neighborhood of $E$ there holds
\begin{equation}
                       \label{2.2b}
\Big\|\big(Q(z)-\Lambda\big)^{-1}\Big\|\le \dfrac{c}{|z-E|}
\text{ for some } c>0.
\end{equation}
\end{thm}

\begin{proof}
Clearly, one can assume that $E$ is real. Denote $Q_0:=Q(E)$,
$Q_1:=Q'(E)$. Both $Q_0$ and $Q_1$ are bounded self-adjoint
operators. By \eqref{Q2} there holds $Q_1=\gamma^*(E)\,\gamma(E)$,
therefore, $Q_1$ is positive definite. Take any $r<\dist(E,\spec H^0\cup\spec
H_\Lambda\setminus\{E\})$. For $|z-E|<r$ we have an expansion
\begin{equation}
                    \label{tb1}
Q(z)=Q_0+(z-E)Q_1+(z-E)^2S(z),
\end{equation}
where $S$ is a holomorphic map from a neighborhood of $E$ to $\bL(\cG,\cG)$.

$(1)\Rightarrow(2).$ Let $E$ be an isolated point of the spectrum
of $H_\Lambda$.
Since $E$ is an isolated point in the spectrum of $H_\Lambda$, the
resolvent $R_\Lambda(z)\equiv(H_\Lambda-E)^{-1}$ has a first order
pole at $z=E$, therefore, as follows from the resolvent
formula~\eqref{res-krein}, the function $z\mapsto\big(Q(z)-\Lambda\big)^{-1}$
also has a first order pole at the same point. Hence, we can suppose that for
$0<|z-E|<r$ there exists the bounded inverse $\big(Q(z)-\Lambda\big)^{-1}$
and, moreover, $\|(z-E)(Q(z)-\Lambda)^{-1}\|\le c$
for some constant $c>0$. This implies the estimate \eqref{2.2b}. By~\eqref{tb1}
we can choose $r$ small enough, such that
$Q_0-\Lambda+(z-E)Q_1$ has a bounded inverse for $0<|z-E|<r$.
Representing
\[
Q_0-\Lambda+(z-E)Q_1=
Q_1^{1/2}\Big(Q_1^{-1/2}\big(Q(E)-\Lambda\big)Q_1^{-1/2}+(z-E)I\Big)Q_1^{1/2}
\]
we see that $0$ is an isolated point in the spectrum
of $Q_1^{-1/2}\big(Q(E)-\Lambda\big)Q_1^{-1/2}$ and hence
of $Q(E)-\Lambda$ in virtue of lemma~\ref{lem22b}.

$(2)\Rightarrow(1).$ Conversely, let $0$ be an isolated point of
the spectrum of $Q(E)-\Lambda$ or, which is equivalent by
lemma~\ref{lem22b}, in the spectrum of
$T:=Q_1^{-1/2}\big(Q(E)-\Lambda\big)Q_1^{-1/2}$.
For sufficiently small $r$ and $0<|z-E|<r$ the operator
$M(z):=T+(z-E)I$ is invertible, and $\|(z-E)M^{-1}(z)\|\le c'$ for
these $z$ for some constant $c'$.
For the same $z$, the operator
$Q_0-\Lambda+(z-E)Q_1\equiv Q_1^{1/2}M(z)Q_1^{1/2}$ is also
boundedly invertible, and
$\Big\|(z-E)\big(Q_0-\Lambda+(z-E)Q_1\big)^{-1}\Big\|\le c''$.
Hence, we can chose $r$ such that
$Q(z)-\Lambda=Q_0-\Lambda+(z-E)Q_1+(z-E)^2S(z)$ is invertible for
$0<|z-E|<r$, which by \eqref{eq-spH} means that $z\notin\res
H_\Lambda$.
\end{proof}

Now we are able to refine the relationship \eqref{eq-spH} between the spectra
of $H^0$ and $H_\Lambda$. This is the main result of the subsection.
\begin{thm}\label{th-dis-ess}
The spectra of $H$ and $H_\Lambda$ are related by
\begin{equation}
         \label{eq-dess}
\spec\nolimits_{\bullet} H_\Lambda\setminus\spec H^0=\Big\{
E\in\res H^0:\, 0\in\spec\nolimits_{\bullet}
\big(Q(E)-\Lambda\big) \Big\}
\end{equation}
with $\bullet\in\{\mathrm{pp},\mathrm{dis},\mathrm{ess}\}$.
\end{thm}

\begin{proof} By theorem~\ref{thm-btgf}(1),
Eq. \eqref{eq-dess} holds for $\bullet=\mathrm{pp}$, moreover, the
multiplicities of the eigenvalues coincide in this case.
Therefore, by theorem~\ref{th22b}, the isolated eigenvalues of
finite multiplicities for $H_\Lambda$ correspond to the isolated
zero eigenvalues for $Q(z)-E$, which proves \eqref{eq-dess} for
$\bullet=\mathrm{dis}$. By duality this holds for the essential
spectra too.
\end{proof}

It is also useful to write down the spectral projector for $H_\Lambda$
corresponding to isolated eigenvalues lying in $\res H^0$.
\begin{prop}\label{th22e} Let $E \in\res H^0$
be an isolated eigenvalue of $H_\Lambda$. Then the eigenprojector
$P_\Lambda$ for $H_\Lambda$ corresponding to $E$ is given by
\[
P_\Lambda=\gamma(E)\big(Q'(E)\big)^{-1/2}\Pi
\big(Q'(E)\big)^{-1/2}\gamma^*(E),
\]
where $\Pi$ is the
orthoprojector on
$\ker (Q'(E)\big)^{-1/2}\big(Q(E)-\Lambda\big)\big(Q'(E)\big)^{-1/2}$
in $\cG$.
\end{prop}

\begin{proof}
Follows from the equality
$P_\Lambda=-{\rm Res}\,\big[R_\Lambda(z);\,z=E\big]$.
\end{proof}

\subsection{Estimates for spectral measures}
In this subsection we are going to obtain some information
on the absolutely continuous, singular continuous, and point spectra
of $H_\Lambda$ using the asymptotic behavior of $\big(Q(x+iy)-\Lambda\big)^{-1}$
for $x\in\mR$ and $y\to 0+$.
To do this, we need first an expression for the resolvent $R_\Lambda$
on the defect subspaces of $S$.

\begin{lem}\label{lem8}
Let $\zeta,z\in\mC\setminus\mR$, $z\ne \zeta$, and $g\in\dom \Lambda$.
For $\varphi=\gamma(\zeta)g$ there holds
$R_\Lambda(z)\varphi=
\dfrac{1}{\zeta-z}\,\Big[\varphi-\gamma(z)\big(Q(z)-\Lambda\big)^{-1}
\big(Q(\zeta)-\Lambda\big)g\Big]$.
\end{lem}

\begin{proof}
Substituting identities \eqref{Gam2} and \eqref{Gam6}
into \eqref{res-krein} we obtain:
\begin{align*}
R_\Lambda(z)\varphi&=R^0(z)\gamma(\zeta)g-\gamma(z)\big[Q(z)-\Lambda\big]^{-1}
\gamma^*(\bar z)\gamma(\zeta)g\\
&=R^0(z)\gamma(\zeta)g-\gamma(z)\big[Q(z)-\Lambda\big]^{-1}
\gamma^*(\bar \zeta)\gamma(z)g\\
&=\dfrac{\gamma(z)-\gamma(\zeta)}{z-\zeta}\,g-\gamma(z)\big[Q(z)-\Lambda\big]^{-1}
\dfrac{Q(z)-Q(\zeta)}{z-\zeta} \,g\\
&= \dfrac{1}{\zeta-z}\bigg[\gamma(\zeta)g-\gamma(z)\Big\{I-
\big[Q(z)-\Lambda\big]^{-1}\\
&\quad{}\times
\big(Q(z)-\Lambda+\Lambda-Q(\zeta)\big)\Big\}g\bigg]\\
&=\dfrac{1}{\zeta-z}\Big[\varphi-\gamma(z)\big(Q(z)-\Lambda\big)^{-1}
\big(Q(\zeta)-\Lambda\big)g\Big]\,.
\end{align*}
\end{proof}

\begin{thm}\label{prop31} Fix $\zeta_0\in\mC\setminus\mR$.
Let $g\in\dom\Lambda$; denote $h:=\big(Q(\zeta_0)-\Lambda\big)g$,
$\varphi:=\gamma(\zeta_0)g$, and let $\mu_\varphi$ be the spectral
measure for $H_\Lambda$ associated with $\varphi$.
\begin{enumerate}
\item[(1)] If $[a,b]\subset \res H^0\cap\mR$ and
$a,b\notin\spec_{\rp\rp}H_\Lambda$, then
\begin{multline*}
\mu_\varphi\big([a,b]\big)\equiv\big\|P_{[a,b]}(H_\Lambda)\varphi \big\|^2\\
=\lim_{y\to+0}\frac{y}{\pi}\int_a^b\dfrac{1}{|\zeta_0-x|^2}\big\|\big(Q'(x)\big)^{1/2}
\big(Q(x+iy) -\Lambda\big)^{-1}h\big\|^2\,dx\,.
\end{multline*}

\item[(2)] For a.e. $x\in \res H^0\cap\mR$ there exists the limit
\[
f(x):=\lim_{y\to+0}y\Big\|\big(Q'(x)\big)^{1/2}\big(Q(x+iy)-\Lambda\big)^{-1}h\Big\|^2,
\]
and the function $\displaystyle F(x):=\dfrac{1}
{\pi|\zeta_0-x|^2}f(x)$ is the Lebesgue density of the measure
$\displaystyle\mu^{\ra\rc}_\varphi$, i.e.
$\mu^{\ra\rc}_\varphi=F(x)\,dx$.

\item[(3)] For $a\in \res H^0\cap\mR$ the limit
\[
\lim_{y\to+0}y^2\Big\|\big(Q'(a)\big)^{1/2}\big(Q(a+iy)-\Lambda\big)^{-1}h\Big\|^2
\]
exists and is equal to $\mu^{\rp}_\varphi(\{a\})$.
\end{enumerate}
\end{thm}

\begin{proof} We start with proving item (2). Using
lemma~\ref{lem8} we get for $y>0$:
\[
R_\Lambda(x+iy)\varphi= \dfrac{1}{\zeta_0-x-iy}\,\varphi-
\dfrac{1}{\zeta_0-x-iy}\,\gamma(x+iy)\big[Q(x+iy)-\Lambda\big]^{-1}h,
\]
therefore
\begin{multline*}
\bigg|\,\Big\|\sqrt{y} R_\Lambda(x+iy)\varphi\Big\|-\Big\|\dfrac{\sqrt{y}}
{\zeta_0-x-iy}\varphi\Big\|\,\bigg|\\
\le\dfrac{\sqrt{y}}{|\zeta_0-x-iy|}\,\Big\|\gamma(x+iy)\big(Q(x+iy)-
\Lambda\big)^{-1}h\Big\|\\
\le\sqrt{y}\big\|R_\Lambda(x+iy)\varphi\big\|+\dfrac{\sqrt{y}}{|\zeta_0-x-iy|}
\|\varphi\|\,.
\end{multline*}
Hence, if $\sqrt{y}\|R_\Lambda(x+iy)\varphi\|$ has a limit (finite
or infinite) as $y\to+0$, then also
$\sqrt{y}\Big\|\gamma(x+iy)\big(Q(x+iy)-\Lambda\big)^{-1}h\Big\|$
does, and in this case
\begin{multline}
                             \label{3.1.1}
\lim_{y\to+0}\sqrt{y}\big\|R_\Lambda(x+iy)\varphi\big\|\\
=\dfrac{1}{|\zeta_0-x|}\,\lim_{y\to+0}\sqrt{y}\|\gamma(x+iy)\big(Q(x+iy)-\Lambda\big)^{-1}h\|\,.
\end{multline}
Let us show that, at fixed $x$, the finiteness of
the limit \eqref{3.1.1} is equivalent to
\begin{equation}
                             \label{3.1.2}
\sup_{0<y<1}\sqrt{y}\Big\|\big(Q(x+iy)-\Lambda\big)^{-1}h\Big\|<\infty\,.
\end{equation}

Indeed, since $\gamma(z)$ is a linear topological isomorphism on
its image and is analytic, for a given $x\in \res H^0$ there exists $c>0$ such
that $c^{-1}\|g\|\le \sup_{0<y<1}\big\|\gamma(x+iy)g\big\|\le
c\|g\|$ for all $g\in\cG$.
This shows that the conditions
\begin{gather*}
\lim_{y\to+0}\sqrt{y}\big\|\gamma(x)\big(Q(x+iy)-\Lambda\big)^{-1}h\big\|=+\infty\\
\intertext{and}
\lim_{y\to+0}\sqrt{y}\big\|\gamma(x+iy)\big(Q(x+iy)-\Lambda\big)^{-1}h\big\|=+\infty
\end{gather*}
are equivalent.
Assume now $\lim_{y\to+0}\sqrt{y}\Big\|\gamma(x+iy)\big(Q(x+iy)-\Lambda\big)^{-1}h\Big\|<+\infty$,
then for all $0<y<1$ one has
\begin{multline*}
\Big|\sqrt{y}\big\|\gamma(x+iy)\big(Q(x+iy)-\Lambda\big)^{-1}h\big\|-
\sqrt{y}\big\|\gamma(x)\big(Q(x+iy)-\Lambda\big)^{-1}h\big\|\,\Big|\\
\le \sqrt{y}\big\|\gamma(x+iy)\big(Q(x+iy)-\Lambda\big)^{-1}h
-\gamma(x)\big(Q(x+iy)-\Lambda\big)^{-1}h\big\|\\
\le c\big\|\gamma(x+iy)-\gamma(x)\big\|\,,
\end{multline*}
where $c=\sup_{0<y<1}\sqrt{y}\big\|\big(Q(x+iy)-\Lambda\big)^{-1}h\big\|<\infty$.
Thus, we have
\begin{equation}
                             \label{3.1.3}
\lim_{y\to+0}\sqrt{y}\Big\|R_\Lambda(x+iy)\varphi\Big\|=
\dfrac{1}{|\zeta_0-x|}\lim_{y\to+0}\sqrt{y}\Big\|\gamma(x)\big(Q(x+iy)-\Lambda\big)^{-1}h\Big\|\,.
\end{equation}
On the other hand there holds
\begin{multline*}
\Big\|\gamma(x)(Q(x+iy)-\Lambda)^{-1}h\Big\|^2\\
= \Big\langle\gamma^*(x)\gamma(x)\big(Q(x+iy)-\Lambda\big)^{-1}h\,\Big|\,
\big(Q(x+iy)-\Lambda\big)^{-1}h\Big\rangle,
\end{multline*}
and, due to identities $\gamma^*(x)\gamma(x)\equiv Q'(x)$ and
$\big\|\gamma(x)\big(Q(x+iy)-\Lambda\big)^{-1}h\big\|^2=
\big\|\big(Q'(x)\big)^{1/2}\big(Q(x+iy)-\Lambda\big)^{-1}h\big\|^2$,
item (2) follows from proposition~\ref{prop-jak}.

The proof of items (1) and (3) is completely similar to that for
item (2); in the case of (1) one should use the norm
\[
\|f\|_2=\Big(\int_a^b\|f(x)\|^2\,dx\Big)^{1/2}
\]
on the space $L^2([a,b];\cH)$ in the above estimates.
\end{proof}

Below we will use the notation
\[
\cH_0:=\Big(\,\bigcup_{\Im\zeta\ne0}\ker(S^*-\zeta)\Big)^\bot\,,\quad\quad
\cH_1:=\cH_0^\bot\,.
\]
For a subspace $L\subset\cG$ we write
$\cH_1(L):=\bigcup_{\Im\zeta\ne0}\gamma(\zeta)X_\zeta$
with $X_\zeta(L)=\big(Q(\zeta)-\Lambda\big)^{-1}L$.
Note that if $\Span L$ is dense in $\cG$, then also $X_\zeta(L)$
is, and the linear hull of $\cH_0\cup\cH_1(L)$ is dense in $\cH$.

If $\psi\in\cH_0$, then $\gamma^*(\zeta)\psi=0$ for all
$\zeta\in\mC\setminus\mR$.
By \eqref{res-krein}, it follows $R_\Lambda(\zeta)\psi=
R^0(\zeta)\psi$, and hence $\mu_\psi(\Omega)=0$ for all Borel sets
$\Omega\subset\res H^0\cap\mR$, where $\mu_\psi$ is the spectral measure
for $H_\Lambda$ associated with $\psi$.

\begin{prop}[cf. Theorem~2 from \cite{GM}]\label{cc1}
Let $a,b\in \res H^0$. Suppose that there exists a subset $
L\subset \cG$ with dense $\Span L$ such that
\[
\sup\Big\{\big\|\big(Q(x+iy)-\Lambda\big)^{-1}h\big\|:\,a<x<b,\,0<y<1\Big\}<\infty
\]
for all $h\in L$. Then $(a,b)\cap\spec H_\Lambda=\emptyset$.
\end{prop}

\begin{proof} We can assume that $a,b\notin\spec_{\rp\rp}H_\Lambda$;
otherwise we consider $(a,b)$ as the
union of a increasing sequence of intervals $(a_n,b_n)$, where
$a_n,b_n\notin \spec_{\rp\rp}H_\Lambda$.

It is sufficient to show that $P_{(a,b)}(H_\Lambda)\cH_1(L)=0$.
Let $\varphi\in \cH_1(L)$, then there is $g\in L$ and $\zeta\in\mC\setminus\mR$
such that $\varphi=\gamma(\zeta)\big(Q(\zeta)-\Lambda\big)^{-1}g$.
Using lemma \ref{lem8} with $z=x+iy$ we get
\begin{equation}
       \label{eq-rl1}
R_\Lambda(x+iy)\varphi=\frac{1}{\zeta-x-iy}\,\Big[
\varphi-\gamma(x+iy)\big(Q(x+iy)-\Lambda\big)^{-1}g\Big].
\end{equation}
Using \eqref{SP1m} we arrive at $P_{(a,b)}(H_\Lambda)\varphi=0$.
\end{proof}

\begin{prop}\label{prop32}
For any $x_0\in \res H^0\cap\mR$ the following two assertions are equivalent:
\begin{enumerate}
\item[(1)] $x_0\notin \spec H_\Lambda$; \item[(2)] there exist
$\varepsilon>0$ and a subset $L\subset \cG$ with dense $\Span L$
such that $(x_0-\varepsilon,x_0+\varepsilon)\subset \res H^0$ and
\[
\lim_{y\to+0}\,y\int_{x_0-\varepsilon}^{x_0+\varepsilon}
\Big\|\big(Q(x+iy)-\Lambda\big)^{-1}h\Big\|^2\,dx=0\,.
\]
for all $h\in L$.
\end{enumerate}
\end{prop}

\begin{proof}
The implication $(1)\Rightarrow(2)$ is
trivial. Let us prove $(2)\Rightarrow(1)$.

It is sufficient to show that
$\|P_{(x_0-\varepsilon,x_0+\varepsilon)}(H^\Lambda)\varphi\|=0$
for all $\varphi\in\cH_1(L)$. For a given
$\varphi\in\gamma(\zeta)X_\zeta(L)$ with $\Im\zeta\ne0$ we take
$h\in L$ such that $h=\big(Q(\zeta)-\Lambda\big)g$,
$\varphi=\gamma(\zeta)g$ for some $g\in\dom \Lambda$. Then the
equality
$\|P_{(x_0-\varepsilon,x_0+\varepsilon)}(H^\Lambda)\varphi\|=0$
follows from theorem~\ref{prop31}(1).
\end{proof}

\begin{prop}\label{cc2} Let $a,b\in \res H^0$.
Suppose that there exists a subset $ L\subset \cG$ with dense
${\rm span}\,L$ such that for all $h\in L$ and $x\in(a,b)$
there holds
\[
\sup\Big\{\sqrt{y}\big\|\big(Q(x+iy)-\Lambda\big)^{-1}h\big\|:\,0<y<1\Big\}<\infty\,.
\]
Then $(a,b)\cap\spec_{\rs} H_\Lambda=\emptyset$.
\end{prop}

\begin{proof} Let $\mu_\varphi$ be the spectral measure associated
with $\varphi$ and $H_\Lambda$. It is sufficent to show that $\mu^\rs_\varphi(a,b)=0$
for all $\varphi\in\cH_1(L)$. Writing any $\varphi\in\cH_1(L)$
in the form $\varphi=\gamma(\zeta)\big(Q(\zeta)-\Lambda\big)^{-1}g$
with $g\in L$ and $\Im\zeta\ne 0$ one arrives again at \eqref{eq-rl1}.
Therefore, for any $x\in(a,b)$ one has
$\sup_{y\in(0,1)}\sqrt{y}\|R_\Lambda(x+iy)\varphi\|<\infty$,
and $\supp\mu^\rs_\varphi\cap(a,b)=\emptyset$ by theorem~\ref{prop-jak}(2).
\end{proof}

\begin{prop}\label{prop33}
Let \it $x_0\in {\rm res} H^0\cap \mR$. Then the following assertions are
equivalent:
\begin{enumerate}
\item[(1)] $x_0\notin \spec_{\ra\rc} H^\Lambda$;
\item[(2)] there exist $\varepsilon>0$ and a subset $ L\subset \cG$ with
dense $\Span L$ such that
$(x_0-\varepsilon,x_0+\varepsilon)\subset \res H^0$ and
$\lim_{y\to+0}\sqrt{y}\big(Q(x+iy)-\Lambda\big)^{-1}h=0$
for all $h\in L$ and for a.e. $x\in(x_0-\varepsilon,x_0+\varepsilon)$.
\end{enumerate}
\end{prop}

\begin{proof} The proof of the implication
$(2)\Rightarrow(1)$ is completely similar to that for
proposition~\ref{prop32}, cf. theorem~\ref{prop31}(1)

To prove $(1)\Rightarrow(2)$ we take $\varepsilon>0$ such that
$(x_0-\varepsilon,x_0+\varepsilon)\cap \spec_{\ra\rc} H_\Lambda=\emptyset$.
According to theorem~\ref{prop31}(2) we have
\[
\lim_{y\to+0}y\Big\|\big(Q'(x)\big)^{1/2}\big(Q(x+iy)-\Lambda\big)^{-1}h\Big\|^2=0
\]
for all $h\in \cG$, and it is sufficient to note that
$\big(Q'(x)\big)^{1/2}$ is a linear topological isomorphism.
\end{proof}

\begin{prop}\label{prop34}
Let \it $x_0\in \res H^0$. Then the following assertions are
equivalent:
\begin{enumerate}
\item[(1)] $x_0\notin \spec_{\rp} H_\Lambda$;
\item[(2)] there exist $\varepsilon>0$ and a subset $ L\subset \cG$ with
dense $\Span L$ such that
$(x_0-\varepsilon,x_0+\varepsilon)\subset \res H^0$ and
$\lim_{y\to+0}y\big(Q(x+iy)-\Lambda\big)^{-1}h=0$
for all $h\in L$ and for every
$x\in(x_0-\varepsilon,x_0+\varepsilon)$.
\end{enumerate}
\end{prop}

\begin{proof}
Similar to the proof of proposition~\ref{prop33} using theorem~\ref{prop31}(3).
\end{proof}
Using propositions~\ref{prop33} and \ref{prop34} we get immediately
\begin{prop}\label{prop35}
Let \it $x_0\in \res H^0\cap\spec H_\Lambda$. If for every
$\varepsilon>0$ there exists $h\in\cG$ such that
\begin{itemize}
\item $\lim_{y\to+0}y\big(Q(x+iy)-\Lambda\big)^{-1}h=0$ for all
$x\in(x_0-\varepsilon,x_0+\varepsilon)$ and
\item
$\lim_{y\to+0}\sqrt{y}\big(Q(x+iy)-\Lambda\big)^{-1}h=0 $ for a.e.
$x\in(x_0-\varepsilon,x_0+\varepsilon)$,
\end{itemize}
then $x_0\in \spec_{\rs\rc} H_\Lambda$.
\end{prop}

\subsection{Special $\cQ$-functions}

In this subsection we assume that the expression $Q(z)-\Lambda$
in the Krein forumula \eqref{res-krein} has the following special
form:
\begin{equation}
        \label{eq-ql}
Q(z)-\Lambda= \dfrac{A-m(z)}{n(z)},
\end{equation}
where
\begin{itemize}
\item $m$ and $n$ are (scalar) analytic functions
at least in $\mC\setminus\mR$,
\item $A$ is a self-adjoint operator in $\cG$.
\end{itemize}
We assume that $m$ and $n$ admit analytic continuation to some
interval $(a,b)\subset\res H^0\cap\mR$, moreover, they both are
real and $n\ne 0$ in this interval.

Below, in subsections~\ref{sec-graphs} and~\ref{sec-array} we provide
examples where such a situation arises.
Our aim is to relate the spectral properties of $H_\Lambda$ in $(a,b)$
to the spectral properties of $A$.
In what follows we denote by $\cJ:=(\inf\spec A,\sup\spec A)$.

\begin{lem}\label{lem-mon}
If $n$ is constant, then $m$ is monoton in $(a,b)$.
If $n$ is non-constant and $m'(x)=0$ for some $x\in (a,b)$, then either
$m(x)<\inf\spec A$ or $m(x)>\sup\,\spec A$.
\end{lem}

\begin{proof}
For any $f\in\dom A$ consider the function
$a_f(x):=\dfrac{1}{n(x)}\Big\langle f\Big|\big(A-m(x)\big)f\Big\rangle$.
Using \eqref{Q2} we write
\[
c\|f\|^2\le \big\langle f\big|Q'(x)f\big\rangle\equiv
a'_f(x)
=-\dfrac{m'(x)}{n(x)}\|f\|^2-
\dfrac{n'(x)}{n^2(x)}\Big\langle f\Big|\big(A-m(x)\big)f\Big\rangle
\]
with some constant  $c>0$ which is independent of $f$.

For constant $n$ one has $n'\equiv0$ and
$-\dfrac{m'(x)}{n(x)}\ge c$, i.e. $m'\ne 0$.

If $n'\ne 0$ and $m'(x)=0$, then
$\dfrac{n'(x)}{n^2(x)}\Big\langle f\Big|\big(A-m(x)\big)f\Big\rangle\ge c\|f\|^2$
for any $f$, i.e. the operator $A-m(x)$ is either positive definite
or negative definite.
\end{proof}

\begin{lem}\label{lem41} Let $K$ be a compact subset of
$(a,b)\cap m^{-1}(\,\overline \cJ\,)$, then
there is $y_0>0$ such that for $x\in K$ and $0<y<y_0$ one has
\begin{equation}
                          \label{4.1.1}
\big(Q(x+iy)-\Lambda\big)^{-1}=
n(x+iy)\,L(x,y)\big[A-m(x)-iym'(x)\big]^{-1},
\end{equation}
where $L(x,y)$ is a bounded operator and $\|L(x,y)-I\|\to 0$
uniformly with respect to $x\in K$ as $y\to0$.
\end{lem}

\begin{proof} We have $\big(Q(x+iy)-\Lambda\big)^{-1}=
n(x+iy)\big(A-m(x+iy)\big)^{-1}$.
Further, $A-m(x+iy)=A-m(x)-iym'(x)+B(x,y)$,
where $\|B(x,y)\|=O(y^2)$ uniformly with respect to $x\in K$.
Since $m'(x)\ne0$ for $x\in K$ by lemma~\ref{lem-mon},
the operator $A-m(x)-iym'(x)$ has a bounded inverse defined everywhere,
and
\[
A-m(x+iy)=
\big(A-m(x)-iym'(x)\big)
\Big[1+(A-m(x)-iym'(x)\big)^{-1}B(x,y)\Big].
\]
It is easy to see that
$\big\|\big(A-m(x)-iym'(x)\big)^{-1}\big\|=O\big(|y|^{-1}\big)$
uniformly with respect to $x\in K$. Therefore, for sufficiently
small $y$,
\[
\big(A-m(x+iy)\big)^{-1}=
\big(1+B_1(x,y)\big)^{-1}
\big[A-m(x)-iym'(x)\big]^{-1}\,
\]
with $\|B_1(x,y)\|=O\big(|y|\big)$ uniformly with respect to $x\in
K$.
\end{proof}

\begin{lem}\label{lem42}
Fix $\zeta_0$ with $\Im\zeta_0\ne0$ and let
$h\in\cG$, $\varphi=\gamma(\zeta_0)\big(Q(\zeta_0)-\Lambda\big)^{-1}h$.
Denote by $\mu$ the spectral measure for the pair
$(H_\Lambda,\varphi)$ and by $\nu$ the spectral measure for the pair
$(A,h)$. There is a constant $c>0$ with the following property:
for any segment $K:=[\alpha,\beta]\subset(a,b)\cap m^{-1}(\,\overline \cJ\,)$
such that $\alpha,\beta\notin\spec_{\rp\rp}H_\Lambda$ there holds
$\mu(K)\le c\nu\big(m(K)\big)$.
\end{lem}

\begin{proof}
Note first that $m'\ne 0$ on $[\alpha,\beta]$.
To be definite, we suppose $m'>0$.
According to theorem~\ref{prop31}(1) and
lemma~\ref{lem41}, we have
\[
\mu(K)=\lim_{y\to+0}\frac{y}{\pi}\int_K\dfrac{n^2(x)}{|\zeta_0-x|^2}\,
\|\big(Q'(x)\big)^{1/2}\big(A-m(x)-iym'(x)\big)^{-1}h\big\|^2\,dx.
\]
Substituting $\xi:=m(x)$ and denoting $\tau(\xi):=m'\big(m^{-1}(\xi)\big)$
we arrive at
\begin{multline*}
\mu(K)=\lim_{y\to+0}\frac{y}{\pi}
\int_{m(K)}\dfrac{n\big(m^{-1}(\xi)\big)^2}{\tau(\xi)\cdot|\zeta_0-m^{-1}(\xi)|^2}\\
{}\times
\Big\|
\Big(Q'\big(\vartheta^{-1}(\xi)\big)\Big)^{1/2}
\big(A-\xi-iy\tau(\xi)\big)^{-1}h\Big\|^2\,
d\xi.
\end{multline*}
Since
\begin{multline*}
\int_{m(K)}\dfrac{n\big(m^{-1}(\xi)\big)^2}{\tau(\xi)\cdot|\zeta_0-m^{-1}(\xi)|^2}
\Big\|
\Big(Q'\big(\vartheta^{-1}(\xi)\big)\Big)^{1/2}
\big(A-\xi-iy\tau(\xi)\big)^{-1}h\Big\|^2\,
d\xi,\\
\le c\int_{m(K)}\big\|\big(A-\xi-iy\tau(\xi)\big)^{-1}h\big\|^2\,
d\xi\,,
\end{multline*}
where $c$ is independent of $K$, we obtain the result with the
help lemma~\ref{lem30}.
\end{proof}

Here is the main result of the subsection.

\begin{thm}\label{thm-equiv} Assume that the term $Q(z)-\Lambda$
in the Krein resolvent formula~\eqref{res-krein} admits the
representation~\eqref{eq-ql}, then
for any $x_0\in\spec H_\Lambda\cap(a,b)$
and any $\bullet\in\{\mathrm{dis}, \mathrm{ess}, \mathrm{pp},
\mathrm{p}, \mathrm{ac}, \mathrm{s}, \mathrm{sc} ,\mathrm{c}\}$
the conditions
\begin{description}
\item[\rm($\bullet$)] $x_0\in\spec_{\bullet} H_\Lambda$,
\item[\rm(m--\,$\bullet$)] $m(x_0)\in\spec_\bullet A$
\end{description}
are equivalent.
\end{thm}

\begin{proof}
For $\bullet=\mathrm{pp}, \mathrm{dis}, \mathrm{ess}$
see theorem~\ref{th-dis-ess}.
As $m$ is a homeomorphism, the same holds for $\spec_\rp\equiv\overline{\mathstrut \spec_{\rp\rp}}$.

For $\bullet=\mathrm{ac}$ use the following sequence of
mutually equivalent assertions:
\begin{itemize}
\item $m(x_0)\notin\spec_{\ra\rc}A$,
\item There is a neighborhood $V$ of $m(x_0)$ such that
$y\|(A-\xi-iy)^{-1})h\|^2\stackrel{y\to0+}{\longrightarrow}0$
for all $\xi\in V$ and $h\in \cG$ (use item 1 of theorem~\ref{prop-jak}),
\item There is a neighborhood $W$ of $x_0$ such that
$y\|(Q(x+iy)-\Lambda)^{-1})h\|^2\stackrel{y\to0+}{\longrightarrow}0$ for all
$\xi\in W$ and $h\in \cG$ (use lemma~\ref{lem41} and replace
$iy m'(x)$ at any fixed $x$  by $iy$),
\item $x_0\notin\spec_{\ra\rc} H_\Lambda$ (proposition~\ref{prop33}).
\end{itemize}

Assume now $m(x_0)\in\spec_{\rs\rc} A$. There exists
a neighborhood $V$ of $m(x_0)$ such that for some
$h\in\cG$ we have $\nu_h^{\ra\rc}(V)=\nu_h^{\rp}(V)=0$, where
$\nu$ stands for the spectral measure for $A$. Using lemma~\ref{lem41}
and theorem~\ref{prop-jak} one can see that there exists
a neighborhood $W$ of $x_0$ such that
$\lim_{y\to+0}y^2\|(Q(x+iy)-\Lambda)^{-1}h\|^2=0$ for all $x\in W$ and
$\lim_{y\to+0}y\|(Q(x+iy)-\Lambda)^{-1}h\|^2=0$ for a.e. $x\in W$.
By proposition~\ref{prop35} this means that $x_0\in\spec_{\rs\rc}(H_\Lambda)$.
Hence, we prove (m-sc)$\Rightarrow$(sc).
Since $\spec_{\rs}A=\spec_{\rp}A\cup\spec_{\rs\rc}A$, we prove
also that (m-s)$\Rightarrow$(s).

Let now $m(x_0)\notin\spec_{\rs}A$. To show that $x_0\notin\spec_{\rs}H_\Lambda$
it is sufficient to consider the case $m(x_0)\in\spec A\setminus\spec_{\rs} A$.
Then by theorem~XIII.20 from~\cite{RS4}, there exist a dense
subset $L\subset\cG$ and a neighborhood $V$ of $m(x_0)$
such that
\[
\sup\big\{\big\|(A-\xi-iy)^{-1}h\big\|:\,0<y<1,\,\,\xi\in
V\big\}<\infty
\]
for all $h\in L$. We can assume without loss of generality that
$m'(x_0)>0$, then by lemma~\ref{lem41} we have for a neighborhood
$W$ of $x_0$ and for some $y_0$, $y_0>0$,
\[
\sup\big\{\sqrt{y}\big\|\big(Q(x+iy)-\Lambda\big)^{-1}h\big\|: \quad 0<y<y_0,\quad
x\in W\big\}<\infty,
\]
and $x_0\notin\spec_{\rs} H_\Lambda$ by proposition~\ref{cc2}.
Thus, the equivalence (s)$\Leftrightarrow$(m-s) is proven.

Now we prove the impication (sc)$\Rightarrow$(m-sc). Assume that
$x_0\in\spec_{\rs\rc}(H_\Lambda)$ but $m(x_0)\notin\spec_{\rs\rc} A$.
Denote the spectral measure for $A$ by $\nu$ and that for
$H_\Lambda$ by $\mu$, then there is an interval $I$ containing $x_0$ such
that for $J=m(I)$ there holds: $\nu^{\rs\rc}_h(J)=0$ for
all $h\in\cG$. According to lemma~\ref{lem42}, if $X$ is a Borel
subset of $I$ such that $\nu_h\big(m(X)\big)=0$ for all $h$, then
also $\mu_\varphi(X)=0$ for all $\varphi\in\cH_1$.
In particular, let $X$ be a Borel subset of
$I$ of zero Lebesgue measure and containing no eigenvalues of $H_\Lambda$.
Then $m(X)$ is a Borel subset of
$J$ which contains no eigenvalues of $A$ and also has the
Lebesgue measure zero. Therefore $\nu_h\big(m(X)\big)=0$, and
hence $\mu_\varphi(X)=0$. We see, that the restriction of
$\mu_\varphi$ to $I$ is mutually singular with each singular
continuous measure on $I$. Hence, it is true for $\mu_\varphi$
with each $\varphi\in\cH$. This contradicts to the assumption
$x_0\in\spec_{\rs\rc} H_\Lambda$, and the implication
(sc)$\Rightarrow$(m-sc) is proven.

The equivalence (c)$\Leftrightarrow$(m-c) follows from
(sc)$\Leftrightarrow$(m-sc) and (ac)$\Leftrightarrow$(m-ac).\,
\end{proof}

We note that theorem~\ref{thm-equiv} may be considered as an abstract version
of the dimension reduction: we reduce the spectrum problem for self-adjoint extensions
to a spectral problem ``on the boundary'', i.e. in the space $\cG$.


\subsection{Spectral duality for quantum and combinatorial graphs}\label{sec-graphs}

We have already mentioned that the theory of self-adjoint extensions
has obvious applications in the theory of quantum graphs.
Here we are going to develop the results of the recent paper~\cite{Pan}
concerning the relationship between the spectra of quantum graphs
and discrete Laplacians using theorem~\ref{thm-equiv}. Actually, this problem was
the starting point of the work.

Let $G$ be a countable directed graph.
The sets of the vertices and of the edges of $G$ will be denoted by $V$ and $E$, respectively. We do not exclude multiple edges and self-loops.
For an edge $e\in E$ we denote by $\iota{e}$ its initial
vertex and by $\tau{e}$ its terminal vertex.
For a vertex $v$, the number of outgoing edges (outdegree) will be denoted
by $\outdeg v$ and the number of ingoing edges (indegree) will be denoted
by $\indeg v$. The degree of $v$ is $\deg v:=\indeg v+\outdeg v$.
In what follows we assume that the degrees of the vertices of $G$
are uniformly bounded, $1\le\deg v\le N$ for all $v\in V$, in particular,
there are no isolated vertices. Note that each self-loop at $v$
counts in both $\indeg v$ and $\outdeg v$.

By identifying each edge $e$ of $G$ with a copy of the segment $[0,1]$, such that
$0$ is identified with the vertex $\iota{e}$ and $1$ is identified with the vertex
$\tau{e}$, one obtain a certain topological space.
A magnetic Schr\"odinger operator in such a structure is defined as follows.
The state space of the graph is $\cH=\bigoplus_{e\in E}\cH_e$,
$\cH_e=L^2[0,1]$, consisting of functions $f=(f_e)$, $f_e\in \cH_e$.
On each edge consider the same scalar potential $U\in L^2[0,1]$.
Let $a_e\in C^1[0,1]$ be real-valued magnetic potentials on the edges $e\in E$.
Associate with each edge a differential
expression $L_e:=(i\partial +a_e)^2+U$. The maximal operator which can be associated with these differential expressions acts as
$(g_e)\mapsto(L_e g_e)$ on functions $g\in\bigoplus H^2[0,1]$. The integration by parts shows that this operator
is not symmetric, and it is necessary to introduce boundary conditions at the vertices to obtain a self-adjoint operator. The standard self-adjoint boundary
conditions for magnetic operators are
\begin{gather*}
g_e(1)=g_b(0)=:g(v)\quad \text{ for all } b,e\in E \text{ with } \iota{b}=\tau{e}=v,\\
\sum_{e:\iota{e}=v} \big(g'_e(0)-ia_e(0)g_e(0)\big)-\sum_{e:\tau{e}=v} \big(g'_e(1)-ia_e(1)g_e(1)\big)=\alpha(v) g(v),
\end{gather*}
where $\alpha(v)$ are real numbers, the so-called coupling constants.
The gauge transformation $g_e(t)=\exp\Big(i\displaystyle\int_0^ta_e(s)ds\Big)f_e(t)$ removes the magnetic
potentials from the differential expressions, $\big((i\partial +a_e)^2+U\big)g_e=-f''_e+Uf_e$,
but the magnetic field
enters the boundary conditions through the parameters $\beta(e)=\displaystyle\int_0^1 a_e(s)\,ds$ in the following way:
\begin{subequations}
  \label{eq-fv}
\begin{gather}
  \label{eq-fv1}
e^{i\beta(e)}f_e(1)=f_b(0)=:f(v)\quad \text{ for all } b,e\in E \text{ with } \iota{b}=\tau{e}=v,\\
  \label{eq-fv2}
f'(v):=\sum_{e:\iota{e}=v} f'_e(0)-\sum_{e:\tau{e}=v} e^{i\beta(e)}f'_e(1)=\alpha(v) f(v).
\end{gather}
\end{subequations}
The self-adjoint operator in $\cH$ acting as $(f_e)\mapsto (-f''_e+Uf_e)$ on functions
$(f_e)\in\bigoplus H^2[0,1]$ satisfying the boundary conditions~\eqref{eq-fv1}
and~\eqref{eq-fv2} for all $v\in V$ will be denoted by $H$. This is our central object.

To describe the spectrum of $H$ let us make some preliminary constructions.
We introduce a discrete Hilbert space $l^2(G)$
consisting of functions on $V$ which
are summable with respect to the weighted scalar product
$\langle f,g\rangle=\sum_{v\in V} \deg v \,\overline{f(v)}g(v)$.
Consider an arbitrary function $\beta:E\to\mR$ and consider
the corresponding discrete magnetic Laplacian in $l^2(G)$,
\begin{equation}
        \label{eq-qmz}
\Delta_G h(v)=\frac{1}{\deg v}\Big(\sum_{e:\iota{e}=v} e^{-i\beta(e)}h(\tau{e})+
\sum_{e:\tau{e}=v} e^{i\beta(e)}h(\iota{e})
\Big).
\end{equation}
This expression defines a bounded self-adjoint operator in $l^2(G)$.

Denote by $D$ the Dirichlet realtizetion of $-d^2/dt^2+U$ on the segment $[0,1]$,
$Df=-f''+Uf$, $\dom D=\{f\in H^2[0,1]:\,f(0)=f(1)=0\}$. The spectrum of $D$ is a discrete set of simple eigenvalues.

For any $z\in\mC$ denote by $s(\cdot;z)$ and $c(x;z)$ the solutions
to $-y''+Uy=zy$ satisfying $s(0;z)=c'(0;z)=0$ and $s'(0;z)=c(0;z)=1$.
Introduce an extension of $H$, $\Pi$, defined by
$\dom\Pi=\{f\in\bigoplus H^2[0,1]:\,\text{Eq.~\eqref{eq-fv1} holds}\}$
and $\Pi(f_e)=(-f''_e+Uf_e)$. The following proposition is proved in~\cite{Pan}.

\begin{prop}\label{prop-graf} The operator $\Pi$ is closed.
For $f\in\dom\Pi$ put
\[
\Gamma_1 f=\big(f(v)\big)_{v\in V},\quad
\Gamma_2 f=\Big(\dfrac{f'(v)}{\deg v}\Big)_{v\in V}
\]
with $f(v)$ and $f'(v)$ given by~\eqref{eq-fv}, then
$\big(l^2(G),\Gamma_1,\Gamma_2\big)$ is a boundary triple for $\Pi$.
The induced $\Gamma$-field $\gamma$ and $\cQ$-function $Q$
are of the form
\[
\big(\gamma(z)h\big)_e(x)=\dfrac{1}{s(1;z)}\,\Big[
h(\iota{e})\,\big(s(1;z)c(x;z)-s(x;z)c(1;z)\big)\\+e^{-i\beta(e)}h(\tau{e}) s(x;z)\Big],
\]
and
\[
Q(z)f(v)=\dfrac{1}{\deg v\, s(1;z)}\Big(\Delta_G
-\big[\outdeg v\, c(1;z)+ \indeg v\,s'(1;z)\big]\Big) f(v).
\]
\end{prop}
Now let us make some additional assumptions.
We will say that the \emph{symmetry condition} is satisfied if \emph{at least one} of the following properties holds:
$\indeg v=\outdeg v$ for all $v\in V$ \emph{or}
$U$ is even, i.e. $U(x)=U(1-x)$.

The following theorem provides a complete description
of the spectrum of the quantum graph $H$ outside
$\spec D$ in terms of the discrete Laplacian $\Delta_G$.

\begin{thm}
Let the symmetry condition be satisfied and the coupling constants $\alpha(v)$
be of the form $\alpha(v)=\dfrac{\deg v}{2}\,\alpha$, then
$\spec_\bullet \Lambda\setminus\spec D=\eta^{-1}(\spec_\bullet \Delta_G)\setminus\spec D$
for $\bullet\in\{
\mathrm{dis}, \mathrm{ess}, \mathrm{pp},
\mathrm{p}, \mathrm{ac}, \mathrm{s}, \mathrm{sc} ,\mathrm{c}
\}$, where $\eta(z)=\dfrac{1}{2}\Big(s'(1;z)+c(1;z)+\alpha s(1;z)\Big)$.
\end{thm}

\begin{proof}
Let the symmetry conditions be satisfied. If $U$ is even, then
$s'(1;z)\equiv c(1;z)$. If $\outdeg v=\indeg v$ for all $v$, then
$\outdeg v=\indeg v=\dfrac{1}{2}\,\deg v$. In both cases
one has $Q(z)=\dfrac{2\Delta_G-s'(1;z)-c(1;z)}{2s(1;z)}$
(see~\cite{Pan} for a more detailed discussion).
The operator $H$ itself
is the restriction of $\Pi$ to the functions $f$ satisfying
$\Gamma_2=\dfrac{\alpha}{2}\Gamma_1f$ with $\Gamma_{1,2}$ from proposition~\ref{prop-graf}.
The restriction $H^0$ of $S$ to $\ker\Gamma_1$ is nothing but the direct sum
of the operators $D$ over all edges. By~theorem~\ref{krein}, the resolvents of $H$
and $H_0$ are related by the Krein resolvent formula and, in particular,
the corresponding term $Q(z)-\Lambda$ has the form
$Q(z)-\Lambda=\dfrac{\Delta_G-\eta(z)}{s(1;z)}$, and we are in the situation of theorem~\ref{thm-equiv}.
\end{proof}

\subsection{Array-type systems}\label{sec-array}

Another situation in which theorem~\ref{thm-equiv} becomes useful
appears when the $\cQ$-function is of scalar type~\cite{ABMN},
i.e. when $Q(z)$ is just the multiplication by a certain complex
function; such functions are of interest in the invesre spectral
problem for self-adjoint extensions~\cite{Bra}. In this case the
representition~\eqref{eq-ql} holds for any self-adjoint operator
$\Lambda$, and one has
\begin{prop}\label{prop-sc} Let $Q$ be of scalar type, then for any
$\Lambda$ there holds
$\spec\nolimits_\bullet H_\Lambda\setminus\spec H^0=Q^{-1}(\spec\nolimits_\bullet \Lambda)\setminus\spec H^0$ with $\bullet\in\{\mathrm{dis}, \mathrm{ess}, \mathrm{pp},
\mathrm{p}, \mathrm{ac}, \mathrm{s}, \mathrm{sc} ,\mathrm{c}\}$.
\end{prop}
In other words, the nature of the spectrum of the ``perturbed''
operator $H_\Lambda$ in the gaps of the ``unperturbed'' operator $H^0$
is completely determined in terms of the parameter $\Lambda$.

Scalar type $\cQ$-functions arise, for example, as follows.
Let $\cH_0$ be a separable Hilbert space and $S_0$ be a closed symmetric operator in $\cH_0$ with the deficiency indices $(1,1)$.
Let $(\mC,\Gamma^0_1,\Gamma^0_2)$ be a boundary triple for the adjoint
$S_0^*$, and $\gamma_0(z)$ and $q(z)$ be the induced $\Gamma$-field and $\cQ$-function.
Let $D$ be the restriction of $S_0^*$ to $\ker \Gamma_1^0$; this
is a self-adjoint operator.

Let $\cA$ be a certain countable set.
Consider the operator $S:=\bigoplus_{\alpha\in \cA} S_\alpha$ in the space  $\cH:=\bigoplus_{\alpha\in \cA} \cH_e$, where $\cH_\alpha\simeq\cH_0$ and $S_\alpha=S_0$. Clearly,
$\big(l^2(\cA), \Gamma_1,\Gamma_2\big)$ with $\Gamma_1 (f_\alpha)=(\Gamma_1^0 f_\alpha)$
and $\Gamma_2 (f_\alpha)=(\Gamma_2^0 f_\alpha)$ becomes a boundary triple for $S^*$.
The induced $\Gamma$-field is $\gamma(z)(\xi_\alpha)=(\gamma_0(z)\xi_\alpha)$
and the $\cQ$-function is scalar, $Q(z)=q(z)\text{id}$.
It is worthy to note that the corresponding operator $H^0$,
which is the restriction of $S^*$ to $\ker\Gamma_1$, is just the direct
sum of the copies of $D$ over the set $\cA$ and, in
particular, $\spec H^0=\spec D$.
Proposition \ref{prop-sc} becomes especially useful if the spectrum
of $D$ is a discrete set, then the spectrum of $H_\Lambda$ is (almost) completely
determined in terms of the parameterizing operator $\Lambda$.

The models of the above type can be used for the construction of
solvable models for array of quantum dots and antidots. One of
pecularities of such arrays is that they involve the miscroscopic
properties of a single point as well as the macropscopic
properties of the whole system. We consider for technical simplicity
two-dimensional periodic arrays in a uniform magnetic field
orthogonal to the plane of the system. For a large class of such
models we refer to~\cite{GPP}.

Let $\ba_1$, $\ba_2$ be linearly independent vectors of $\mR^2$
and $\cA$ be the lattice spanned by them, $\cA:=\mZ\ba_1+\mZ\ba_2$.
Assume that each note $\alpha$ of the lattice is occupied by a certain object
(quantum dot) whose state space is $\cH_\alpha$ with a Hamiltonian $H_\alpha$
(their concrete form will be given later). We assume that all quantum dots are identical,
i.e. $\cH_\alpha:=\cH_0$, $H_\alpha=H_0$. The system is subjected
to a uniform field orthogonal to the plane.

In our case, the inner state space $\cH_0$ will be $L^2(\mR^2)$.
The Hamiltonian $H_0$ will be taken in the form
\[
H_0=-\dfrac{1}{2}\Big[
\big(\dfrac{\partial }{\partial x}+\pi i\xi y\big)^2+
\big(\dfrac{\partial }{\partial y}-\pi i\xi x\big)^2
\Big]+\dfrac{\omega^2}{2}\big(x^2+y^2\big).
\]
Here $\xi$ is the number of magnetic flux quanta thorugh a unit area segment of the plane,
and $\omega$ is the strength of the quantum dot potential. Note that the spectrum of $H^0$
is pure point and consists of the infinite degenerate eigenvalues $E_{mn}$,
\[
E_{mn}=\dfrac{1}{2}\,(n+m+1)\Omega+(n-m)\xi,\quad
\Omega:=2\sqrt{\pi^2\xi^2+\omega^2},\quad m,n\in\mZ,\quad m,n\ge 0.
\]

The Hamiltonian $H:=\oplus_{\alpha\in\cA}H_\alpha$, describe the array of non-interacting quantum dots. To take into account the interdot interaction we use the restriction-extnesion procedure. Namely denote by $S_\alpha$ the restriction of $H_\alpha$
to the functions vanishing at the origin. As we have shown in subsubsection~\ref{ss-ppm},
these operators are closed and have deficiency indices $(1,1)$.
Respectively, one can construct the corrsponding boundary triples for $S_\alpha^*$.
Namely, for $f_\alpha\in\dom S_\alpha^*$ we denote
\[
a(f_\alpha):=-\lim_{r\to 0} \dfrac{\pi}{\log |r|} f_\alpha(r),
\quad
b(f_\alpha):=\lim_{r\to 0}\Big[
f(r)+a( f_\alpha) \dfrac{1}{\pi} \log |r|
\big].
\]
Accoriding to the constructions of subsubsection~\ref{ss-ppm},
$(\mC,a,b)$ form a boundary triple for $S_\alpha^*$, and the corresponding $\cQ$-function
is
\[
q(z)=-\dfrac{1}{2\pi}\,\Big[
\psi\big(\dfrac12-\dfrac z\Omega\big)+\log \dfrac{\Omega}{2\pi}+2C_E
\Big],
\]
where $\psi$ is the logarithic derivative of the $\Gamma$ function
and $C_E$ is the Euler constant.

Respectively, the triple $\big(l^2(\cA),\Gamma_1,\Gamma_2\big)$ with
\[
\Gamma_1(f_\alpha):=\big(a(f_\alpha)\big),\quad
\Gamma_2(f_\alpha):=\big(b(f_\alpha)\big),
\]
is a boundary triple for the operator $S^*$,
$S:=\bigoplus S_\alpha$, and the induced $\cQ$-function
is the multiplication by $q(z)$.

The above defined operator $H$ corresponds exactly
to the boundary condition $\Gamma_1 f=0$.
For a self-adjoint operator $L$  in $l^2(\cA)$ denote
by $H_L$ the self-adjoint extension of $S$ corresponding to the boundary conditions
$\Gamma_2f=L\Gamma_1 f$. This operator will be considered
as a Hamiltonian of interacting quantum dots, and the way how
different nodes interact with each other is determined by the operator $L$.
To avoid technical difficulties we assume that $L$ is bounded.
Furthermore, $L$ must satisfy some additional assumptions
in order to take into account the nature of the problem.

First, any reasonable definition of a periodic system with magnetic field
must include the invariance under the magnetic translation group.
In our case this means that the matrix of $L$
in the standard basis of $l^2(\cA)$ satisfies
\[
L(\alpha,\alpha+\beta)=e^{\pi i \xi \alpha\wedge\beta} L(0,\beta)
\quad\text{for any}\quad \alpha,\beta\in\cA.
\]
Second, we assume that only the nearest neighbors
interact with each other, i.e.
\[
L(\alpha,0)=\begin{cases}
\lambda_1, & \alpha=\pm \ba_1,\\
\lambda_2, & \alpha=\pm \ba_2,\\
0, & \text{otherwise,}
            \end{cases}
\quad \lambda_1,\lambda_2\in\mR\setminus\{0\}.
\]
Roughly speaking, the above assumptions mean the following:
each node interact $\alpha$ with the four nearest nodes
$\alpha\pm \ba_j$, $j=1,2$, and the interaction is independent
of $\alpha$. For further analysis it is useful to idenitfy
$l^2(\cA)$ with $l^2(\mZ^2)$ by
$(f_{n_1\ba_1+n_2\ba_2})\sim \big(f(n_1,n_2)\big)$, $n_1,n_2\in\mZ$.
Then the operator $L$ acts as follows:
\begin{multline*}
L f(n_1,n_2)\equiv L(\eta)f(n_1,n_2)=
\lambda_1\big[e^{i\pi \eta n_2}f(n_1-1,n_2)+
e^{-i\pi \eta n_2}f(n_1+1,n_2)\big]\\
+\lambda_2\big[
e^{-i\pi \eta n_1}f(n_1,n_2-1) +e^{i\pi \eta n_1}f(n_1,n_2+1)\big],
\quad \eta=\xi \ba_1\wedge\ba_2.
\end{multline*}
This operator $L(\eta)$ is well-known and is called the
\emph{discrete magnetic Laplacian}, and using
proposition~\ref{prop-sc} we can transfer the complete spectral
information for $L$ to the Hamiltonian of quantum dots $H_L$. One
of interesting moments in the spectral analysis of $L$ is the
relationship with the almost Mathieu operator in the space
$l^2(\mZ)$~\cite{Shu},
\[
M(\eta,\theta) f(n)=
\lambda_{1}\big[f(n-1)+f(n+1)\big]
+2\lambda_2\cos\big(2\pi\eta n+\theta\big)\, f(n), \quad \theta\in[-\pi,\pi).
\]
In particular,
\[
\spec L(\eta)=\overline{\bigcup_{\theta\in[-\pi,\pi)} \spec M(\eta,\theta)}.
\]
Elementary constructions of the Bloch analysis show that
the spectrum of $L(\eta)$ is absolutely continuous and has a band structure.
At the same time, for irrational $\eta$ the spectrum of $M(\eta,\theta)$
is independent of $\theta$ and hence coincides with the spectrum of $L(\eta)$.
It was shown only recently that
the spectrum of $M(\eta,\theta)$ is a Cantor set for all irrational $\eta$
and non-zero $\lambda_1$, $\lambda_2$, see \cite{AJ}.
Using our analysis we can claim that, up to the discrete
set $\{E_{m,n}\}$ (a more precise analysis shows that these eigenvalues are all in the spectrum
of the array) we can transfer the spectral information for $L(\eta)$
to the array of quantum dots; in particular,
we obtain a Cantor spectrum for irrational $\eta$ due to the analyticity of the $\cQ$-function.

\section{Isolated eigenvalues}\label{sec5}

\subsection{Problem setting}

In the previous sections we have analyzed the part of the spectrum
of the ``perturbed'' operator $H_\Lambda$ lying in the resolvent
set of the ``unperturbed'' operator $H^0$. If $E\in\spec H^0$,
then, in general, it is difficult to determine whether or not
$E\in\spec H_\Lambda$. Nevertheless, if $E$ is an isolated
eigenvalue of $H^0$, then the question whether $E$ in the spectrum of $H_\Lambda$
becomes easier in comparison with the general case. (Examples
of subsections \ref{sec-graphs} and \ref{sec-array} show that
this situation is rather typical for applications.)
In this section we give a
necessary and sufficient condition for such an $E$ to be an isolated
eigenvalue of $H_\Lambda$ and completely describe the
corresponding eigensubspace of $H_\Lambda$ (theorem~\ref{isol}).
For simplicity, we consider only the case of \emph{bounded}
self-adjoint operator $\Lambda$ in $\cG$.

In addition to the notation given in subsection~\ref{sec-spec}, in
this section $\varepsilon^0$ denotes an eigenvalue of $H^0$ with
the eigensubspace $\cH^0$ (which can be infinite-dimensional),
$P^0$ denotes the orthoprojector on $\cH^0$. We denote by
$\cV(\varepsilon^0)$ the set of all open balls $\cO$ centered at
$\varepsilon^0$ and such that $\spec
H^0\cap\cO=\{\varepsilon^0\}$. By $\bG\bL(\cG)$ we denote the set
of bounded linear operators in $\cG$ having a bounded inverse. If
$\cO\in\cV(\varepsilon^0)$, then $\bK(\cO;\cG)$ denotes the space
of all analytic mappings $V:\, \cO \rightarrow \bG\bL(\cG)$ such
that $V(\varepsilon^0)=I$ and $V^*(\bar z)=V^{-1}(z)$ (the latter
condition is equivalent to the following one: $V(z)$ is a unitary
operator for $z\in\mR\cap\cO$).

\subsection{Auxiliary constructions}

Further we need the following lemma.

\begin{lem} \label{lem51}
For any $z,\zeta \in \res H^0$ there holds:

\begin{enumerate}
\item[(1)] $P^0 \cN_{z}=P^0\cN_{\zeta}$;

\item[(2)] $\cH^0\cap\dom H_\Lambda=\cH^0\cap\cN_{z}^{\bot}=
\cH^0\ominus\ran P^0\gamma(\zeta)$;

\item[(3)] $\ker \gamma^*(z)P^0\gamma(z)=\ker P^0\gamma(\zeta)$,
i.e., the restriction of $\gamma^*(z)$ to $\ran
 P_0\gamma(\zeta)$ is an injection.
In particular, $\dim\ran\gamma^*(z)P^0\gamma(z)=
\dim\ran P^0\gamma(z)$.
\end{enumerate}
\end{lem}

\begin{proof} (1)
Recall that $P^0=-i\lim_{\delta\to+0}\,\delta R^0(\epsilon^0+i\delta)$
in the weak operator topology. By \eqref{Gam2}, for any $\delta>0$ one has
\begin{equation}
                            \label{aa2.1}
\gamma(z)+(\epsilon^0+i\delta-z)R^0(\epsilon^0+i\delta)\gamma(z)=
\gamma(\zeta)+(\epsilon^0+i\delta-\zeta)R^0(\epsilon^0+i\delta)\gamma(\zeta).
\end{equation}
Multiplying \eqref{aa2.1} with $\delta$ and sending $\delta$ to $0$ we arrive at
\begin{equation}
          \label{eq-ppg0}
(\varepsilon^0-z)P^0\gamma(z)=(\varepsilon^0-\zeta)P^0\gamma(\zeta).
\end{equation}
Now it is sufficient to recall that $\cN_z=\ran\gamma(z)$
for all $z\in\res H^0$.

(2) Let $\phi\in\cH^0\cap\dom H_\Lambda$ and $\psi\in
\cN_z$. As $H^0$ and $H_\Lambda$ are disjoint, $\phi\in\dom S$
and $S\phi=\epsilon^0\phi$.
There holds $(\varepsilon^0- z)\langle \phi|\psi \rangle=
\big\langle (S-\overline z)\phi\big|\psi \big\rangle= \big\langle \phi\big|(S^*-z)\psi
\big\rangle=0$. Hence  $\phi\perp\cN_z$.

Conversely, let $\phi\in\cH^0\cap\cN_z^{\bot}$. By~\eqref{Gam5},
$\gamma^*(z)\phi=0$. As follows from the Krein resolvent
formula~\eqref{res-krein},  $(\varepsilon^0-\bar
z)^{-1}\phi=R^0(\bar z)\phi=R_\Lambda(\bar z)\phi\in\dom
H_\Lambda$. Hence, $\phi\in\dom H_\Lambda$, and the first equality
is proved. The second equality follows immediately from the
relations: (a) for any $\phi\in\cH^0$ and $\psi\in\cN_z$ one has
$\langle\phi|\psi\rangle=\langle \phi,P^0\psi\rangle$, (b)
$\cN_z=\ran\gamma(z)$, (c) $\ran P^0\gamma(z)=\ran
P^0\gamma(\zeta)$.

(3) Let $\gamma^*(z)P^0\gamma(\zeta)g=0$. By~\eqref{Gam5},
$P^0\gamma(\zeta)g\perp \cN_z$.
According to~\eqref{eq-ppg0}, $P^0\gamma(\zeta)g\perp \cN_\zeta$.
It follows from the second equality in item (2) that
$P^0\gamma(\zeta)g\perp\ran\,P^0\gamma(\zeta)$. Hence
$P^0\gamma(\zeta)g=0$.
\end{proof}

The item (3) of lemma~\ref{lem51} can be generalized as follows.

\begin{lem}\label{lem52}
Let $\epsilon_j$, $j=1,\ldots,m$, be distinct eigenvalues of
$H^0$, $P^j$ be orthoprojectors on the corresponding
eigensubspaces and
\[
P:=\sum\limits_{j=1}^m P_j.
\]
Then
$(I-P)\gamma(z)$ is an injection for any $z\in\res H^0$.
\end{lem}

\begin{proof} Let $(I-P)\psi=0$ where $\psi=\gamma(z)\phi$ for
some $z\in\res H^0$, $\phi\in \cG$. Then $\psi=P\psi\in \dom H^0$ and,
therefore, $H^0\psi=z\psi$. Hence $\psi=0$ and $\phi=0$.
\end{proof}

In what follows $z_0$ denotes a fixed number from $\res H^0$, $x_0:=\Re
z_0$, $y_0:=\Im z_0$, $L:=\gamma(z_0)$. Recall that $L$ is a linear topological
isomorphism on the deficiency subspace $\cN:=\cN_{z_0}\subset\cH$.

Since, by definition, $\gamma(z)=L+(z-z_0)R^0(z)L$ for any
$z\in\res H^0$, the point $\varepsilon^0$ is either a regular
point for $\gamma$ or a simple pole with the residue
\begin{equation}
                     \label{resgam}
\mathop{\rm
Res}[\gamma(z):\,z=\varepsilon^0]=(z_0-\varepsilon^0)P^0L\,.
\end{equation}
Similarly, as $Q(z)=C+(z-x_0)L^*L+(z-z_0)(z-\bar z_0)L^*R^0(z)L$,
with a bounded self-adjoint operator $C$ (see
proposition~\ref{prop-qu}), the point $\varepsilon^0$ is either a
regular point for $Q$ or a simple pole with the residue:
\begin{equation}
            \label{eq-45}
\mathop{\rm Res}[Q(z):\,z=\varepsilon^0]
=-|\varepsilon^0-z_0|^2L^*P^0L.
\end{equation}From the equality $\|P^0L\phi\|^2=\langle
L^*P^0L\phi|\,\phi\rangle$ one easily sees that $\ker P^0L=\ker
L^*P^0L$ (see also Lemma~\ref{lem51}(3)). In particular, $P^0L=0$
if and only if $L^*P^0L=0$, and there are simple examples where
$P^0L=0$. Moreover, the following lemma holds.

\begin{lem}\label{lem53}
Let $\cH_1$ and $\cH_2$ be two Hilbert spaces and
$A:\cH_1\rightarrow\cH_2$ be a bounded linear operator. Then
the two conditions below are equivalent:
\begin{itemize}
\item[(1)] $\ran A$ is closed;
\item[(2)] $\ran A^*A$ is closed.
\end{itemize}
In particular, $\ran P^0L$ is closed
if and only if $\ran L^*P^0L$ is closed.
\end{lem}

\begin{proof} Condition (1) is satisfied if and only if there is a
constant $c>0$ such that $\|A\phi\|\ge c\|\phi\|$ for all
$\phi\in(\ker A)^{\bot}$. On the other hand, condition (2) is
satisfied if and only if there is a constant $c'>0$ such that
$\langle A^*A\phi|\,\phi\rangle\ge c'\|\phi\|^2$ for all
$\phi\in(\ker A^*A)^{\bot}$. Since $\ker A^*A=\ker A$, we get the
result.
\end{proof}

Now we denote by $\cG_r:=\ker L^*P^0L\subset\cG$,
$\cG_1:=\cG^\bot$. The orthoprojectors of $\cG$ on $\cG_r$
(respectively, on $\cG_1$) are denoted by $\Pi_r$ (respectively,
by $\Pi_1$). If $A$ is a bounded operator in $\cG$, then we write
$A_r:=\Pi_r A\Pi_r$, and this will be considered as an operator in
$\cG_r$. If $z\in\res H^0$, then $\gamma_r(z)$ denotes the
operator $(I-P^0)\gamma(z)\Pi_r$ acting from $\cG_r$ to $\cH$ (to
avoid a confusion with the previous notation, we suppose without
loss of generality $\cG\ne\cH$). Further, we denote by $\cH_r$ the
subspace $(I-P^0)\cH$ and by $H^0_r$ the part of $H^0$ in $\cH_r$;
clearly, $\varepsilon^0\in \res H^0_r$, and both the mappings
$\gamma_r$ and $Q_r$ have analytic continuation to $\epsilon^0$.
Finally, denote
$\cG_3=\ker\big(Q_r(\varepsilon^0)-\Lambda_r\big)$, and $\cG_2=
\cG_r\ominus\cG_3$.

\begin{lem}\label{lem54} There exists a closed symmetric
densely defined restriction $S_r$ of
$H^0_r$ such that $\gamma_r$ is a Krein
$\Gamma$-field for the triple $(S_r,H^0_r,\cG_r)$, and $Q_r$ is a Krein
 $\cQ$-function associated
with this triple and $\gamma_r$.
\end{lem}

\begin{proof} We use proposition~\ref{prop-gamma1}.
Since $P^0$ and $R^0(z)$ commute for all $z\in\res H^0$, it is
clear that $\gamma_r$ satisfies the condition
\eqref{Gam2}. Further, $z_0$ belongs to $\res H^0_r$ and
$\gamma_r(z_0)=(I-P_0)L\Pi_r$.

Let us show that the subspace $\cN':=\ran \gamma_r(z_0)$ is
closed. Let $(\phi_n)\in\cG_r$ such that $\psi_n:=(I-P^0)L\phi_n$
converge to some $\psi\in\cH_r$. Since $\phi_n\in \cG_r$, one has
$L^*P^0L\phi_n=0$, hence $P^0L\phi_n=0$. On the other hand,
$L\phi_n\in\cN$ by definition of $L$. Denote the orthoprojector of
$\cH$ onto $\cN$ by $P$, then we have $P\psi_n=L\phi_n$, hence
$L\phi_n$ converge to $P\psi$. Therefore, the sequence
$(L^*L\phi_n)$ converges to $L^*P\psi$ in $\cG$. Since $L^*L$ is a
linear topological automorphism of $\cG$, there exists $\lim
\phi_n$ and this limit belongs to $\cG_r$ because $\cG_r$ is
closed. Thus, $\psi\in\cN'$ and $\cN'$ is closed.

By lemma~\ref{lem51}(3), $\gamma_r(z_0)$ is injective. By the
closed graph theorem, $\gamma_r(z_0)$ is a linear topological
isomorphism of $\cG_r$ onto $\cN'$.

Now we show that $\cN'\cap\dom H^0_r=0$. It is sufficient to show
that $\big((I-P^0)\cN\big)\cap\dom H^0=0$. Let
$\psi\in\big((I-P^0)\cN\big)\cap\dom H^0$. As $\psi\in(I-P^0)\cN$,
we have $\psi= \phi-P^0\phi$ for some $\phi\in\cN$. Since
$\psi,P^0\phi\in\dom H^0$, $\phi\in\dom H^0$. Hence $\phi=0$ and
$\psi=0$. Thus, by proposition~\ref{prop-gamma1}, there exists a
closed symmetric densely defined restriction of $H^0_r$ such that
$\gamma_r$ is a $\Gamma$-field for the triple $(S_r,H^0_r,\cG_r)$.

Since $Q(z)=C-iy_0L^*L+(z-\bar z_0)L^*\gamma(z)$
with a bounded self-adjoint operator $C$ in $\cG$ (proposition~\ref{prop-qu}),
we have
\begin{align*}
Q_r(z)&=\Pi_r C\Pi_r-iy_0\Pi_r
L^*L\Pi_r+(z-\bar z_0)\Pi_r L^*\gamma(z)\Pi_r\\
&=\Pi_r C\Pi_r -iy_0\Pi_r L^*(I-P^0)L\Pi_r\\
&\quad+(z-\bar z_0)\Pi_r L^*(I-P^0)\gamma(z)\Pi_r
-iy_0\Pi_r L^*P^0L \Pi_r\\
&\quad+(z-\bar z_0)\Pi_r L^*P^0\gamma(z)\Pi_r\,.
\end{align*}
Now we use the equations
\begin{equation}
                          \label{dop}
\Pi_r L^*P^0L\Pi_r=0\,\quad\quad \Pi_r L^*P^0\gamma(z)\Pi_r=0\,.
\end{equation}
The first one follows from definition of $\Pi_r$, to prove the
second one we note that $\gamma(z)=L+(z-z_0)R^0(z)L$, therefore
\[
\Pi_r
L^*P^0\gamma(z)\Pi_r=\dfrac{\varepsilon^0-z_0}{\varepsilon^0-z}\,\Pi_r
L^*P^0L\Pi_r=0\,.
\] From (\ref{dop}) we obtain
\[
Q_r(z)=C'-iy_0\,\gamma^*_r(z_0)\,\gamma_r(z_0)+(z-\bar z_0)\gamma^*_r(z_0)\,\gamma_r(z)\,,
\]
where $C'=\Pi_rC\Pi_r$ is a self-adjoint bounded
operator in $\cG_r$. Hence, $Q_r$ is the Krein
$\cQ$-function associated with the $\Gamma$-field $\gamma_r$.
\end{proof}

To prove the main result of the section we need the following
lemma.

\begin{lem}\label{lem55} Let $S$  be an analytic function in the disk
$\bD=\{z\in\mC:\,|z|<r\}$ with values in the Banach space of all
bounded linear operators $\bL(\cG)$ such that there is a bounded
inverse $S^{-1}(z)$ for all $z$ from the punctured disk
$\bD\setminus\{0\}$ and the function $S^{-1}(z)$ is meromorphic.
If $\ker\,S(0)=0$, then $S_0:=S(0)$ has the bounded inverse (and,
therefore, $S^{-1}$ has an analytic continuation to the point $0$
of the disk). If  $S_0$ is self-adjoint and $0$ is a pole at most
of first order for $S^{-1}(z)$, then $\ran S_0$ is closed, i.e.
there is a punctured neighborhood of $0$ which has no point of
$\spec S_0$.
\end{lem}

\begin{proof}
Consider the Laurent expansion
\[
S^{-1}(z)=\sum\limits_{n=-m}^{\infty}T_nz^n\,.
\]
where $m$ is a natural number. If $m\le 0$, the lemma is trivial.
Suppose $m>0$. Since $S(z)S^{-1}(z)=I$ for all $z$, we have
$S_0T_{-m}=0$. Let $\ker\,S_0=0$, then $T_{-m}=0$, and by
recursion, $T_n=0$ for all $n<0$. Then $S_0T_0=T_0S_0=I$ and the
first part of the lemma is proved.

Let now $m=1$. Then $S_0T_{-1}=0$ and $T_{-1}S_1+T_0S_0=I$, where
$S_1=S'(0)$. This implies $S_0T_0S_0=S_0$. Let $x\in\ran S_0$,
then $S_0T_0x=x$. Since $\ran S_0\subset (\ker S_0)^\bot$, there
is a linear operator $A:\ran S_0\rightarrow\ran S_0$ such that
$AS_0x=x$ for all $x\in\ran S_0$. From $S_0T_0x=x$ we have
$A=T_0$, i.e. $A$ is bounded. Hence, there is $c>0$ such that
$\|x\|\le c\|S_0x\|$ for all $x\in\ran S_0$ and hence for all
$x\in(\ker S_0)^\bot$.
\end{proof}

\begin{rem} If $0$ is a second order pole for
$S^{-1}(z)$, then the range of $S_0$ can be non-closed.
For example,
let $A$ be a self-adjoint operator in a Hilbert space $\cH$ such
that $\ran A$ is non-closed. Let $\cG=\cH\oplus\cH$, and $S(z)$ is
defined as follows
\[
S(z)=\left[
\begin{array}{cc}
A&z\\
z&0\\
\end{array}
\right]\,.
\]
Then
\[
S^{-1}(z)=\frac{1}{z^2}\left[
\begin{array}{cc}
0&z\\
z&-A\\
\end{array}
\right]\,.
\]
\end{rem}

\subsection{Description of eigensubspace}

\begin{thm}\label{isol}
Let $\varepsilon^0$ be an isolated eigenvalue of $H^0$ and
$\ran P^0L$ be closed. Then the following assertions are mutually
equivalent.
\begin{enumerate}
\item[(1)] There exists a punctured neighborhood of
$\varepsilon^0$ that contains no point of $\spec H_\Lambda$ (in
particular, if $\varepsilon^0\in\spec H_\Lambda$, then
$\epsilon^0$ is an isolated point in the spectrum of $H_\Lambda$).

\item[(2)] The operator $Q(z)-\Lambda$ has a bounded inverse for
all $z$ from a punctured neighborhood of $\varepsilon^0$.

\item[(3)] $\ran\big(Q_r(\varepsilon^0)-\Lambda_r\big)$ is closed.

\item[(4)] There is a punctured neighborhood of $0$ which contains
no point from the spectrum of the operator
$Q_r(\varepsilon^0)-\Lambda_r$.
\end{enumerate}

Let one of the condition $(1)$--$(4)$ be satisfied. Then the
eigensubspace $\cH^0_\Lambda:=\ker (H_\Lambda-\varepsilon^0)$ is
the direct sum, $\cH^0_\Lambda=\cH_{\rm old}\oplus \cH_{\rm new}$,
where $\cH_{\rm old}=\cH^0\cap\dom H_\Lambda =\cH^0\cap\dom S$,
$\cH_{\rm new}=\gamma_r(\varepsilon^0)\ker\big[Q_r(\varepsilon^0)-\Lambda_r\big]$ and
$\dim \cH^0\ominus\cH_{\rm old}=\dim \cG\ominus\cG_r$.
Therefore, $\varepsilon^0\in\spec H_\Lambda$
if and only if at least one of the following two
conditions is satisfied:
\begin{itemize}
\item  $\cH^0\cap\dom H_\Lambda\ne\{0\}$,
\item  $\ker\big[Q_r(\varepsilon^0)-\Lambda_r\big] \ne \{0\}$.
\end{itemize}
\end{thm}

\begin{rem}\label{rem1} Since
$\cH^0\cap\dom H_\Lambda=\cH^0\cap\dom S$, the component $\cH_{\rm
old}$ of $\ker(H_\Lambda-\epsilon^0)$ is independent of
$\Lambda$, i.e. this part is the same for all extensions of
$S$ disjoint to $H^0$. On the other hand, the component $\cH_{\rm new}$ depends on
$\Lambda$.
\end{rem}

\begin{rem} Clearly, $\ran P^0L$ is
closed, if the deficiency index of $S$ or $\dim \cH^0$  are finite
(this simple case is very important in applications of
theorem~\ref{isol}). To show that the assumptions are essential
for infinite deficiency indices, we provide here
an example when the range of $P^0L$ is not closed.

Let $\cH_k=l^2(\mN)$ for $k=0,1,\ldots$ and let
$(e^{(k)}_n)_{n\ge0}$ be the standard basis in $\cH_k$:
$e_n^{(k)}=(\delta_{mn})_{m\ge0}$. Denote by $H^0_k$ the
self-adjoint operator in $\cH_k$ which is determined by $H^0_k
e_n^{(k)}=(n+1/2)e_n^{(k)}$. Choose $a\in\cH_0$ such that
$\|a\|=1$, $\langle a|\,e_0^{(0)}\rangle=0$, $a\notin\cD(H^0_0)$,
and set $a^{(k)}=a$. Consider in $\cH_k$ the one-dimensional
subspace $N_k$ generated by $e_0^{(k)}+(k+1)a^{(k)}$. Fix
$z_0\in\mC\setminus\mR$. By Proposition~\ref{prop-gamma1} there
exists a symmetric restriction $S_k$ of $H^0_k$ such that
$\cN_{z_0}(S_k)=N_k$. Let now $\cH=\bigoplus\cH_k$, $H^0=\bigoplus
H^0_k$, $S=\bigoplus S_k$. Then the eigensubspace $\cH^0$ of $H_0$
corresponding to the eigenvalue $\varepsilon^0=1/2$ is the closed
linear span of $(e_0^{(k)})$, $k=0,1,\ldots$, and $\cN_{z_0}(S)$
is the closed linear span of $(e_0^{(k)}+(k+1)a^{(k)})$,
$k=0,1,\ldots$. We can choose $\cG:=\cN_{z_0}$, $\gamma(z_0)=L=I$
where $I$ is the identical embedding of $\cN_{z_0}$ into $\cH$. It
is clear, that the image of $P^0L$ is the set $M$ of all vectors
$x$ from $\cH^0$ having the form $x=\sum\lambda_k e_0^{(k)}$ where
$\sum(k+1)^2|\lambda_k|^2<\infty$. Obviously, $M$ is dense in
$\cH^0$ but $M\ne\cH^0$, hence $M$ is not closed.
\end{rem}

\begin{proof}[\bf Proof of theorem~\ref{isol}]
The equivalence $(1)\Leftrightarrow(2)$
follows from theorem~\ref{th22b}, and the equivalence $(3)\Leftrightarrow(4)$
is trivial.

Let us prove the implication $(1) \Rightarrow (3)$. Choose
$\cO\in\cV(\varepsilon^0)$ such that $Q(z)-\Lambda$ has a bounded
inverse for all $z\in\cO\setminus\{\varepsilon^0\}$ and for
$z\in\cO\setminus\{\varepsilon^0\}$ consider the mapping
$T(z)=(z-\varepsilon^0)\big(Q(z)-\Lambda\big)$. Note that
\begin{itemize}
\item  $T$ has an analytic continuation to  $\varepsilon^0$ by setting
$T(\varepsilon^0)=-|\varepsilon^0-z_0|^2L^*P^0L$, see Eq.~\eqref{eq-45}, and
\item $T$ has a bounded inverse in $\cO\setminus\{\varepsilon^0\}$.
\end{itemize}
Since the operator $L^*P_0L$ has the closed range, we can apply a
result of Kato (\cite{Kat}, Sections~VII.1.3 and VII.3.1).
According the mentioned result, there is a mapping $V$,
$V\in\bK(\cO;\,\cG)$, such that the operator $V(z)T(z)V^{-1}(z)$
has the diagonal matrix representation with respect to the
decomposition $\cG=\cG_1\oplus \cG_r$:
\begin{equation}
                   \label{2.6.1}
V^{-1}(z)T(z)V(z)=\left[
\begin{array}{cc}
\hat T_{11}(z)&0\\
0&\hat T_{rr}(z)
\end{array}
\right]\,.
\end{equation}
Because the left-hand side of Eq.~(\ref{2.6.1}) has a bounded
inverse for $z\in\cO\setminus\{\varepsilon^0\}$, the same is true,
in particular for the operator $S(z):=(z-\varepsilon^0)^{-1}\hat
T_{rr}(z)=\Pi_rV^{-1}(z)[Q(z)-\Lambda]V(z)\Pi_r$
considered in the space $\cG_r$.

Our next aim to prove that
\begin{equation}\label{aim}
\|S^{-1}(z)\|\le c|z-\varepsilon^0|^{-1}
\end{equation}
with a constant $c>0$ for all $z$ in a punctured neighborhood of
$\varepsilon^0$. For this purpose we consider together with the
decomposition $\cG=\cG_1\oplus\cG_r$ of the space $\cG$, the
decomposition $\cH=\cH_1\oplus\cH_r$, where $\cH_1=\cH^0$,
$\cH_r=(I-P^0)\cH^0$. In virtue of to the Krein resolvent
formula~\eqref{res-krein},
\begin{multline*}
(z-\varepsilon^0)R_\Lambda(z)=(z-\varepsilon^0)R^0(z)-(z-\varepsilon^0)^2\gamma(z)T^{-1}(z)
\gamma^*(\bar z)\\
 =(z-\varepsilon^0)R^0(z)-(z-\varepsilon^0)^2[\gamma(z)V(z)]V^{-1}(z)
T^{-1}(z)V(z)[\gamma(\bar z)V(\bar z)]^*\,.
\end{multline*}
Represent the operator $\gamma(z)V(z)$ according to the above
mentioned representations of $\cH$ and $\cG$ in the matrix form:
\begin{equation}
                   \label{aim2}
\gamma(z)V(z)=\left[
\begin{array}{cc}
\hat \gamma_{11}(z)&\hat \gamma_{1r}(z)\\
\hat \gamma_{r1}(z)&\hat \gamma_{rr}(z)\\
\end{array}
\right]\,.
\end{equation}
Since $(z-\varepsilon^0)R_\Lambda(z)$ and
$(z-\varepsilon^0)R^0(z)$ are analytic functions in a neighborhood
of $\varepsilon^0$, all the matrix term in
$[\gamma(z)V(z)]V^{-1}(z) T^{-1}(z)V(z)[\gamma(\bar z)V(\bar
z)]^*$ are also analytic in the same neighborhood. In particular,
we can chose $\cO$ in such a way that the function
\[
z\mapsto(z-\varepsilon^0)^2\left(\hat \gamma_{r1}(z)\hat
T^{-1}_{11}(z)\hat \gamma^*_{r1}(\bar z)+ \hat \gamma_{rr}(z)\hat
T^{-1}_{rr}(z)\hat \gamma^*_{rr}(\bar z)\right)
\]
is analytic in $\cO$. Since $\hat
T_{11}(\varepsilon^0)\equiv-|z_0-\varepsilon^0|^2L^*P^0L$ has a
bounded inverse in $\cG_1$, the function $\hat T^{-1}_{11}(z)$ is
analytic in a neighborhood of $\varepsilon^0$. Therefore, we can
chose $\cO$ such that $(z-\varepsilon^0)^2\hat \gamma_{rr}(z)\hat
T^{-1}_{rr}(z)\hat \gamma^*_{rr}(\bar z)$ is analytic in $\cO$.
Further $\hat \gamma_{rr}(\varepsilon^0)=\gamma_r(\varepsilon^0)$.
In virtue of Lemma~\ref{lem54} and definition of the
$\Gamma$-field, we can find a constant $c'>0$ such that
$\|\gamma_r(\varepsilon^0)g\|\ge c'\|g\|$ for all $g\in\cG_r$.
Therefore we can chose $\cO$ so small that $\|\hat
\gamma_{rr}(z)g\|\ge c''\|g\|$ for all $z\in\cO$, $g\in\cG_r$ with
some $c''>0$. Since $\hat \gamma^*_{rr}(\bar z)$ is an isomorphism
of $\ran\gamma_r(\bar z)$ on $\cG_r$, we see that
$(z-\varepsilon^0)^2\hat T^{-1}_{rr}(z)$ is bounded in a
neighborhood of $\varepsilon^0$. Hence, we obtain (\ref{aim}) in a
punctured neighborhood of $\varepsilon^0$. By Theorem~3.13.3 from
\cite{HP}, $S^{-1}(z)$ has at point $\varepsilon^0$ a pole of the
order $\le1$. Therefore, $(1) \Rightarrow (3)$ by
Lemma~\ref{lem55}.

Now we prove $(4)\Rightarrow(2)$. Choose
$\cO\in\cV(\varepsilon^0)$ such that $Q(z)-\Lambda$ has no
spectrum in  $\cO\setminus\{\varepsilon^0\}$. Moreover, we can use
again the representation \eqref{2.6.1}. Since
$V(z)=I+O(z-\varepsilon^0)$,
the function $S(z):=\Pi_r
V^{-1}(z)\Pi_r[Q(z)-\Lambda]\Pi_rV(z)\Pi_r$ has an analytic
continuation at $\varepsilon^0$ with the value
$S(\varepsilon^0)=Q_r(\varepsilon^0)-\Lambda_r$. To proceed
further, we need the following auxiliary result.

\begin{lem}\label{42} The operator
$S'(\varepsilon^0)$ is strictly positive on
$\ker[Q_r(\varepsilon^0)-\Lambda_r]$.
\end{lem}

\begin{proof}[\bf  Proof of lemma~\ref{42}] Since $V^{-1}(x)=V^*(x)$ for
$x\in\cO\cap\mR$, for the derivative of $S$ one has:
\begin{equation}
                       \label{der}
\begin{aligned}
S'(\varepsilon^0)=&\Pi_r (V')^*(\varepsilon^0)\Pi_r[Q(\varepsilon^0)-\Lambda]\Pi_r\\
&+\Pi_r[Q(\varepsilon^0)-\Lambda]\Pi_rV'(\varepsilon^0)\Pi_r+\Pi_r Q'(\varepsilon^0)\Pi_r
\end{aligned}
\end{equation}
(note that $\Pi_r Q(\varepsilon^0)$ and
$Q(\varepsilon^0)\Pi_r$ are well defined). Let
now $\phi\in\ker[Q_r(\varepsilon^0)-\Lambda_r]$.
Then we have from (\ref{der}) that
$\langle\phi|S'(\varepsilon^0)\phi\rangle=\langle\phi|Q'(\varepsilon^0)\phi
\rangle$. Since $S'(\varepsilon^0)$ is a self-adjoint operator, we
have that $S'(\varepsilon^0)\phi=Q'(\varepsilon^0)\phi$ on
$\ker\,[Q_r(\varepsilon^0)-\Lambda_r]$. Therefore,
by Lemma~\ref{lem54} and~\eqref{Q2},
\begin{equation}
                         \label{deri}
S'(\varepsilon^0)\phi=\gamma_r^*(\varepsilon^0)\gamma_r(\varepsilon^0)\phi
\text{ for all } \phi\in\ker\,[Q_r(\varepsilon^0)-\Lambda_r],
\end{equation}
hence $S'(\varepsilon^0)$ is strictly positive on
$\ker[Q_r(\varepsilon^0)-\Lambda_r]$.
\end{proof}

To prove the required implication $(4)\Rightarrow(2)$, it is now
sufficient to show that $S(z)$ has a bounded inverse in a
punctured neighborhood of $\varepsilon^0$. Since $S(z)$ is
analytic, it suffice to prove that the operator
$J(z):=S(\varepsilon^0)+S'(\varepsilon^0)(z-\varepsilon^0)$ has a
bounded inverse in a punctured neighborhood of $\varepsilon^0$
with the estimate $\big\|J(z)^{-1}\big\|\le
c|z-\varepsilon^0|^{-1}$. For this purpose we represent
$S'(\varepsilon^0)$ in the matrix form
\[
S'(\varepsilon^0)=\begin{bmatrix}
S'_{22}&S'_{23}\\
S'_{32}& S'_{33}
\end{bmatrix}
\]
according to the representation $\cG_r=\cG_2\oplus\cG_3$.
Then $J$ has the matrix representation
\[
J(z)= \begin{bmatrix}
S_0+(z-\varepsilon^0)S'_{22}&(z-\varepsilon^0)S'_{23}\\
(z-\varepsilon^0)S'_{32}&(z-\varepsilon^0)S'_{33}.
\end{bmatrix}
\]
where $S_0:=S(\varepsilon^0)$. By the assumption of item~(4),
$S_0$ has a bounded inverse in $\cG_2$, and by \eqref{deri} the
operator $S'_{22}$ has a bounded inverse in $\cG_3$. Now we use
the Frobenius formula for the inverse of a block-matrix \cite{HJ}:
\begin{equation}
                                        \label{Frob}
\begin{gathered}
\begin{bmatrix}
A_{11} & A_{12}\\
A_{21} & A_{22}
\end{bmatrix}^{-1}\\
=
\begin{bmatrix}
[A_{11}-A_{12}A_{22}^{-1}A_{21}]^{-1}&
A_{11}^{-1}A_{12}[A_{21}A_{11}^{-1}A_{12}-A_{22}]^{-1}\\
[A_{21}A_{11}^{-1}A_{12}-A_{22}]^{-1}A_{21}A_{11}^{-1} &
[A_{22}-A_{21}A_{11}^{-1}A_{12}]^{-1}
\end{bmatrix}
\end{gathered}
\end{equation}
which is valid if all the inverse matrices on the right-hand side
exist. Using~\eqref{Frob} it is easy to see that $J^{-1}(z)$
exists for all $z$ in a punctured neighborhood of $\varepsilon^0$
and obeys the estimate $\big\|J(z)^{-1}\big\|\le
c|z-\varepsilon^0|^{-1}$ with some $c>0$. Thus, the implication
$(4)\Rightarrow(2)$ and, hence, the equivalence of all the items
(1)~-- (4) are proven.

Now suppose that the conditions of items (1)~-- (4) are satisfied.
To determine the eigenspace $\cH^0_\Lambda$ we find the
orthoprojector $P^0_\Lambda$ on this space calculating the residue
of the resolvent, $P^0_\Lambda=-{\rm
Res}\,[R_\Lambda(z):z=\varepsilon^0]=P^0+ {\rm
Res}\,[M(z):z=\epsilon_0]$, where
$$
M(z):=\gamma(z)[Q(z)-\Lambda]^{-1}\gamma^*(\bar z)\,.
$$
Using the conditions of item~(4), we find
$\cO\in\cV(\varepsilon^0)$ and $V\in\bK(\cO,\cG)$ such that for
$z$ in $\cO\setminus\{\varepsilon^0\}$
\begin{equation*}
V^{-1}(z)[Q(z)-\Lambda]V(z)= \left[
\begin{array}{cc}
S_1(z)&0\\
0& S_r(z)
\end{array}
\right]\,,
\end{equation*}
according to the decomposition $\cG=\cG_1\oplus\cG_r$ where $S_1$
and $S_r$ have the following properties: $S_r$  is analytic in
$\cO$ with $S_r(\varepsilon^0)=Q_r(\varepsilon^0)-\Lambda_r$ and
\begin{equation}
            \label{P1}
S_1(z)=-|\varepsilon^0-z_0|^2\,\frac{L^*P^0L}{z-\varepsilon^0}+
F_1(z), \text{ where } F_1 \text{ is analytic in } \cO\,.\\
\end{equation}

Using Lemma~\ref{42}, we find a function $W\in\bK(\cO,\cG_r)$ such
that for $z$ in $\cO\setminus\{\varepsilon^0\}$ one has
\[
W^{-1}(z)S_r(z)W(z)= \left[
\begin{array}{cc}
S_2(z)&0\\
0& S_3(z)
\end{array}
\right]\,,
\]
according to the decomposition $\cG_r=\cG_2\oplus\cG_3$
where $S_2$ and $S_3$ have the properties:
\begin{gather}
          \label{P3}
\parbox{110mm}{$\ker\,S_2(\varepsilon^0)=0$ and
$S_2(\varepsilon^0)\phi=[Q_r(\varepsilon^0)-\Lambda_r]\phi$  for $\phi\in\cG_2$,}\\
          \label{P4}
\parbox{110mm}{$S_3$ is analytic in $\cO$ and has the form
$S_3(z)=(z-\varepsilon^0)T(z)$ where $T_0:=T(\varepsilon^0)$ is a
strictly positive operator in $\cG_3$.}
\end{gather}
Denote now
\[
U(z):= V(z) \left[
\begin{array}{cc}
I_1&0\\
0& W(z)
\end{array}
\right]\,,
\]
where the matrices are decomposed according to the representation
$\cG=\cG_1\oplus\cG_r$ and $I_1$ is the
identity operator on $\cG_1$. Further, denote $\hat
Q(z)=U^{-1}(z)[Q(z)-\Lambda]U(z)$, $\hat \gamma(z)=\gamma(z)U(z)$,
then $M(z)=\hat \gamma(z)\hat Q^{-1}(z)\hat\gamma^*(\bar z)$, and for $z\in\cO\setminus\{\varepsilon^0\}$
one has
\[
\hat Q^{-1}(z)=\left[
\begin{array}{ccc}
S^{-1}_1(z)&0&0\\
0& S^{-1}_2(z)&0\\
0&0&S^{-1}_3(z)\\
\end{array}
\right]\,.
\]
An important property of $\hat \gamma$ we need is follows
\begin{equation}
    \label{P5}
\Hat
\gamma(z)=\frac{z_0-\varepsilon^0}{z-\varepsilon^0}P^0LU(z)+(I-P^0)\gamma(z)U(z),
\end{equation}
and $(I-P^0)\gamma$ is analytic in $\cO$. Represent $M$ as the sum
$M(z)=A_1(z)+A_2(z)+A_3(z)$, where $A_j(z)=\hat
\gamma(z)\Pi_jS^{-1}_j(z)\Pi_j\hat\gamma^*(\bar z)$; here $\Pi_j$
denote the orthoprojectors of $\cG$ onto $\cG_j$, $j=1,2,3$.

It is clear from \eqref{P1}--\eqref{P5} that at the point
$z=\varepsilon^0$, the function $A_j(z)$ has a pole at most of
$j$-th order. Let
\[
A_j(z)=A_j^{(-j)}(z-\varepsilon^0)^{-j}+
A_j^{(-j+1)}(z-\varepsilon^0)^{-j+1}+\ldots
\]
be the Laurent expansion for $A_j$ at the point $\varepsilon^0$.
According to the definition of $A_j(z)$ and formulas
\eqref{P1}--\eqref{P5} we have
\[
A_j^{(-j)}=C_jB_jC_j^*\,,\quad\,
A_j^{(-j+1)}=C_jB_jD_j^*+D_jB_jC_j^*+C_jB'_jC_j^*\,,
\]
where
\begin{gather*}
C_j=(z_0-\varepsilon^0)P^0L\Pi_j,\quad B_1=|\varepsilon^0-z_0|^{-2}\,(\Pi_1L^*P^0L\Pi_1)^{-1},\\
B_2=(\Pi_2S(\varepsilon^0)\Pi_2)^{-1},\quad
B_3=(\Pi_3T_0\Pi_3)^{-1},
\end{gather*}
and $B'_j$, $C_j$, $D_j$ are some
bounded operators (we need no concrete form of them). By
definition of the spaces $\cG_j$, we have $\Pi_jL^*P^0L\Pi_j=0$
for $j=2,3$, and hence, $P^0L\Pi_j=0$ for the same $j$'s. As a
result we have that $A_2(z)$ has no pole at $z=\varepsilon^0$,
i.e.,
\begin{equation}
                        \label{ress}
{\rm Res}\,[A_2(z):\,z=\varepsilon^0]=0\,,
\end{equation}
and $A_3(z)$ has at this point a pole at least of first order.
Using~\eqref{P5} and taking into consideration $P^0L\Pi_3=0$, we
obtain
\begin{equation}
                        \label{res3}
\begin{aligned}
{\rm Res}\,[A_3(z):\,z=\varepsilon^0]=:P_3&=(I-P^0)\gamma(\varepsilon^0)\Pi_3T_0^{-1}\Pi_3\gamma^*(\varepsilon^0)(I-P^0)\\
&=\gamma_r(\varepsilon^0)\Pi_3T_0^{-1}\Pi_3\gamma^*_r(\varepsilon^0)\,.
\end{aligned}
\end{equation}
Now we have according to~\eqref{P1} and \eqref{P5}\begin{equation}
       \label{eq-A1z}
{\rm
Res}\,[A_1(z):\,z=\varepsilon^0]=:-P_1=-P^0L\Pi_1(\Pi_1L^*P^0L\Pi_1)^{-1}\Pi_1L^*P^0\,.
\end{equation}
As a result, we have from \eqref{ress}, \eqref{res3}, and
\eqref{eq-A1z} $P^0_\Lambda=P^0-P_1+P_3$.

Eq.~\eqref{eq-A1z} shows that $P_1$ is an orthoprojector with
$\ran P_1\subset \ran P^0$. Therefore, $P^0-P_1$ is an
orthoprojector on a subspace of $\cH^0$. Eq.~\eqref{res3} shows
that $\ran P_3\subset \ran (I-P^0)$, therefore $(P^0-P_1)P_3=0$.
Since $P_3$ is self-adjoint, $P_3(P^0-P_1)=0$. Using
$(P^0_\Lambda)^2=P^0_\Lambda$ we see that $P^2_3=P_3$, hence $P_3$
is an orthoprojector and $P_3\perp P^0$.

By Lemma~\ref{lem51}, $\ran(P^0-P_1)=\cH^0\cap\dom H_\Lambda\equiv\cH_\text{old}$. The
relation
$\ran P_3=\gamma_r(\varepsilon^0)\ker\big[Q_r(\varepsilon^0)-
\Lambda_r\big]\equiv\cH_\text{new}$
follows from~\eqref{res3} and the definition of $\cG_3$.
Theorem~\ref{isol} is proved.
\end{proof}

\section{Acknowledgments}

The work was supported in part by the Deutsche
Forshungsgemeinschaft (PA 1555/1-1 and 436~RUS~113/785/0-1), the
SFB 647 ``Space, Time, Matter'', and the German Aerospace Center
(Internationales B\"uro, WTZ Deutschland--Neuseeland NZL 05/001).
In course of preparing the manuscript the authors had numerous
useful discussions with Sergio Albeverio, Jussi Behrndt, Johannes
Brasche, Yves Colin de Verdi\`ere, Pavel Exner, Daniel Grieser, Bernard Helffer,
Peter Kuchment,
Hagen Neidhardt, Mark Malamud, Boris Pavlov, Thierry Ramond,
Henk de Snoo, and Pavel \v{S}\v{t}ov\'{\i}\v{c}ek, which are gratefully acknowledged.

\end{document}